\documentclass[12pt, nofootinbib,showkeys,prd,aps]{revtex4-1}

\usepackage[brazil, english]{babel}
\usepackage[utf8]{inputenc}
\usepackage{graphicx}
\usepackage{epsfig}
\usepackage{amssymb}
\usepackage{amsmath} 
\usepackage{slashed}
\usepackage{physics}
\usepackage{subcaption}
\usepackage{caption}
\usepackage[unicode=true,bookmarks=false,breaklinks=false,pdfborder={0 0 1},colorlinks=true]
 {hyperref}
\hypersetup{
 citecolor=blue,linkcolor=blue,urlcolor=blue}

\graphicspath{%
    {converted_graphics/}
    {/}
}
\begin{document}

\title{Proton Structure Functions from Holographic Einstein-Dilaton Models}
\author{Ayrton da Cruz Pereira do Nascimento$^{1,2}$}
\email{ayrton@pos.if.ufrj.br}
\author{Henrique Boschi-Filho$^1$}
\email{boschi@if.ufrj.br}  
\author{Jorge Noronha$^2$}
\email{jn0508@illinois.edu}
\affiliation{$^1$Instituto de F\'\i sica, Universidade Federal do Rio de Janeiro, 21.941-909, Rio de Janeiro, RJ, Brazil\\
$^2$Illinois Center for Advanced Studies of the Universe \& Department of Physics,
University of Illinois Urbana-Champaign, Urbana, IL 61801-3003, USA}

\begin{abstract}
We study the proton structure functions $F_1$ and $F_2$ in the context of holography. We develop a general framework that extends previous holographic calculations of $F_1$ and $F_2$ to the case where the bulk geometry stems from bottom-up Einstein-Dilaton models, which are commonly used in the literature to describe some properties of QCD in the strong coupling regime. We focus on a choice of the dilaton potential that leads to a holographic model able to reproduce known lattice QCD results for the glueball masses at zero temperature and pure Yang-Mills thermodynamics above deconfinement. Once the parameters of the background holographic model are fixed, we introduce probe fermionic and gauge fields in the bulk {\it a la} Polchinski and Strassler to determine the corresponding structure functions. This particular realization of the model can
successfully describe the proton mass and provide results for $F_2$ at large $x$ in very good agreement with experimental data.  
\end{abstract}


\maketitle



\section{Introduction}
Deep inelastic scattering (DIS) is a powerful tool to understand the nucleon structure. In the late 1960s, experiments at the Stanford Linear Accelerator (SLAC) \cite{Breidenbach:1969kd,Bloom:1969kc} (see also \cite{Whitlow:1991uw} for a more recent account) used electron-proton ($ep$) scattering to obtain the first estimates for the structure functions of nucleons. Quantum chromodynamics (QCD) then emerged \cite{Gross:1973id,Politzer:1973fx} and was consolidated as the fundamental theory of strong interactions between quarks and gluons inside protons and neutrons. A QCD-based formalism known as the Dokshitzer-Gribov-Lipatov-Altarelli-Parisi (DGLAP) evolution equations \cite{Dokshitzer:1977sg,Gribov:1972ri,Altarelli:1977zs} was then developed to describe how parton distribution functions\footnote{Parton distribution functions describe the longitudinal momentum distribution of quarks and gluons (collectively called partons) inside nucleons, and are closely related to structure functions. However, in this work, we will only consider structure functions, not PDFs.} (PDFs) evolve with momentum transfer $Q$ and large Bjorken $x$. This allowed QCD to be used to analyze experimental data (see, e.g.
\cite{Gross:1976xt, Jaroszewicz:1982gr, Manohar:1992tz, Gelis:2010nm, Accardi:2012qut}). 
Later, experiments at HERA \cite{Habib:2010zz} enabled unpolarized $ep$ scattering to be analyzed at wider kinematic ranges. Future facilities, such as the Electron Ion Collider \cite{Accardi:2012qut}, are expected to greatly improve our understanding of nucleon structure in the next decade.

Because of confinement, perturbative calculations become unreliable at low transverse-momentum scales, where QCD becomes strongly coupled. This requires alternative approaches to describe QCD phenomena in that particular regime. Of particular relevance to this work are holographic models constructed within the framework of the gauge/gravity duality \cite{Maldacena:1997re,Gubser:1998bc,Witten:1998qj,Aharony:1999ti,Gubser:2002tv}, which have been extensively used to obtain insight into general phenomena displayed by strongly coupled gauge theories. In that context, following the pioneer work of Polchinski and Strassler \cite{Polchinski:2002jw}, different approaches to DIS in holography have been proposed using both \textit{top-down} and \textit{bottom-up} models (see, for example,  \cite{Borsa:2023tqr,Kovensky:2018xxa,Hatta:2007he, BallonBayona:2007rs,BallonBayona:2007qr,Cornalba:2008sp,Pire:2008zf,Albacete:2008ze,BallonBayona:2008zi,Gao:2009ze,Taliotis:2009ne,Yoshida:2009dw,Hatta:2009ra,Avsar:2009xf,Cornalba:2009ax,Bayona:2009qe,Cornalba:2010vk,Brower:2010wf,Gao:2010qk,BallonBayona:2010ae,Braga:2011wa,Koile:2013hba,Koile:2014vca,Gao:2014nwa,Capossoli:2015sfa,FolcoCapossoli:2020pks,Koile:2015qsa,Jorrin:2016rbx,Kovensky:2016ryy,Kovensky:2017oqs,Amorim:2018yod,Watanabe:2019zny,Jorrin:2020cil,Amorim:2021ffr,Mamo:2021cle,Tahery:2021xsj,Jorrin:2022lua,Bigazzi:2023odl,Mayrhofer:2024vnb}). In particular, such models are devised to reproduce confinement, among other properties of strong interactions.  

A particular set of holographic bottom-up models, originally investigated in \cite{Gursoy:2007cb, Gursoy:2007er}, involve the metric and a dynamical dilaton scalar field in 5-dimensional asymptotically anti-de Sitter (AdS) spacetime. In this dynamical framework, a bulk action describing the general interactions between the five-dimensional metric $g_{mn}$, dual to the energy-momentum tensor of the boundary gauge theory, and a scalar field $\phi$, responsible for breaking conformal invariance and dual to the glueball operator $\textit{Tr}\,F^2$, is used to determine the coupled set of equations of motion describing the bulk geometry and the scalar, which may then be solved to determine the properties of the dual gauge theory. This is different than other AdS/QCD approaches, such as \cite{Erlich:2005qh,Karch:2006pv}, where the metric is given and it is not necessarily a solution of equations of motion stemming from a well-defined action. The unknown in this bottom-up model is the dilaton potential $V(\phi)$, which is chosen to reproduce the properties of the problem at hand, such as linear confinement and glueball spectra, see \cite{Gursoy:2007cb, Gursoy:2007er} (see also Refs.\ \cite{Ballon-Bayona:2017sxa,Ballon-Bayona:2023zal,Ballon-Bayona:2024yuz} for related work on  
hadronic spectroscopy with or without chiral symmetry breaking at zero temperature). Once the potential is chosen, no other parameters are needed, and the corresponding results obtained from this approach can be seen as predictions of the model. Einstein-Dilaton models have also been very useful to investigate thermodynamic and transport properties of QCD at nonzero temperature \cite{Gursoy:2008za, Gursoy:2008bu,Gubser:2008ny,
Gubser:2008yx,
Gubser:2008sz,Noronha:2009ud,Gursoy:2010fj,Finazzo:2013efa,Finazzo:2014zga} and baryon density\footnote{Nonzero baryon density requires including in the holographic model a dynamical gauge field in the bulk.} \cite{DeWolfe:2010he, DeWolfe:2011ts, Rougemont:2015wca,Rougemont:2015ona,Critelli:2017oub,Ballon-Bayona:2020xls, Grefa:2021qvt,Grefa:2022sav,Hippert:2023bel}, for a comprehensive review, see \cite{Rougemont:2023gfz}. Finally, we remark that a generalization of the Einstein-Dilaton model to describe strong-coupled gauge theories in the Veneziano limit (where both the number of colors, $N$, and the number of flavors, $N_f$, are taken to infinity, though $N_f/N$ is finite), known as V-QCD, were developed in \cite{Jarvinen:2011qe} (see \cite{Deng:2025fpq} for an application in the context of nuclear structure).

In this work, we present for the first time a framework to study DIS of (unpolarized) targets in Einstein-Dilaton models. The background fields are obtained by solving the equations of motion following from the Einstein-Dilaton model. We follow Polchinski and Strassler \cite{Polchinski:2002jw} and introduce probe fermionic and gauge fields in this dynamically determined bulk. The probe fermionic fields in this model obey a Schr\"odinger-like equation with a confining potential, as demonstrated in this work. 
We then show how to use the solution of the equations of motion of the probe fields to calculate the $F_1$ and $F_2$ structure functions of the proton. As an application of this approach,  
we use the particular realization of the Einstein-Dilaton model worked out in Ref.\ \cite{Finazzo:2014zga}, which provided a good description of the tensor glueball spectrum \cite{Lucini:2001ej,Lucini:2010nv} and the thermodynamic properties \cite{Panero:2009tv} of large $N$ pure Yang-Mills theory. This way, the background is fixed and no new parameters are introduced in that regard. We numerically solve the equations of motion of the background and use this background to numerically solve the equations of motion of the probe fields needed for the holographic DIS calculation. This particular realization of
the model can successfully describe the proton mass and provide results for $F_2$ at
large $x$ that are in very good agreement with experimental data. The model also reproduces some other general features common to other holographic models previously used to study DIS.

This paper is organized as follows. In Section \ref{secdis}, we review how the structure functions of an unpolarized fixed target show up from the DIS amplitude through the definition of the hadronic tensor. In Section \ref{Happroach}, we briefly describe the holographic approach of \cite{Polchinski:2002jw} to compute the hadronic tensor through a probe interaction action in the bulk spacetime. Next, in Sect.\ \ref{defdis}, we present the details of the Einstein-Dilaton model we use and how we solve its equations of motion. Section \ref{probes} is dedicated to studying the fermionic and gauge probe fields in this geometry, in particular their equations of motion and their solutions. We also present the potentials and wave functions of the fermionic fields involved. Section \ref{disinteraction} is devoted to the evaluation of the interaction action between these fields and to the computation of the hadronic tensor and the further extraction of the desired structure function. In Section \ref{results} we present our numerical results for the proton structure function $F_2$, and the comparison to experimental data, for the specific choice of the dilaton potential used in this work. Finally, in \ref{conc} we discuss these results and draw conclusions about the use of Einstein-Dilaton models to model DIS problems at large $x$.

Our notation is the following: $g_{mn}$ is the five-dimensional metric of the asymptotic AdS spacetime with determinant $g = \det(g_{mn})$ and $\eta_{\mu\nu}$ is the Minkowski metric for the flat four-dimensional spacetime, with $\mu,\nu=0,1,2,3$. We use natural units $\hbar=c=k_B=1$.


\section{DIS structure functions}\label{secdis}

\begin{figure}[!ht]
\centering
  \includegraphics[scale = 0.45]{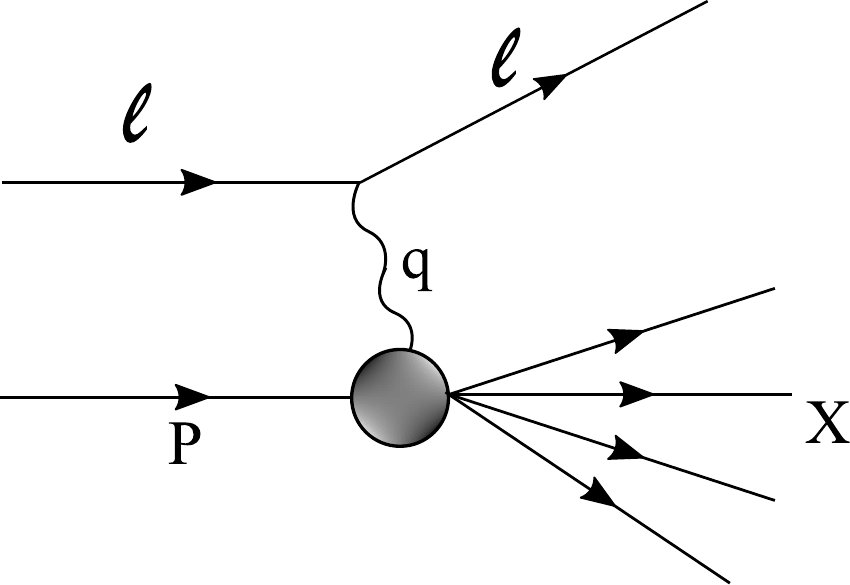} 
\caption{\sl Schematic description of a DIS process. A lepton with momentum $\ell$ interacts with a proton with momentum $P$ through the exchange of a virtual photon with momentum $q$. The final hadronic state is represented by multiple particles generically called $X$.}
\label{dis}
\end{figure}

Deep inelastic scattering plays an essential role in the description of nucleon structure. In particular, to study the internal structure of a proton, one can make use of lepton-proton inelastic scattering in which an incoming lepton hits the proton and interacts with it through the exchange of a virtual photon of momentum $q$. With energies high enough above $1$ GeV, the virtual photon can probe the internal substructure of a proton. This process is depicted in Fig.\ \ref{dis}. The state $X$ represents the final state, which is no longer the proton.

  %
%
%
%

%
\vskip0.5cm 

%
%
The internal structure of a proton can be described by the hadronic tensor in four dimensions, defined as 
\begin{equation}\label{f3}
W^{\mu \nu} = \frac{1}{4\, \pi} \sum_s \int d^4 y~ e^{i q.y} \langle P, s|\left[J^{\mu}(y)\,, J^{\nu}(0)\right]|P, s\rangle, 
\end{equation}
where $J^\mu$ represents the electromagnetic hadronic current and $s$ the spin of the proton. In this work we are going to deal with an unpolarized target, for which the most general tensor decomposition of $W^{\mu\nu}$ can be written as \cite{Manohar:1992tz}
%
%
%
\begin{equation}\label{f4}
W^{\mu \nu} = F_{1}(x,q) \left( \eta^{\mu \nu} - \frac{q^{\mu} q^{\nu}}{q^2} \right) + \frac{2x}{q^2} F_2 (x,q) \left( P^{\mu} + \frac{q^{\mu}}{2x} \right)\left( P^{\nu} + \frac{q^{\nu}}{2x} \right), 
\end{equation}
where $x = - {q^2}/{(2 P \cdot q)}\,$ is the momentum fraction carried by the parton (that is, the parameter Bjorken $x$), and $F_{1,2}(x,q)$ are the unpolarized DIS structure functions, with $0\le x \le 1$ and $q^2>0$. The lower the value of $x$, the more inelastic the process is. The limit $x\to 1$ corresponds to the elastic regime. 

Inserting a complete set of final states $\ket{X}$ in (\ref{f3}), one can write $W^{\mu\nu}$ as
        \begin{equation}\label{Wmunu}
            W^{\mu\nu}=\frac{1}{4}\sum_s\sum_{X} \delta^4\left(M_{X}^2-(P+q)^2\right)\bra{P,s}J^\mu(0)\ket{X}\bra{X}J^\nu(0)\ket{P,s}.
        \end{equation}
The task is then to compute the matrix element in Eq.\ (\ref{Wmunu}). In the following section, we describe the holographic prescription put forward by \cite{Polchinski:2002jw} to perform this calculation. Later on, in Sec.\ \ref{disinteraction}, we proceed with the full calculation of this quantity in our Einstein-Dilaton model.
\section{Holographic approach for determining structure functions}\label{Happroach}

As discussed in the previous section, the hadronic tensor is given in terms of the matrix element
\begin{equation}
    \langle X\left|J^\nu(0)\right|P,s\rangle.\label{matrixE}
\end{equation}
To calculate (\ref{matrixE}) using holographic tools, we follow steps similar to the approach used by  Polchinski and Strassler in \cite{Polchinski:2002jw}. The idea is to consider the matrix element above given by an interaction action defined as
\begin{equation}\label{PP}
    \langle X\left|J^\nu(0)\right|P,s\rangle=S_{\text{int}}= \int{dz\,d^4y\,\sqrt{-g}\,e^{F(z)}{\cal A}^m\, \bar{\Psi}_X\,\Gamma_m\,\Psi_i},
\end{equation}
where $z$ is the holographic coordinate, $g$ the determiant of the metric to be specified in next section, $F(z)$ is an effective scale dependent coupling  between the photon field ${\cal A}^m$ and the initial and final fermionic states $\Psi_i$ and $\Psi_X$, involved in the DIS process, respectively. For instance, in the particular case of the soft wall model, the function $F(z)$ is usually taken as the dilaton field \cite{Abidin:2009hr, Braga:2011wa}. All these fields live in the bulk of the five-dimensional asymptotic AdS space, and $\Gamma_m$ are the four-component Dirac matrices in those dimensions.

Polschinski and Strassler \cite{Polchinski:2002jw}, and other related works (see, e.g. 
 \cite{Borsa:2023tqr,Kovensky:2018xxa,Hatta:2007he,BallonBayona:2007rs,BallonBayona:2007qr,Cornalba:2008sp,Pire:2008zf,Albacete:2008ze,BallonBayona:2008zi,Gao:2009ze,Taliotis:2009ne,Yoshida:2009dw,Hatta:2009ra,Avsar:2009xf,Cornalba:2009ax,Bayona:2009qe,Cornalba:2010vk,Brower:2010wf,Gao:2010qk,BallonBayona:2010ae,Braga:2011wa,Koile:2013hba,Koile:2014vca,Gao:2014nwa,Capossoli:2015sfa,FolcoCapossoli:2020pks,Koile:2015qsa,Jorrin:2016rbx,Kovensky:2016ryy,Kovensky:2017oqs,Amorim:2018yod,Watanabe:2019zny,Jorrin:2020cil,Amorim:2021ffr,Mamo:2021cle,Tahery:2021xsj,Jorrin:2022lua,Bigazzi:2023odl,Mayrhofer:2024vnb})
compute (\ref{PP}) by first solving the equations of motion for the fields, derived from actions of the form
\begin{equation}
   \int dz\,d^4x \sqrt{-g}\,e^{F(z)}\,\mathcal{L},
\end{equation}
with or without the scale dependent  coupling ${F(z)}$, where the Lagrangian $\mathcal{L}$ includes the probe fermionic and gauge fields, and then inserts the solutions back into the matrix element to read off the structure functions. In previous works, the form of the asymptotic $AdS_5$ space metric was given, with properties such that some relevant QCD phenomenology could be reproduced, e.g., confinement. In this work, however, we go further and consider a new approach in this context. Instead of \emph{choosing} an ad hoc particular geometry to reproduce some QCD property, we start from an Einstein-Dilaton general action involving the metric and the scalar field and solve the equations of motion for these fields to determine the bulk geometry. Once the on-shell solution for the background is known, we solve the equations for the probe fields on this background to extract the structure functions. This way, one has greater systematic control about the properties of the bulk geometry, and its regime of validity (for example, concerning its behavior near singularities \cite{Gubser:2000nd}). In the following section, we derive the equations of motion of the background Einstein-Dilaton model, and explain how we solve them numerically.

\section{The Einstein-Dilaton model} \label{defdis}

In this section we give the details about the Einstein-Dilaton model used in this work.
We follow Ref.\ \cite{Finazzo:2014zga}, which investigated how the choice of the dilaton potential affects the Debye mass computed from Einstein-Dilaton models. The particular choice of potential used in this work corresponds to model B1 (see Sect.\ \ref{ScalarP}) in \cite{Finazzo:2014zga}, which led to a good description of the tensor glueball spectrum and the thermodynamic properties of large $N$ pure Yang-Mills theory. 

We start with a general Einstein-Dilaton action (with at most two derivatives) in the Einstein frame to describe the bulk
\begin{equation}\label{edaction}
    S=\frac{1}{\kappa_{5}^{2}}\int d^5x\sqrt{-g}\left[\mathcal{R}-\frac{1}{2}\left(\partial_{\mu} \phi\right)\left(\partial^{\mu}\phi\right)-V(\phi)\right],
\end{equation}
where $\kappa_{5}^{2}=8\pi G_5$, with $G_5$ being Newton's constant in five dimensions, coming from a constant coupling $F(z_0)$. We consider here the case at zero temperature, and follow the Gubser-Nellore Ansatz \cite{Gubser:2008ny} for the metric given by
\begin{equation}\label{gs}
ds^2 = g_{mn} dx^m dx^n= e^{ 2{A(\phi)}} \,\eta_{\mu \nu}dx^\mu dx^\nu+e^{2B(\phi)}d\phi^{2}  .\,\; 
\end{equation}
Above, the holographic radial coordinate is given by the scalar field $\phi$ itself. The functions $A(\phi)$ and $B(\phi)$ depend on the coordinate $\phi$, as well as the potential $V(\phi)$ to be specified in Sect.\ \ref{ScalarP}. In the following, we will derive the equations of motion and solve them in order to determine the functions $A(\phi)$ and $B(\phi)$ to fully specify the background spacetime (\ref{gs}). 

\subsection{Equations of motion}
The equations of motion obtained from (\ref{edaction}) are the Einstein field equations
\begin{equation}
    \mathcal{R}_{mn}-\frac{1}{2}g_{mn}\mathcal{R}=   8\pi G_5 T_{mn},\label{einstein}
\end{equation}
where $T_{mn}$ is the stress-energy tensor for the scalar field
\begin{equation}
    T_{mn}=\frac{1}{2}\partial_m \phi\partial_n \phi-\frac{1}{2} g_{mn} \left(\frac{1}{2}(\partial_\alpha \phi)^{2}+V(\phi)\right),
\end{equation}
and the equation for the scalar field is
\begin{equation}
    \partial_\mu\partial^\mu \phi-V^{\prime}=0.\label{scalar}
\end{equation}
Here $V^{\prime}=dV/d\phi$ (throughout this work primes will always denote a derivative with respect to $\phi$). 
Equations (\ref{einstein}) and (\ref{scalar}) with the metric given by (\ref{gs}) lead to the following equations of motion \cite{Finazzo:2014zga}
\begin{eqnarray}
    A^{\prime\prime}-A^{\prime}B^{\prime}+\frac{1}{6}=0\label{1}\\
    24(A^{\prime})^{2}-1+2e^{2B}V=0\label{2}\\
    4A^{\prime}-B^{\prime}- e^{2B}V^{\prime}=0\label{3}.
\end{eqnarray}

\subsection{Obtaining the geometry}
A schematic way to obtain the geometry is done here following the master equation procedure pursued in Ref.\ \cite{Gubser:2008ny}, adapted here for the case where the temperature is zero \cite{Finazzo:2014zga}. The idea is to obtain a first-order equation for $G(\phi)=A^{\prime}(\phi)$ and solve it to determine the functions $A$ and $B$. Using this definition and combining equations (\ref{2}) and (\ref{3}), we find
\begin{equation}
    \frac{V}{V^{\prime}}=\frac{24G^2-1}{-8G+2B^\prime}.
\end{equation}
Using (\ref{1}) to eliminate $B^{\prime}$ from this last equation, one obtains the following 
\begin{equation}
    G+\frac{V}{3V^\prime}=-\frac{6G^\prime G}{24G^2-6G^\prime G-1}.\label{Geq}
\end{equation}
In order to solve (\ref{Geq}) for a given potential $V(\phi)$, we have to specify a boundary condition for $G$.

For the boundary conditions for $G$, we must look at how the potential $V(\phi)$ behaves in deep in the bulk ($\phi\rightarrow\infty$) and at the boundary $(\phi\rightarrow 0)$. To obtain a confining model, the potential must display non-analytical behavior at large $\phi$, see \cite{Gursoy:2007er}. In our model, this requirement can be met \cite{Gubser:2008sz} if our potential has the following asymptotic form in the infrared
\begin{equation}
    V(\phi)\thicksim -\phi^{1/2}\, e^{\beta\phi},
\end{equation}
with $\beta$ being a positive real number. So, deep in the bulk we have $\lim_{\phi \to \infty}V(\phi)/V^{\prime}(\phi)=1/\beta$. Eq.\ (\ref{Geq}) then implies
\begin{equation}
    G(\phi\rightarrow\infty)=-\frac{1}{3\beta}.
\end{equation}

At the UV boundary ($\phi\rightarrow0$) the potential asymptotes to the case of a free massive scalar field plus a negative cosmological constant,
\begin{equation}
    V(\phi)\thicksim -\frac{12}{R^2}+\frac{1}{2}m^2\phi^2, 
\end{equation}
leading to an asymptotic AdS$_5$  spacetime with curvature $R$. Its UV asymptotic geometry is given by
\begin{align}
    A(\phi\rightarrow 0)&=A_0=\frac{\log\phi}{\Delta-4}+\mathcal{O}(1),\\
    B(\phi\rightarrow 0)&=B_0=-\log \phi+\mathcal{O}(1),
\end{align}
where $\Delta$ is the UV scaling dimension of the gauge theory operator associated with $\phi$. According to the holographic dictionary, this quantity is related to bulk parameters by the following relation
\begin{equation}
    \Delta(\Delta-4)=m^2R^2,\label{mass}
\end{equation}
with $m$ being the mass of the scalar field $\phi$ in the bulk AdS$_5$.

With all this at hand, we can obtain the functions $A(\phi)$ and $B(\phi)$. Recalling that $G(\phi)=A^{\prime}(\phi)$, we have
\begin{equation}
    A(\phi)=A_0+\int_{\phi_0}^{\phi}d\Tilde{\phi}\;G(\Tilde{\phi}).\label{A}
\end{equation}
Also, from (\ref{1}), we can write down the answer for $B$ as follows
\begin{equation}
    B(\phi)=B_0+\int_{\phi_0}^{\phi}d\Tilde{\phi}\;\frac{G^{\prime}(\Tilde{\phi})+1/6}{G(\Tilde{\phi})}.\label{B}
\end{equation}
By numerically solving the master equation for $G$ in Eq.\ (\ref{Geq}), and then performing the required integrals to determine $A(\phi)$ and $B(\phi)$, the complete geometry in \eqref{gs} can be found corresponding to the on-shell solution of the Einstein-Dilaton model. The last step in this direction is to choose the scalar potential to be used.


\subsection{Dilaton potential $V(\phi)$}\label{ScalarP}
\begin{figure}[h]
    \centering
    \includegraphics[width=0.8\linewidth]{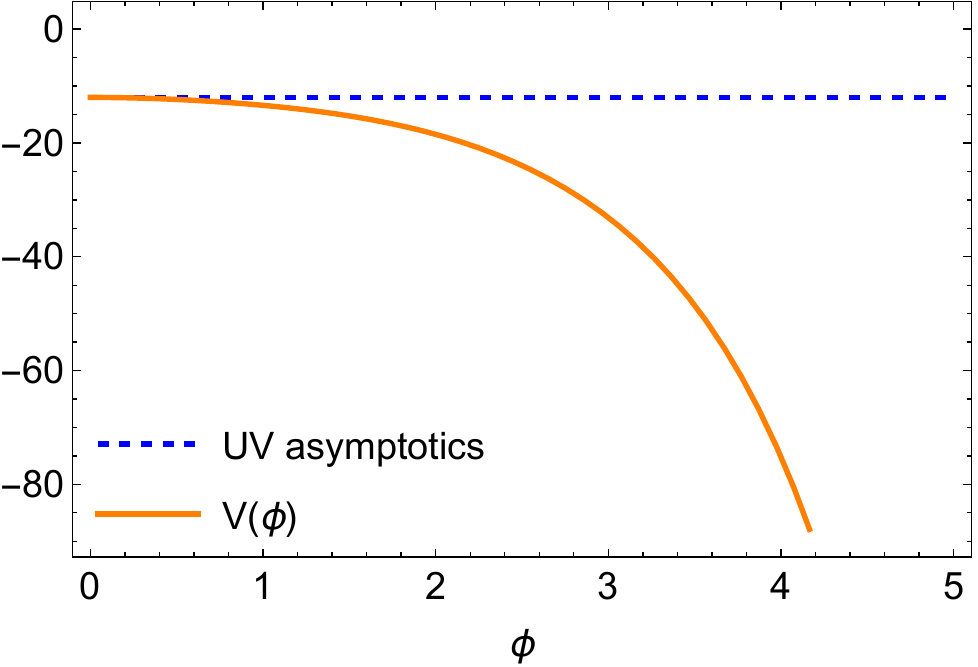}
    \caption{\sl The Einstein-Dilaton model potential, $V(\phi)$, defined in Eq.\ \eqref{Pot}, with $R=1$ and parameters fixed in Table \ref{table:1}, corresponds to the solid orange curve. Its UV asymptotic expression is plotted in dashed blue for comparison.}
    \label{V}
\end{figure}

The construction above and the choice of the scalar potential is inspired by the work in Ref.\ \cite{Finazzo:2014zga}, where the authors use an Einstein-Dilaton model to investigate the thermodynamics and the Debye mass of pure SU($N$) Yang-Mills theory. We use the B1 potential considered in \cite{Finazzo:2014zga}, which is given by
\begin{equation}
    V(\phi)=-\frac{12}{R^2}(1+a\phi^2)^{1/4}\cosh({\beta\phi})+\frac{b_2\phi^2+b_4\phi^4+b_6\phi^6}{R^2}.\label{Pot}
\end{equation}
This choice of potential interpolates between the UV behavior $V\thicksim -\frac{12}{R^2}+\frac{1}{2}m^2\phi^2$ and the IR behavior $V\thicksim -\phi^{1/2}\,e^{\beta\phi}$. Besides being confining at zero temperature, at finite temperature a potential with these properties implies that the gauge theory displays a first-order phase transition \cite{Gursoy:2008bu}.
The free parameters in (\ref{Pot}) are given in Table \ref{table:1} and the corresponding 
potential $V(\phi)$ in Fig.\ \ref{V}.

\begin{table}[h!]
\centering
\begin{tabular}{||c c c c c ||} 
 \hline
 $a$ & $\beta$ & $b_2$ & $b_4$ & $b_6$ \\ [0.5ex] 
 \hline
 $1$ & $\sqrt{2/3}$ & $5.5$ & $0.3957$ & $0.0135$\\ [1ex] 
 \hline
\end{tabular}
\caption{Parameters of the potential $V(\phi)$ (using units where the AdS radius $R=1$).}
\label{table:1}
\end{table}

\begin{figure}[h]
    \centering
    \includegraphics[width=0.5\linewidth]{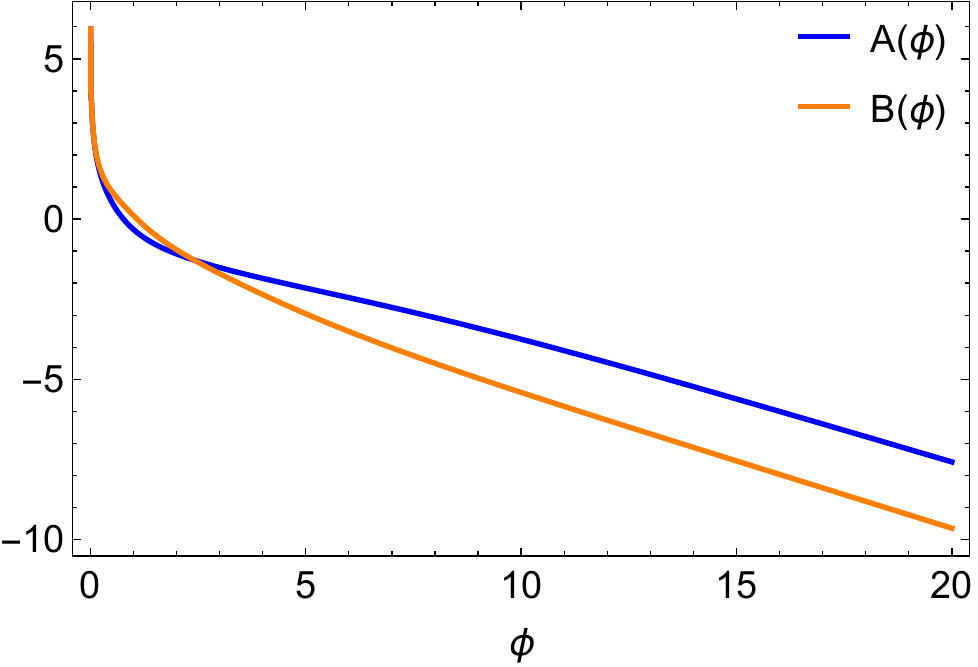}
    \caption{\sl The $A(\phi)$ (blue curve) and $B(\phi)$ (orange curve) functions, given by Eqs.\ \eqref{A} and \eqref{B}, respectively,  which fully determine the geometry of the asymptotic $AdS_5$ spacetime given by Eq.\ (\ref{gs}), after fixing the potential $V(\phi)$ in Eq.\ \eqref{Pot} and solving the master equation for $G(\phi)$ given in Eq.\ \eqref{Geq}.}
    \label{A&B}
\end{figure}

The UV ($\phi\rightarrow 0$) $m^2$ of the effective bulk action can be extracted from Eq.\ (\ref{Pot})
\begin{equation}
    m^2=\frac{2\,(-3a-6\beta^2+b_2)}{R^2}.
\end{equation}
With the values in Table \ref{table:1}, we obtain $m^2R^2=-3$, which satisfies the Breitenlohner-Freedman bound \cite{Breitenlohner:1982bm,Breitenlohner:1982jf,Mezincescu:1984ev}
\begin{equation}
    m^2R^2\geq-4.
\end{equation}
Therefore, using the AdS/CFT dictionary entry in Eq.\ \eqref{mass}, we find that the conformal dimension is $\Delta=3$. Following the reasoning laid out in \cite{Gubser:2008sz}, we see that in this model asymptotic freedom is replaced by conformal invariance in the UV regime.  Also, from now on, we use units where the AdS radius $R=1$. 

Although the potential in \eqref{Pot} was motivated by the investigation of finite temperature properties, the authors of \cite{Finazzo:2014zga} also applied it to the vacuum case. In fact, in this limit, they studied the spectra of scalar $J^{PC}=0^{++}$, tensor $J^{PC}=2^{++}$, and pseudoscalar $J^{PC}=0^{-+}$ glueballs. In particular, the authors found reasonable agreement between the tensor (and axial) glueball spectra and the lattice results for pure Yang-Mills. Inspired by these results in this confining vacuum model, here we employ this zero-temperature setup to study deep inelastic scattering.

With this potential at hand one can completely determine the functions $A(\phi)$ and $B(\phi)$ in Eqs.\ \eqref{A} and \eqref{B}, respectively, and use the resulting geometry to obtain the equations of motion and solutions for the fields involved in the DIS problem. This allows us to calculate the hadronic tensor $W_{\mu\nu}$ and predict $F_2$ in this configuration. The numerical solution for the profiles of $A$ and $B$ that determine the metric, with the parameters fixed in Table \ref{table:1}, are presented in Fig.\ \ref{A&B}. We have checked that we recover the results originally shown in \cite{Finazzo:2014zga}.


\section{Probe Fields}\label{probes}

In the DIS process, the particles involved are probe fields in the bulk geometry we have obtained in the last section. We now proceed to obtain their equations of motion. This is done considering the actions for each field, separately.


\subsection{Virtual photon field}\label{sect.:photon}

The virtual photon is described by the following action
\begin{equation}
    S_\mathrm{photon}=-\int d^5x\sqrt{-g}F^{mn}F_{mn},\label{photon}
\end{equation}
with 
\begin{equation}
    F^{mn}=\partial^m {\cal A}^n-\partial^n {\cal A}^m.
\end{equation}
The equations of motion from the action (\ref{photon}) are
\begin{equation}
    \partial_m\left[\sqrt{-g}F^{mn}\right]=0,\label{photon1}
\end{equation}
where 
\begin{align}
    \sqrt{-g}&=e^{4A+B}.
\end{align}

The case $m=\mu=0,1,2,3$ leads to the following equations 
\begin{eqnarray}
    \square {\cal A}^\nu-\partial_\mu\partial^\nu{\cal A}^\mu=0,\label{um}\\
    \square {\cal A}^\phi-\partial_\phi (\partial_\mu{\cal A}^\mu)=0\label{dois}, 
\end{eqnarray}
while the case $m=\phi$ leads to
\begin{equation}
    (4A^\prime +B^\prime) e^{4A+B}(\partial^\phi{\cal A}^m-\partial^m {\cal A}^\phi)+e^{4A+B}(\partial_{\phi}^{2}{\cal A}^m-\partial_\phi\partial^m{\cal A}^\phi)=0,
\end{equation}
which implies
\begin{align}
\partial_{\phi}^{2}{\cal A}^\mu&=\partial_{\phi}\partial^\mu{\cal A}^\phi,\label{tres}\\
    \partial^\phi{\cal A}^\mu&=\partial^\mu{\cal A}^\phi.\label{quatro}
\end{align}

\begin{figure}[htbp]
    \centering
    \begin{subfigure}[b]{0.45\textwidth}
        \centering
        \includegraphics[width=\textwidth]{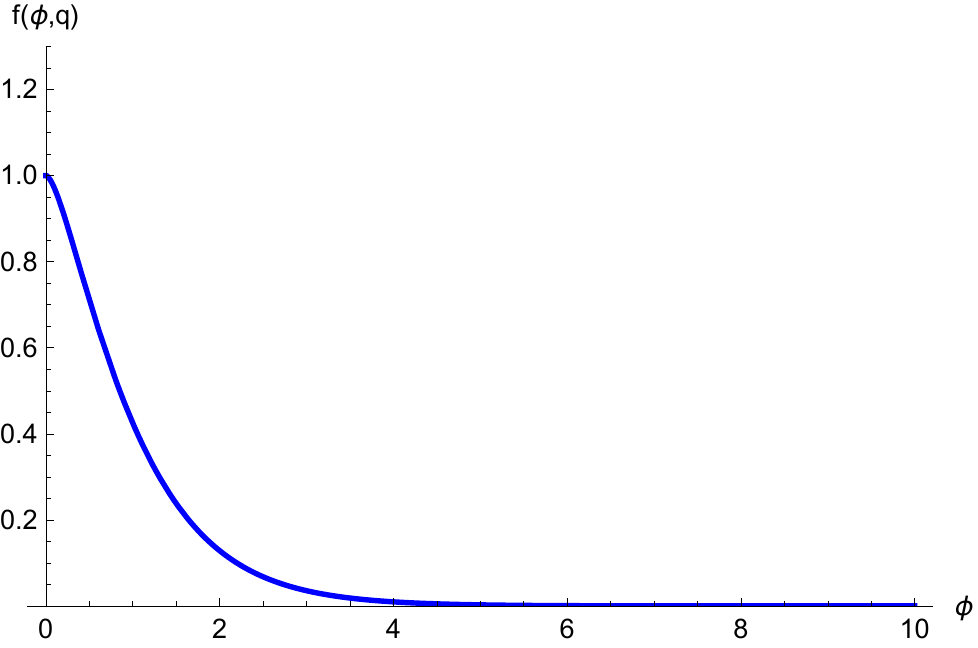}
        \caption{$q^2=0.0$ GeV$^2$}
    \end{subfigure}
    \hfill
    \begin{subfigure}[b]{0.45\textwidth}
        \centering
        \includegraphics[width=\textwidth]{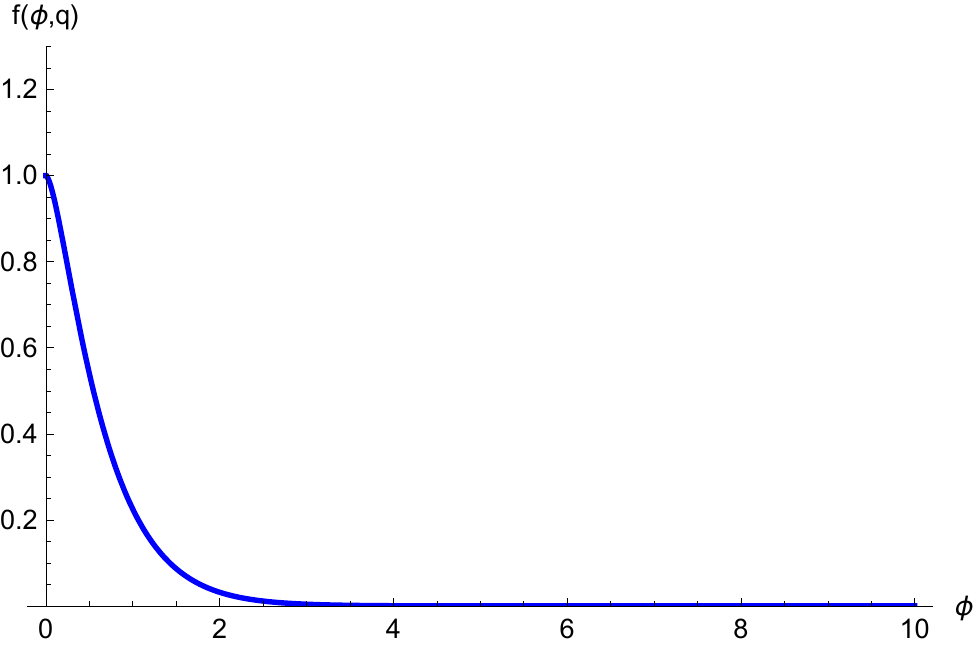}
        \caption{$q^2=4.7$ GeV$^2$}
    \end{subfigure}

    \vskip\baselineskip
    \begin{subfigure}[b]{0.45\textwidth}
        \centering
        \includegraphics[width=\textwidth]{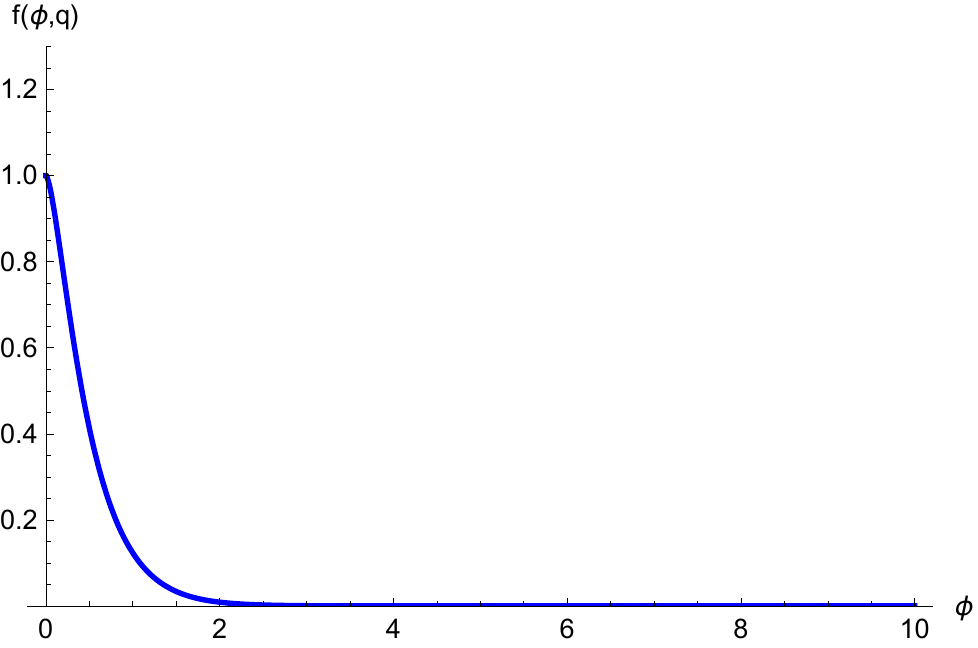}
        \caption{$q^2=11$ GeV$^2$}
    \end{subfigure}
    \hfill
    \begin{subfigure}[b]{0.45\textwidth}
        \centering
        \includegraphics[width=\textwidth]{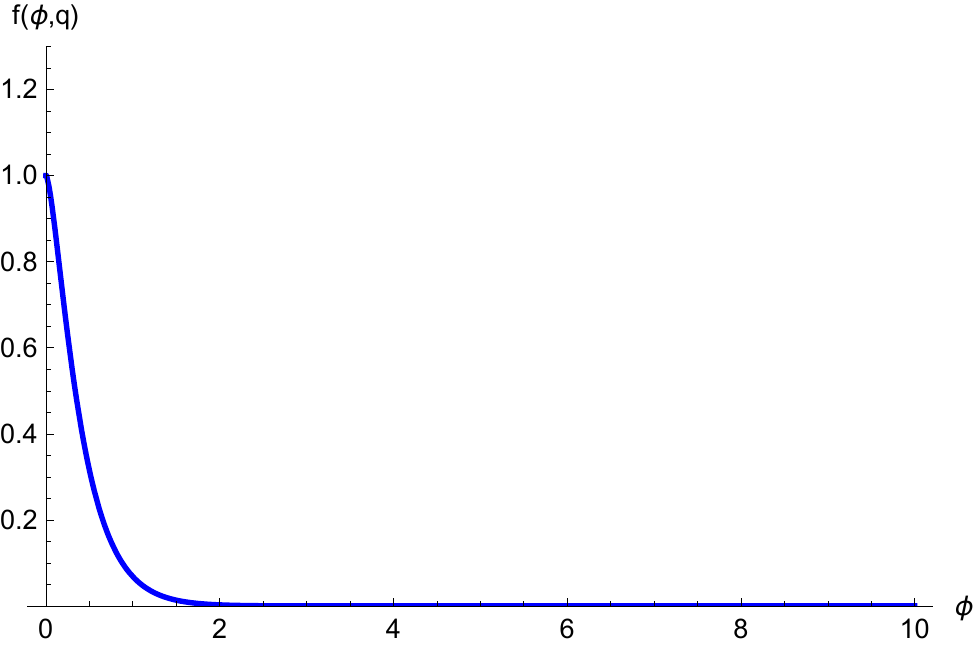}
        \caption{$q^2=18$ GeV$^2$}
    \end{subfigure}

    \caption{\sl The photon field $f(\phi,q)$, Eq. \eqref{phimunorm}, obtained from numerical solutions of Eq. \eqref{eqfphiq}, for the first four values of $q^2$.}
    \label{fofphi}
\end{figure}

Using the gauge-fixing
\begin{equation}
    \partial_\mu{\cal A}^\mu+e^{-A}\partial_\phi(e^{A}{\cal A}_\phi)+e^{-B}\partial_\phi(e^B{\cal A}_\phi)=0\label{gaugefix},
\end{equation}
along with (\ref{tres}) and (\ref{quatro}), one can write (\ref{um}) as
\begin{equation}
    \square{\cal A}^\mu+(A^\prime+B^\prime)\partial_\phi{\cal A}^\mu+2\partial_{\phi}^{2}{\cal A}^\mu=0\label{peq}.
\end{equation}
Furthermore, we will consider a photon with a particular polarization $\eta_\mu$ such that $\eta_\mu q^\mu=0$. In this case, only the components ${\cal A}^\mu$ will contribute to the scattering process, disregarding ${\cal A}^\phi$. Thus, we just have to solve  Eq.\ (\ref{peq}) for the virtual photon. Also, by imposing the boundary condition,
\begin{equation}
    {\cal A}_\mu(\phi,y)|_{\phi=0}=\eta_\mu e^{iq\cdot y},
\end{equation}
we shall consider solutions of the form
\begin{equation}\label{phimunorm}
    {\cal A}_\mu=\eta_\mu e^{iq\cdot y}f(\phi,q).
\end{equation}
Inserting (\ref{phimunorm}) into Eq.\ (\ref{peq}),  we find the following equation of motion for $f(\phi,q)$
\begin{equation}
    2f^{\prime\prime}(\phi,q)+(A^{\prime}+B^\prime)f^{\prime}(\phi,q)-q^2f(\phi,q)=0.\label{eqfphiq}
\end{equation}
This equation will be solved numerically for $f(\phi,q)$, with $A$ and $B$ given as in the previous section, to determine the virtual photon profile in this model. The result is displayed in Fig.\ \ref{fofphi}.

It is also possible to obtain approximate analytical solutions for the photon field. First, if we consider the limit $\phi\to 0$, then, with $\Delta=3$, the functions $A$ and $B$ both reduce to $-\log\phi$, so that $A'+B'\approx -2/\phi$, in this limit. The solution in the case $q^2>0$ is 
\begin{equation}
    f(\phi,q)= c_1 \;\phi \;J_1\left(\frac{iq\phi}{\sqrt{2}}\right)
    + c_2 \;\phi \;Y_1\left(-\frac{iq\phi}{\sqrt{2}}\right),\label{Bessel>0}
\end{equation}
which is nicely written in terms of Bessel functions of the first and second kind. This is just the solution in pure AdS$_5$, as expected. Note that this solution heavily depends on the sign of $q^2$, which defines the virtuality of the photon. In the opposite case, $q^2<0$, one finds
\begin{equation}
    f(\phi,q)= c_1 \;\phi \;J_1\left(\frac{q\phi}{\sqrt{2}}\right)
    + c_2 \;\phi \;Y_1\left(\frac{q\phi}{\sqrt{2}}\right). \label{Bessel<0}
\end{equation}
Thus, we see that solution in Eq.\ \eqref{Bessel>0} is a sum of monotonically increasing and decreasing functions, while the solution \eqref{Bessel<0} is oscillatory. 

Going further in the approximation for the functions $A$ and $B$, now including the contribution of linear terms, one finds
\begin{eqnarray}
    A(\phi)&=&-\log\phi -a_1\phi \label{A1} \\
    B(\phi)&=& -\log\phi -b_1\phi,\label{B1}
\end{eqnarray}
where $a_1\approx 3/8$ and $b_1\approx 5/9$, as can be extracted from Fig.\ \ref{A&B}, corresponding to the values of the functions $A$ and $B$ with the parameters given in Table \ref{table:1}. Substituting the expressions for $A$ and $B$ given by Eqs. \eqref{A1} and \eqref{B1} into equation \eqref{eqfphiq}, one finds the general solution for $q^2\not=0$, $a_1>0$, and $b_1>0$: 
\begin{eqnarray}
    f(\phi,q)=c_{1} \;\phi^2 \;e^{k_{1}\phi/4}\;
U\left(a_{2},3,\frac{1}{2}\,k\phi\right) + c_{2} \;\phi^2 \;e^{k_{1}\phi/4}\;
    L^2_{-a_{2}}\left(\frac{1}{2}\,k\phi\right),
    \label{f_2}
\end{eqnarray}
where $c_{1}$ and $c_{2}$ are arbitrary constants to be fixed by boundary conditions, while $U(a,b,x)$ and $L_a^b(x)$ are the confluent hypergeometric function of the second kind and the associated Laguerre function, respectively. Above, we defined the constants
\begin{eqnarray}
k_{1}&=&a_1+b_1-k\,,\\
k&=&\sqrt{(a_1+b_1)^2+8|q^2|}\,,\\
a_{2}&=&\frac{a_1+b_1+3k}{2k}\,.
\end{eqnarray} 
Note that $k>a_1+b_1>0$ for any $q^2\not=0$, so that $k_1<0$, and $a_2>2$, implying decreasing exponential factors in both terms in Eq.\ \eqref{f_2}. 
This behavior is clearly found in Fig.\ \ref{fofphi} for some values of $q^2$, including the case $q^2=0$, for which a particular solution of Eq.\ \eqref{eqfphiq} can be also found.


\subsection{Baryonic states}
 \label{barstates}
The fermionic fields representing the initial and final hadrons are described by the action
\begin{equation}\label{diracfield}
S_\mathrm{fermion} =  \int d^{5} x\sqrt{g} \; \bar{\Psi}({\slashed D} - \tilde{m}_5 ) \Psi, 
\end{equation}
where $\Psi$ represents a spin 1/2 field with a scale dependent mass $\tilde{m}_5\equiv \tilde{m}_5 (\phi)$ as a function of the coordinate $\phi$ in five-dimensional space, and ${\slashed D}$ is the five-dimensional Dirac operator defined as:
\begin{equation}\label{slash}
{\slashed D} \equiv g^{mn} e^{a}_n \gamma_a \left( \partial_{m} + \frac{1}{2} \omega^{bc}_{m}\, \Sigma_{bc} \right) = e^{-B(\phi)} \gamma^\phi \partial_\phi + e^{-A(\phi)} \gamma^{\mu}  \partial_{\mu} + 2 e^{-B(\phi)} A(\phi)'\gamma^\phi, 
\end{equation}
\noindent where again the prime denotes the derivative with respect to $\phi$, the Dirac matrices $\gamma_a = (\gamma_{\mu}, \gamma_\phi)$, with $\gamma_\phi=\gamma_5$, 
$\left\lbrace \gamma_a, \gamma_b \right\rbrace = 2 \eta_{ab} $, and $\Sigma_{\mu \phi} = \frac{1}{4} \left[ \gamma_{\mu}, \gamma_5\right]$ following the convention of Refs.\ \cite{Henningson:1998cd, Mueck:1998iz, Kirsch:2006he, Abidin:2009hr}. The indices $m, n, p, q, \dots,$ represent curved spacetime labels, referring to the coordinates in the bulk geometry. $a, b, c, \dots,$ represent the quantities of the locally flat tangent space, while $\mu, \nu,\dots,$ represent the Minkowski space indices.
Thus, the vielbein are given by:
\begin{equation}\label{tetra}
e^{a}_\phi = e^{B(\phi)} \delta^{a}_\phi, \; \;\; e^{a}_{\mu}= e^{A(\phi)}\delta^{a}_{\mu} , \; \; \; {\rm with} \; \;\;\;\;m = 0, 1, 2, 3, 5,
\end{equation}
where the fifth coordinate is the radial coordinate $\phi$.

For the spin connection $\omega^{\mu \nu}_{m}$, one finds
\begin{equation}
\omega^{a b}_{m}  = e^a_n \partial_m e^{nb} + e^a_n e^{pb} \Gamma^n_{pm}, 
\end{equation}
where the Christoffel symbols are written as:
\begin{equation}\label{Cris}
\Gamma_{m n}^p = \frac{1}{2} g^{pq}(\partial_n g_{mq} + \partial_m g_{nq} - \partial_q g_{mn}), \;\;\; {\rm with} \;\;\;g_{mn} = e^{a}_{m}e^{b}_{n}\eta_{ab}.
\end{equation}
The only non-vanishing $\Gamma_{m n}^p $ for the asymptotic AdS$_5$ space are
\begin{equation}\label{LV}
\Gamma_{\mu \nu}^\phi = -A(\phi)' e^{2(A-B)} \eta_{\mu \nu}, \; \; \Gamma_{5 5}^5 = B(\phi)'  \; \; {\rm and} \; \; \Gamma_{\phi\nu}^{\mu} = -A(\phi)' \delta^{\mu}_{\nu},
\end{equation}
\noindent so that 
\begin{equation}\label{spin}
\omega^{\phi \nu}_{\mu}=- A^{\prime}(\phi) e^{A-B} \delta^{\nu}_{\mu}, 
\end{equation}
\noindent and all other components of the spin connection vanish. Then, the Dirac operator becomes 
\begin{equation}\label{slash}
{\slashed D} = e^{-B(\phi)} \gamma^\phi \partial_\phi + e^{-A(\phi)} \gamma^{\mu}  \partial_{\mu} + 2 e^{-B(\phi)} A(\phi)'\gamma^\phi. 
\end{equation}

From the action \eqref{diracfield}, one can derive the five-dimensional Dirac equations of motion 
\begin{equation}\label{fermioneom}
({\slashed D} - \tilde m_5 ) \Psi  = 0, 
\end{equation}
which can be written as follows
\begin{equation}\label{newdirac}
\left(  \gamma^\phi \partial_\phi + e^{B(\phi)-A(\phi)} \gamma^{\mu}  \partial_{\mu} + 2 A(\phi)'\gamma^\phi - \tilde{m}_5 e^{B(\phi)}\right) \Psi = 0\,. 
\end{equation}
The four-component Dirac spinor $\Psi$ defined in the bulk of the five-dimensional space can be decomposed into left- and right-handed chiral components as:
\begin{equation}
    \Psi(x^{\mu}, \phi)=\Psi_{L}(x^{\mu}, \phi)+\Psi_{R}(x^{\mu}, \phi),\label{Pcompts}
\end{equation}
where 
\begin{equation}
    \Psi_{L/R}(x^\mu,\phi)=\frac{1\mp \gamma_\phi}{2}\Psi(x^\mu,\phi).
\end{equation}
Since  $\gamma^\phi=\gamma^5$, the following relations hold
\begin{align}
\gamma^5\Psi_{L/R}&=\mp \Psi_{L/R}\\
    \gamma^\mu\partial_\mu\Psi_R & = M\Psi_L,\label{RLm}
\end{align}
where $M$ is the four-dimensional fermionic mass.

Assuming the Kaluza-Klein modes are dual to the chiral spinors, we have
\begin{equation}
    \Psi_{L/R}(x^\mu,\phi)=\sum_n \varphi^{n}_{L/R}(x^\mu)\chi^{n}_{L/R}(\phi).\label{kk}
\end{equation}

Now, using (\ref{kk}) along with (\ref{Pcompts}) into (\ref{newdirac}), we arrive at the following set of coupled equations
\begin{align}
    \left(\partial_\phi +2A^\prime + \tilde{m}_5 \, e^B\right)\chi_{L}^n(\phi)&=+M_{n} e^{B-A}\chi^n_{R}(\phi)\label{chi1},\\
    \left(\partial_\phi +2A^\prime - \tilde{m}_5 \, e^B\right)\chi_{R}^n(\phi)&=-M_{n} e^{B-A}\chi^n_{L}(\phi)\label{chi2}, 
\end{align}
where $M_n$ are the discrete four dimensional fermionic masses with $n=0,1,2,\cdots$, once we find a confining potential for these fermions. 

From now on, we consider the approximation $A(\phi) \approx B(\phi)+0.7$, supported by the numerical results for these functions, as shown in Fig. 
 \ref{A&B}, in the relevant regime where $\phi \approx 5$. Then, Eqs. 
\eqref{chi1} and \eqref{chi2} reduce to
\begin{align}
    \left(\partial_\phi +2A^\prime + \tilde{m}_5 \, e^B\right)\chi_{L}^n(\phi)&=+\frac{M_{n}}{2} \chi^n_{R}(\phi)\label{chi1a},\\
    \left(\partial_\phi +2A^\prime - \tilde{m}_5 \, e^B\right)\chi_{R}^n(\phi)&=-\frac{M_{n}}{2} \chi^n_{L}(\phi)\label{chi2a}.  
\end{align}
To simplify these equations, we perform the transformation
\begin{equation}
    \chi^n_{L/R}(\phi)=\psi^n_{R/L}(\phi)e^{-{2}A(\phi)},
\end{equation}
so that 
\begin{align}
    \left(\partial_\phi + \tilde{m}_5 \, e^B\right)\psi_{L}^n(\phi)&=+\frac{M_{n}}{2} \psi^n_{R}(\phi)\label{chi1b},\\
    \left(\partial_\phi  - \tilde{m}_5 \, e^B\right)\psi_{R}^n(\phi)&=-\frac{M_{n}}{2} \psi^n_{L}(\phi)\label{chi2b}. 
\end{align}

\begin{figure}[h]
    \centering
    \includegraphics[width=0.5\linewidth]{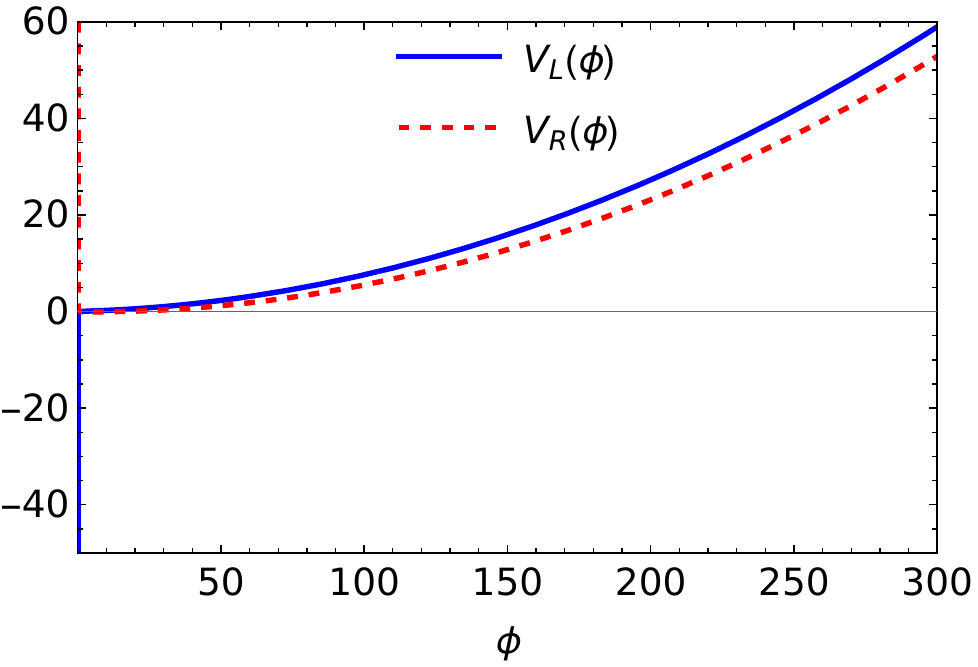}
    \caption{\sl $V_L(\phi)$ and $V_R(\phi)$ represent the potentials of the fermionic probes in the DIS problem, Eq. \eqref{PotSch1}. Since we choose $\phi$ to represent the holographic coordinate, we have $\phi\thicksim1/E$. We then see that deep in the IR, besides largely overlapping, the potentials show a confinement profile in this region, which corresponds to a very low energy scale.}
    \label{PotentialP}
\end{figure}

Decoupling Eqs. (\ref{chi1b}) and (\ref{chi2b}), one finds a Schrödinger-like equation 
\begin{equation}
    -\psi^{n\,\prime\prime}_{R/L}(\phi)+V_{L/R}(\phi)\psi^n_{R/L}(\phi)=\,\left(\frac{M_{n}}{2}\right)^2\psi^n_{R/L}(\phi),\label{Schrodinger}
\end{equation}
with the right and left potentials given by
\begin{equation}
    V_{R/L}(\phi)= \tilde{m}_5  e^{B}
    \left(\tilde{m}_5  e^{B} \pm B^\prime
    \right)
    .\label{PotSch}
\end{equation}
Inspired by previous results \cite{Abidin:2009hr, Ballon-Bayona:2024yuz}, 
 and in order to get a confining potential for fermions, we put: 
\begin{equation}
     \tilde{m}_5 e^{B} =m_5\, B\,, 
     \label{m5}
\end{equation}
where $m_5$ is a constant, 
so the potentials become
\begin{equation}
    V_{R/L}(\phi)=
    {m}_5 B \left( {m}_5 B \pm  B^\prime \right)
    \,.\label{PotSch1}
\end{equation}
These potentials are shown in Figure \ref{PotentialP}, for the function $B(\phi)$ shown in Fig. \ref{A&B}. 

\begin{figure}[h]
    \centering
    \includegraphics[width=0.5\linewidth]{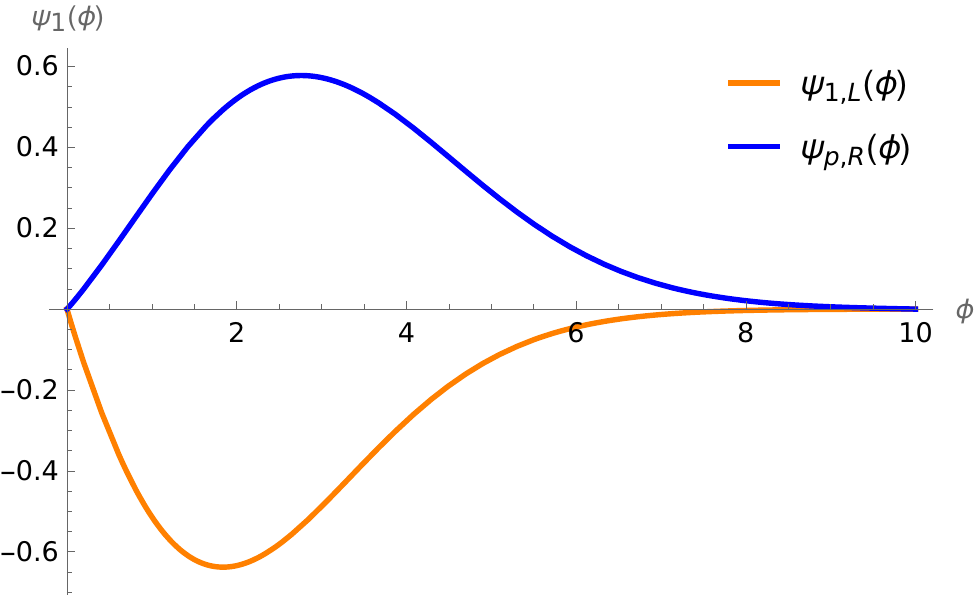}
    \caption{\sl The blue curve corresponds to the left chiral initial state $\psi_{1L}$ and the orange one to the right chiral initial state $\psi_{1R}$, of Eq.~(\ref{Schrodinger}) with the potential (\ref{PotSch1}).}
    \label{Wavef1}
\end{figure}


\begin{figure}
     \centering
         \centering
         \includegraphics[width=0.5\linewidth]{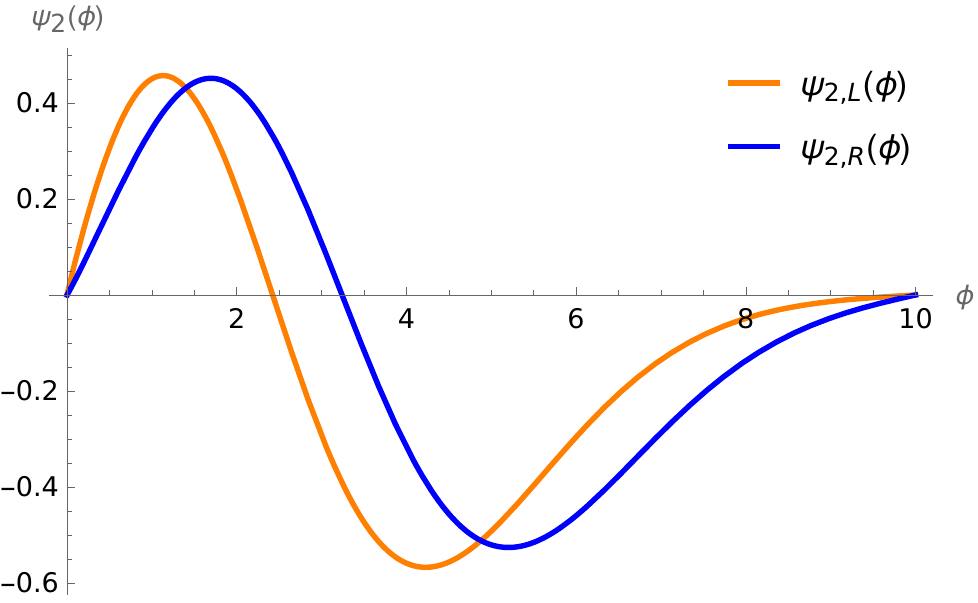}
         \centering
         \includegraphics[width=0.5\linewidth]{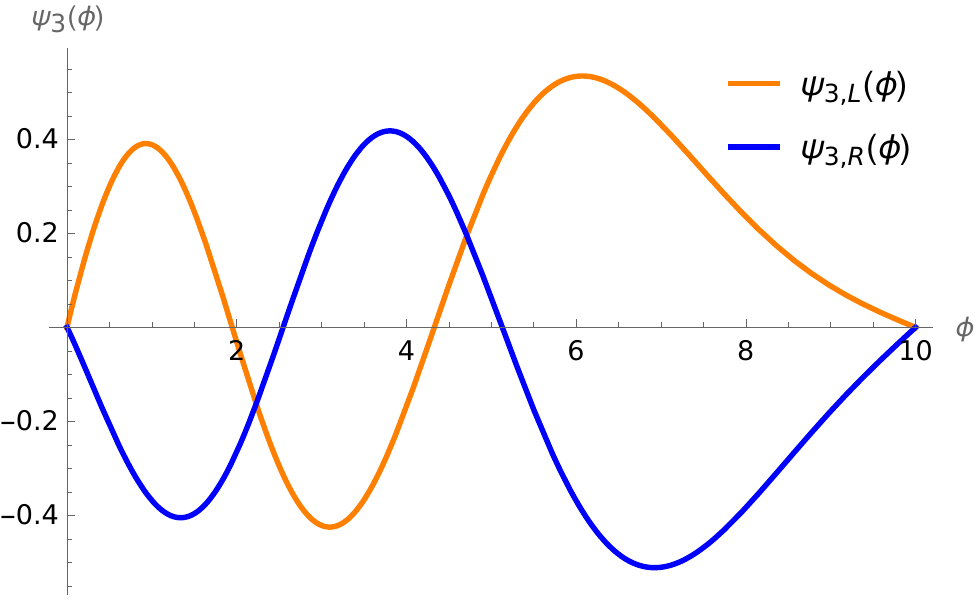}
    \includegraphics[width=0.5\linewidth]{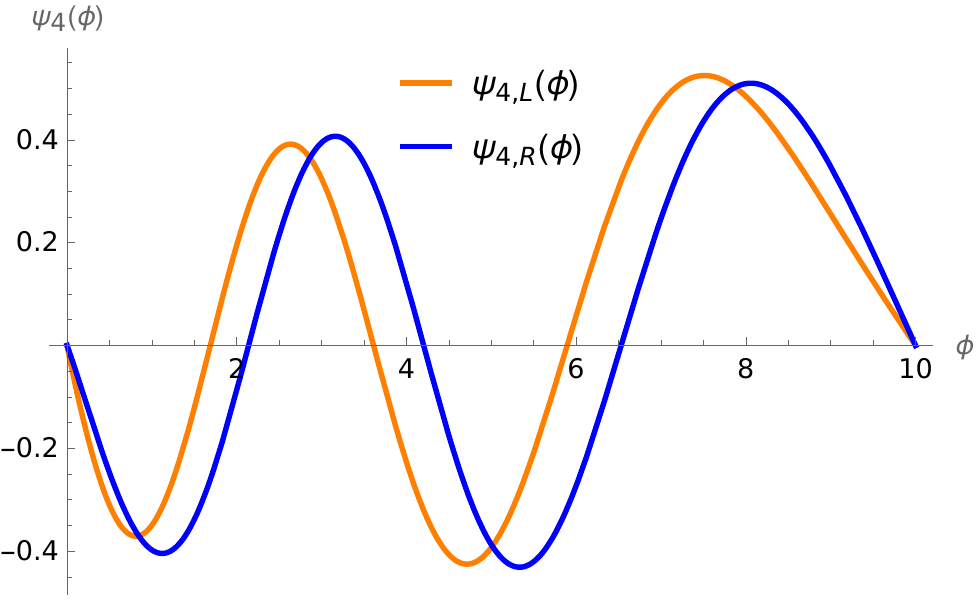}
        \caption{\sl Wave functions of the excited final states of the Schrödinger-like equation, Eq.~(\ref{Schrodinger}). The blue and orange curves' behavior reflects the sign of the last term in the corresponding chiral potential, Eq.  (\ref{PotSch}).
}
        \label{fig:three graphs}
\end{figure}

Then, we can write the initial $(i)$ and final $(X)$ baryonic states as linear combinations of solutions $\psi_{R/L}$: 
\begin{align}
    \Psi_i&=e^{i P\cdot y}e^{-2A}\left[\left(\frac{1+\gamma^5}{2}\right)\psi^i_{L}(\phi)+\left(\frac{1-\gamma^5}{2}\right)\psi^i_{R}(\phi)\right]u_{s_{i}}(P)\label{istate},\\
    \Psi_X&=e^{i P_X\cdot y}e^{-2A}\left[\left(\frac{1+\gamma^5}{2}\right)\psi^X_{L}(\phi)+\left(\frac{1-\gamma^5}{2}\right)\psi^X_{R}(\phi)\right]u_{s_{X}}(P_X)\label{fstate}.
\end{align}

Before we move on to compute the interaction action, we comment on the role of adding an anomalous dimension to this problem. When solving Eq.\ (\ref{Schrodinger}), we want to fix the ground state to be the proton. According to the holographic dictionary, the bulk fermionic mass $m_5^{\text{AdS}}$ is related to the canonical scaling dimension $\Delta_{\text{can}}$ of a fermionic operator $\mathcal{O}$ at the boundary by the following:
\begin{equation}
    \left|m_5^{\text{AdS}}\right|=\Delta_{\text{can}}-2.
\end{equation}
For $\Delta_{\text{can}}$, we consider the initial state of our problem to be the proton, and regard it as single particle. This choice will be justified by the results we get for $F_2$ for large values of the transferred momentum. Then, we take $\Delta_{\text{can}}=3/2$, the canonical scaling dimension for fermions. However, this procedure  does not reproduce the proton mass as the initial state of the eigenvalue problem. Phenomenologically, it is necessary to consider an anomalous dimension $\gamma$ to fit the exact value of the proton mass. We thus solve the Schrödinger-like equation (\ref{Schrodinger}) considering
\begin{equation}
m_5=\left|m_5^{\text{AdS}}\right|=\Delta_{\text{can}}+\gamma-2.
\end{equation}
Choosing an appropriate $\gamma$, we can obtain the proton mass as the ground state for one of the chiral solutions, as discussed in Sec. \ref{results}. The profiles of the first fermionic solutions are shown in Figs. \ref{Wavef1} and \ref{fig:three graphs}.



\section{The DIS interaction action}\label{disinteraction}

In this section, we will explicitly compute the DIS interaction action in Eq.\ \eqref{PP}, where we identify the scale dependent coupling $F(z=\phi)=kB(\phi)$, with $k$ a convenient constant
\begin{equation}
    S_{\text{int}}= \int{d\phi\,d^4y\,\sqrt{-g}\,e^{kB(\phi)}
    {\cal A}^m\, \bar{\Psi}_X\,\Gamma_m\,\Psi_i}\,.
    \label{S_int}
\end{equation}
Using the definitions of the vielbein given by Eqs.\ \eqref{tetra} and \eqref{Cris}, associated with the geometry of the metric \eqref{gs}, one can rewrite the interaction action $S_{\rm int}$, disregarding the component ${\cal A}^\phi$ as discussed in Sect.\ \ref{sect.:photon}, one finds 
\begin{eqnarray}\notag
S_{\rm int}&=& \int{d\phi\,d^4y\,\sqrt{-g}\,e^{kB(\phi)} g^{\mu\,\nu}{\cal A}_\mu\, \bar{\Psi}_X\,e^\alpha_\nu\, \gamma_\alpha\,\Psi_i}\nonumber \\
&=& \int{d\phi\,d^4y\,e^{3A+(k+1)B}
{\cal A}^\mu\, \bar{\Psi}_X\, \gamma_\mu\,\Psi_i}.  \quad
\end{eqnarray}

\noindent  
The initial and final spinor states, $\Psi_i$ and $\Psi_X$, are given, respectively, by Eqs.\ \eqref{istate}
and \eqref{fstate}.  
We also note that
\begin{equation}
 \bar{\Psi}_X=e^{-i\,P_X\cdot\,y}\,e^{-2A}\,\bar{u}_{s_X}(P_X)\left[\left(\frac{1+\gamma_5}{2}\right)\psi^X_L(\phi)+\left(\frac{1-\gamma_5}{2}\right)\,\psi^X_R(\phi)\right].   
\end{equation}
With these results and the gauge field ${\cal A}_{\mu}$ given by Eq.\ \eqref{phimunorm}, using $f\equiv f(\phi,q)$, we can write the interaction action as follows,
\begin{eqnarray}\notag 
&S_\text{int}&\cr
&=& \left(\frac 12\right)^4 \int{d^4y\,
d\phi\,e^{-i\left(P_x-P-q\right)\cdot\,y}\,e^{\tilde k B}\,\eta^\mu \left[\bar{u}_{s_X}\left(\hat{P}_L\,\psi^X_L+\hat{P}_R\,\psi^X_R\right)\gamma_\mu \left(\hat{P}_L\,\psi^i_L+\hat{P}_R\,\psi^i_R\right)\,u_{s_i}\right]f}\\ \notag
&=&  \pi^4\,\delta^4(P_X-P-q)\,\,\eta^\mu\,\int{d\phi\,e^{\tilde k B}\left[\bar{u}_{s_X}\,\gamma_\mu\,\hat{P}_R\,u_{s_i}\psi_L^X\,\psi_L^i\,f+\bar{u}_{s_X}\,\gamma_\mu\,\hat{P}_L\,u_{s_i}\psi_R^X\,\psi_R^i\,f\right]}\\ \label{int-term-2}
&=& \pi^4\,\delta^4(P_X-P-q)\,\,\eta^\mu\,\left[\bar{u}_{s_X}\,\gamma_\mu\,\hat{P}_R\,u_{s_i}\,\mathcal{I}_L+\bar{u}_{s_X}\,\gamma_\mu\,\hat{P}_L\,u_{s_i}\mathcal{I}_R\right],\label{sintfinal}
\end{eqnarray}

\noindent where the integrals $\mathcal{I}_{R/L}$ are defined in terms of the solutions of the chiral fermions and the solution of the photon field $f\equiv f(\phi,q)$, so that
\begin{equation}\
\mathcal{I}_{R/L}=\int{d\phi\,e^{\tilde k B}f(\phi,q)\,\psi_{R/L}^X(\phi,P_X)\,\psi_{R/L}^i(\phi,P)}\,. 
\label{I_RL}
\end{equation}
From Eqs.\ \eqref{PP} and \eqref{sintfinal}, one finds
\begin{eqnarray}
\eta_\mu\langle P_X\left|J^\mu(q)\right|P \rangle&=& \left(2\,\pi\right)^4\,\delta^4(P_X-P-q)\,\eta_\mu\langle P+q\left|J^\mu(0)\right|P \rangle\nonumber\\ \label{geff}
&=& g_{\rm eff}\, \pi^4 \, \delta^4(P_X-P-q)\,\eta_\mu \,\left[\bar{u}_{s_X}\,\gamma^\mu\,\hat{P}_R\,u_{s_i}\,\mathcal{I}_L+\bar{u}_{s_X}\,\gamma^\mu\,\hat{P}_L\,u_{s_i}\mathcal{I}_R\right], 
\\
\eta_\mu\langle P\left|J^\mu(q)\right|P_X \rangle&=&\left(2\,\pi\right)^4\,\delta^4(P_X-P-q)\,\eta_\mu\langle P\left|J^\mu(0)\right|P+q\rangle \nonumber\\
&=&g_{\rm eff}\,\pi^4\,\delta^4(P_X-P-q)\,\eta_\nu \,\left[\bar{u}_{s_i}\,\gamma^\mu\,\hat{P}_R\,u_{s_X}\,\mathcal{I}_L+\bar{u}_{s_i}\,\gamma^\mu\,\hat{P}_L\,u_{s_X}\mathcal{I}_R\right],
\end{eqnarray}
\noindent where $g_{\rm eff}$ is an effective coupling constant related to the choice of $\tilde kB(\phi)$. 
Contracting the photon polarization with the hadronic tensor, Eq.\ \eqref{Wmunu}, one obtains
\begin{eqnarray}
\eta_ \mu \eta_ \nu W^{\mu \nu}&=& \frac{\eta_{\mu \nu}}{4} \sum_{M_x^2}\sum_{s_i, s_X} \left(\pi^4 \, g_{\rm eff}\right)^2 \,\delta(M^2_X-(P+q)^2) \left[\bar{u}_{s_X}\,\gamma^\mu\,\hat{P}_R\,u_{s_i}\, \bar{u}_{s_i}\,\gamma^ \nu\,\hat{P}_R\,u_{s_X}\,{\cal I}_L^2\right.\nonumber\\ &+&\left. \bar{u}_{s_X}\,\gamma^ \mu\,\hat{P}_R\,u_{s_i}\, \bar{u}_{s_i}\,\gamma^ \nu\,\hat{P}_L\,u_{s_X}\, {\cal I}_L\,{\cal I}_R\, + \,\bar{u}_{s_X}\,\gamma^ \mu\,\hat{P}_L\,u_{s_i}\,\bar{u}_{s_i}\,\gamma^\nu\,\hat{P}_R\,u_{s_X}\,{\cal I}_R\,{\cal I}_L \right. \nonumber\\ &+& \left. \bar{u}_{s_X}\,\gamma^ \mu\,\hat{P}_L\,u_{s_i}\,\bar{u}_{s_i}\,\gamma^\nu\,\hat{P}_L\,u_{s_X}\, {\cal I}_R^2\right]\,,
\end{eqnarray}
where $M_X$ is the mass of the final hadronic state. 
As we are interested in a spin-independent scenario, using the following property, we perform a summation over spin, so that
\begin{equation}
\sum_s  (u_s)_ \alpha (p) \, (\bar u_s)_ \beta (p) = (\gamma^ \mu p_ \mu + M)_{\alpha \beta}\,. 
\end{equation}{}
Thus, one finds
\begin{eqnarray}\label{mn}
\eta_ \mu \eta_ \nu W^{\mu \nu}&=& \frac{\eta_\mu\,\eta_ \nu}{4} \left(\pi^4 g_{\rm eff}\right)^2\,\sum_{s_X,\,s_i}\sum_{M_x^2}\delta(M^2_X-(P+q)^2) \, \left[\bar{u}_{s_X}\,\gamma^\mu\,\hat{P}_R\,(\slashed p + M_0)\,\gamma^ \nu\,\hat{P}_R\,u_{s_X}\,{\cal I}_L^2\right.\nonumber\\ &+&\left. \bar{u}_{s_X}\,\gamma^ \mu\,\hat{P}_R\,(\slashed p + M_0)\,\gamma^ \nu\,\hat{P}_L\,u_{s_X}\, {\cal I}_L\,{\cal I}_R\, + \,\bar{u}_{s_X}\,\gamma^ \mu\,\hat{P}_L\,(\slashed p + M_0)\,\gamma^\nu\,\hat{P}_R\,u_{s_X}\,{\cal I}_R\,{\cal I}_L \right. \nonumber\\ &+& \left. \bar{u}_{s_X}\,\gamma^ \mu\,\hat{P}_L\,(\slashed p + M_0)\,\gamma^\nu\,\hat{P}_L\,u_{s_X}\, {\cal I}_R^2\right]\,,
\end{eqnarray}
\noindent where $M_0$ is the fermionic ground state mass that we identify with that of the proton. Applying some trace engineering in Eq.\ \eqref{mn}, we obtain
\begin{multline}
\eta_ \mu \eta_ \nu W^{\mu \nu}=\left(\pi^4\, g_\text{eff}\right)^2\,\sum_{M_X^2}\,\delta(M_X^2-(P+q)^2)\,\Bigg\{(\mathcal{I}_L^2+\mathcal{I}_R^2)\left[(P\cdot \eta)^2-\frac{1}{2}\eta\cdot\eta(P^2+P\cdot q)\right]\\
+\mathcal{I}_L\,\mathcal{I}_R\,M_X^2\,M_0^2\,\eta\cdot\eta\Bigg\}.
\end{multline}
In order to obtain the expression for the structure function $F_2$, we need to sum over the outgoing states $P_X$, as presented in Eq.\ \eqref{Wmunu}. Carrying on this sum to the continuum limit, we can evaluate the invariant mass delta function. Following Ref.\ \cite{Polchinski:2002jw}, this integration will be related to the functional form of the mass spectrum of the produced particles with the excitation number $n$,
\begin{equation*}
    \delta(M_{X}^2-(P+q)^2)\propto \left(\frac{\partial\,M^2_n}{\partial\,n}\right)^{-1},
\end{equation*}
that for the soft and hard wall models  accounts for the lowest state produced at the collision, since the spectrum is linear with $n$  \cite{Polchinski:2002jw, BallonBayona:2007qr}. In our case, this delta will account for $1/ M_X^2$.

Taking into account our choice of transversal polarization ($\eta\cdot q=0$), the hadronic tensor has the following form
\begin{equation}\label{tensorhadron}
\eta_\mu\,\eta_\nu\,W^{\mu\nu}=\eta^2\,F_1(q^2,x)+\frac{2\,x}{q^2}\,(\eta\cdot P)^2\,F_2^2(q^2,x).    
\end{equation}
From this equation, one can explicitly construct the baryonic DIS structure functions. We find
\begin{eqnarray}
F_1(q^2,x)&=&\frac{\left(\pi^4\,{g}_\text{eff}\right)^2}{M_{X}^2}\,\left[M_0\,\sqrt{M_0^2+q^2\left(\frac{1-x}{x}\right)}\,\mathcal{I}_L\,\mathcal{I}_R+\left(\mathcal{I}_L^2+\mathcal{I}_R^2\right)\left(\frac{q^2}{4\,x}+\frac{M_0^2}{2}\right)\right]\label{F1}\\
F_2(q^2,x)&=&\frac{\left(\pi^4\,{g}_\text{eff}\right)^2}{2}\frac{q^2}{x}\left(\mathcal{I}_L^2+\mathcal{I}_R^2\right)\frac{1}{M_{X}^2},\label{F2}
\end{eqnarray}
\noindent where the hadronic masses of the initial and final states, $M_0$ and $M_{X}\equiv M_{X}(q^2,x)$, respectively, are related by 
\begin{equation}
    M_{X}^2(q^2,x) = M_0^2+q^2\,\left(\frac{1-x}{x}\right)\,. 
\end{equation}

Note that from Eqs. \eqref{F1} and \eqref{F2}, 
the two structure functions $F_1$ and $F_2$ are related by
\begin{eqnarray}
F_1(q^2,x)
&=&\left(\pi^4\,{g}_\text{eff}\right)^2\,\frac{M_0}{M_{X}^2}\,\sqrt{M_0^2+q^2\left(\frac{1-x}{x}\right)}\,\mathcal{I}_L\,\mathcal{I}_R 
+ \frac{1}{2} F_2(q^2,x)\left(1+\frac{2xM_0^2}{q^2}\right)\,,
\end{eqnarray}
so that, in the limit of $M_X$ and $q \gg M_0$, with $x\to 1$, one finds
\begin{eqnarray}\label{lcg}
F_1(q^2,x)
\approx 
 \frac{1}{2} F_2(q^2,x)\,,
\end{eqnarray}
which recovers the Callan-Gross relation $2xF_1=F_2$, for $x\to 1$, as expected from the model's asymptotic conformal invariance. Finally, we note that the ratio between the structure functions does not depend on the parameter ${g}^2_\text{eff}$, although it would still depend on the anomalous dimension $\gamma$.


\section{Numerical results for the Structure Function $F_2$}
\label{results}

In this section, we show the numerical results for the structure function $F_2(x,q^2)$ as a function of $q^2$ for two different values of the Bjorken parameter: $x=0.55$, 0.65, 0.75,  and $0.85$. These are the highest values of $x$ measured for $F_2(x,q^2)$ \cite{Whitlow:1991uw}. The kinematical range for $x$ considered above is near the elastic regime ($x$ close to $1$). In this large domain $x$, it is reasonable to consider the scattering of the entire proton rather than by its constituent partons. To fix the fermionic states we choose $\gamma=0.817$, such that $m_5=0.317$ and the ground state of the right chiral mode has a mass 0.937 GeV identified with the proton, while the ground state of the left mode has a mass of 1.76 GeV. 

\begin{table}
    \centering
    \begin{tabular}{|c|c|c|}
    \hline
     $x$ & $g_{\text{eff}}$ & $\tilde k$ \\
         \hline
      \; 0.55 \; & \; 0.715 \; & \; 1.5 \; \\
        0.65 & 0.317 & 1.5 \\
        0.75 & 0.213 & 0.5 \\
        0.85 & 0.126 & 1.3 \\ 
        \hline
    \end{tabular}
    \caption{\sl The values of the free parameters $g_{\text{eff}}$ and $\tilde k$ used to adjust the lines calculated using  Eq.\ (\ref{F2}) to compare with the  available experimental data points for the proton structure function $F_2$.}
    \label{table:geff}
\end{table}

\begin{figure}[ht!]
    \includegraphics[width=0.62\linewidth]{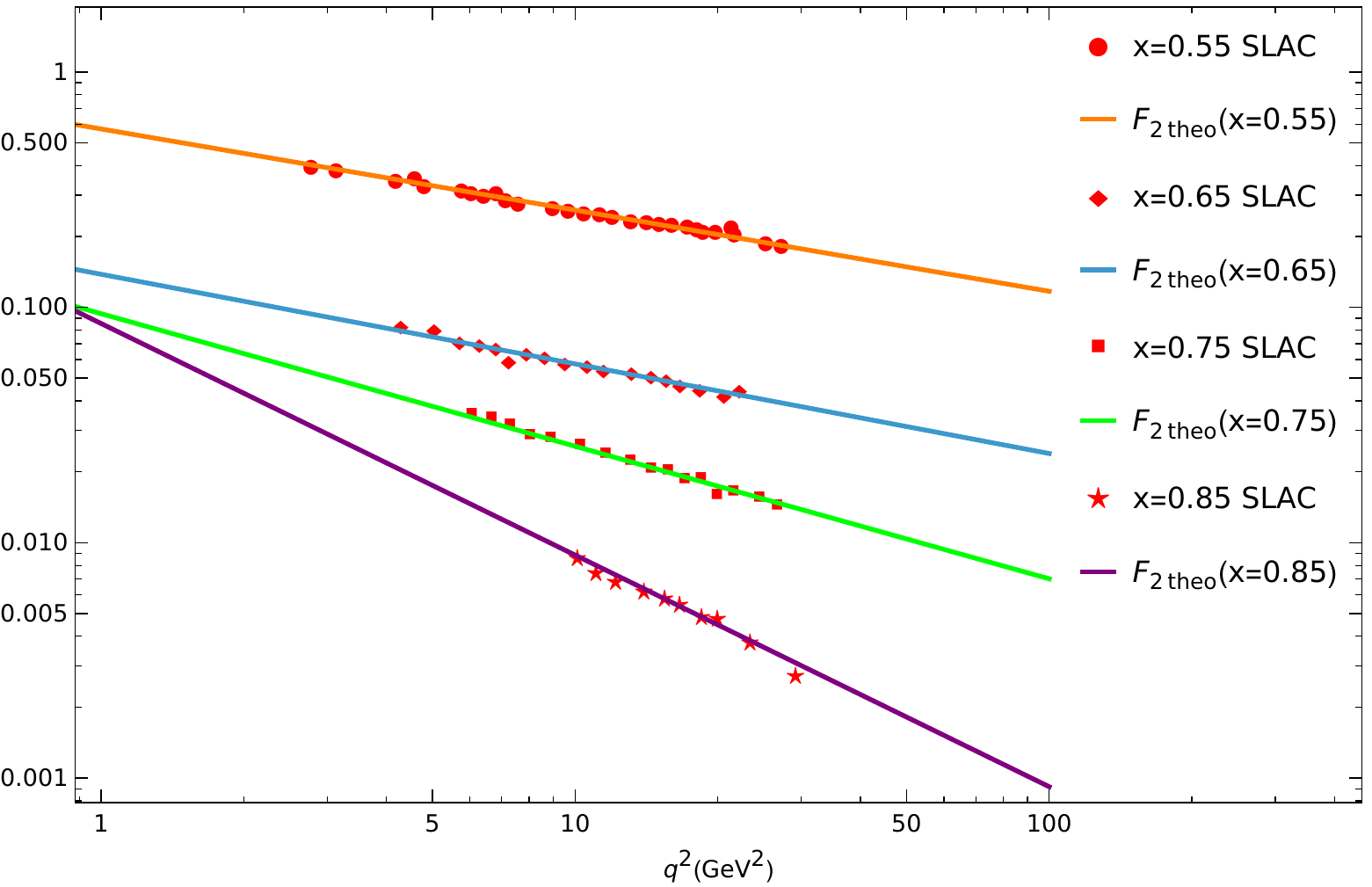}
    \caption{\sl Proton structure function $F_2$ as a function of $q^2$ for fixed values of $x$. Dots ($x=0.55$), diamonds ($x=0.65$), squares ($x=0.75$) and stars ($x=0.85$) correspond to SLAC data \cite{Whitlow:1991uw}, while the solid lines are our results, Eq.\ \eqref{F2}, for each value of $x$.}
    \label{AllF2V3}
\end{figure}

\begin{figure}[ht!]
    \includegraphics[width=0.47\linewidth]{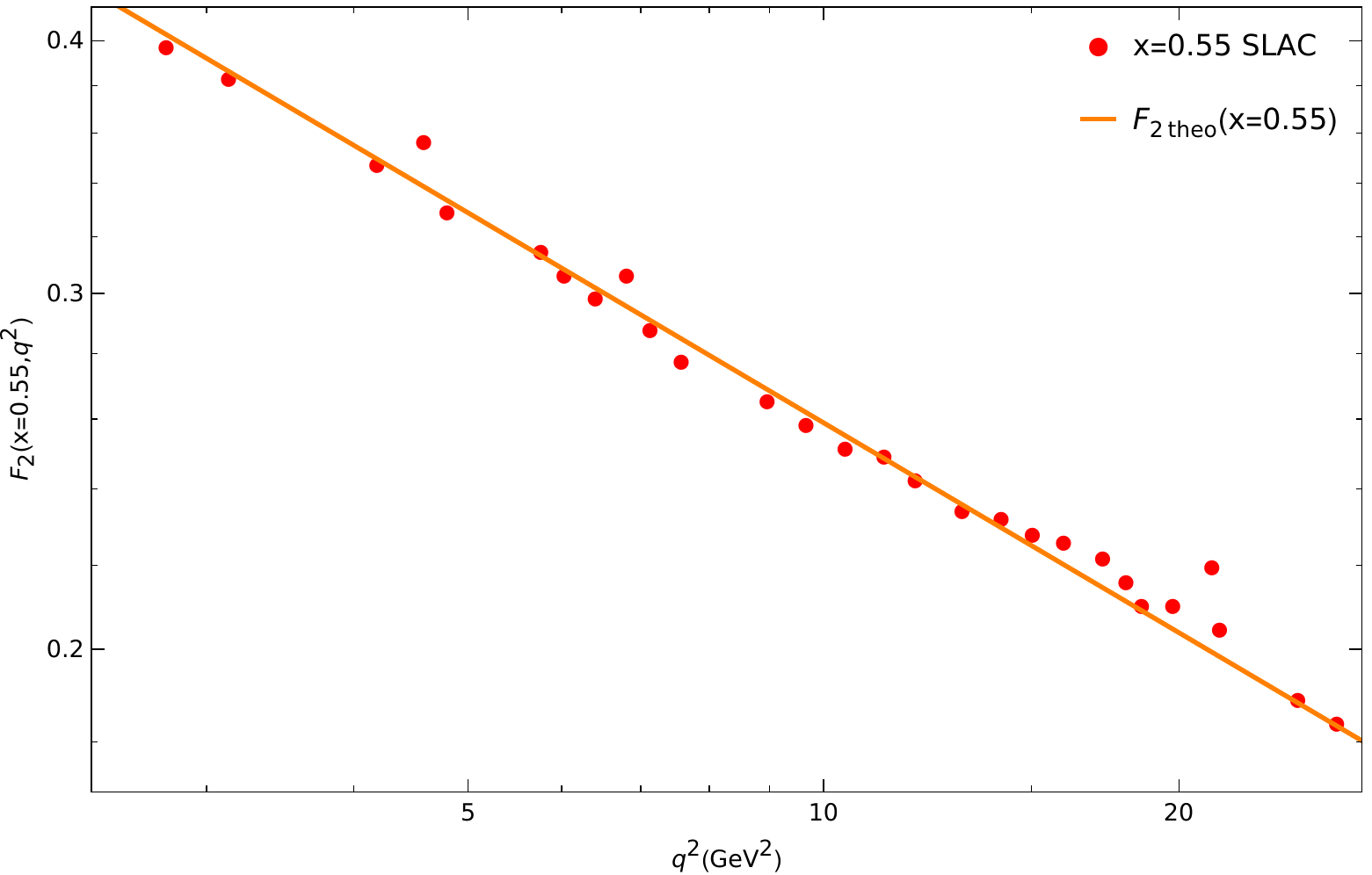}
    \includegraphics[width=0.47\linewidth]{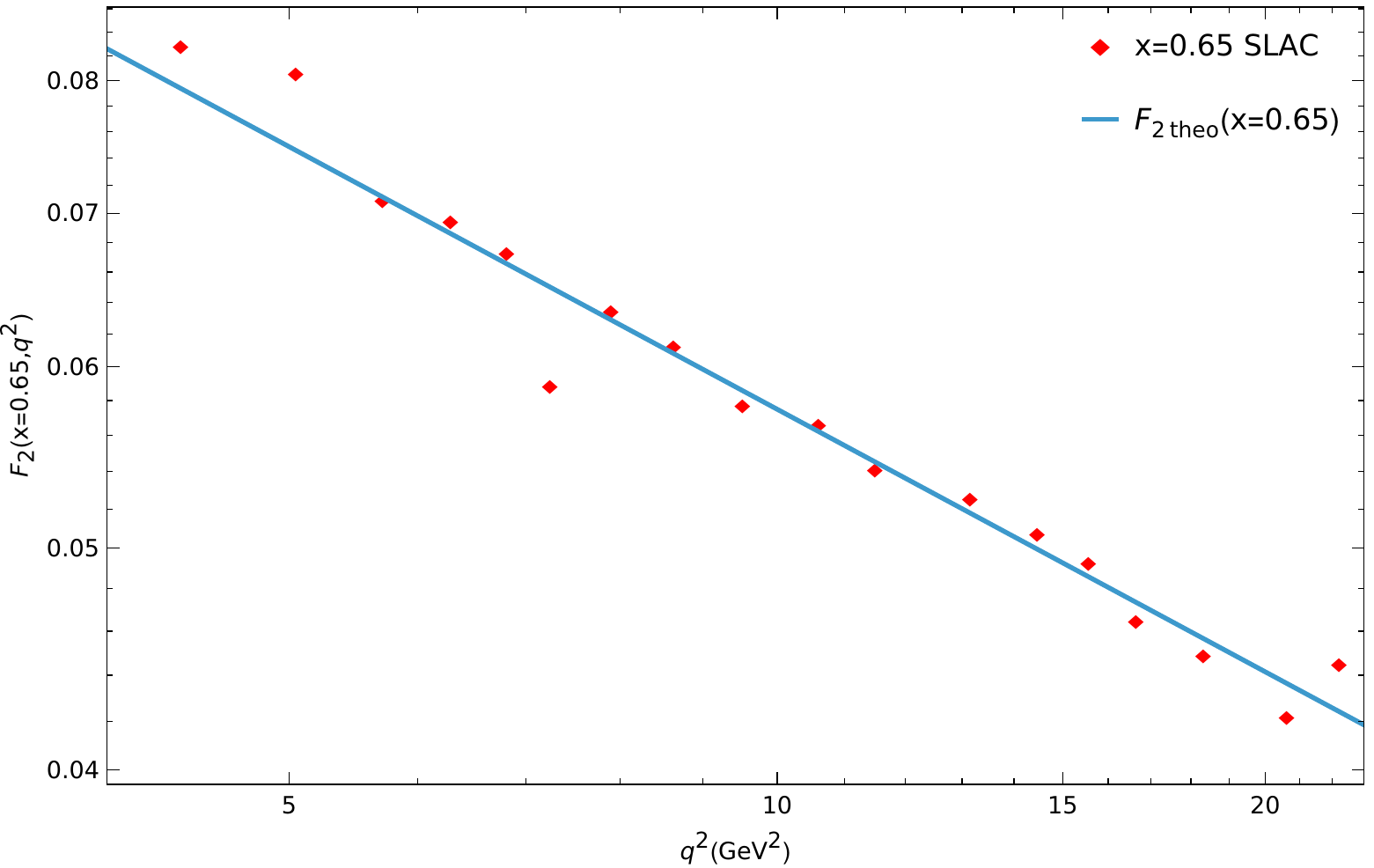}
    
    \bigskip 
    \includegraphics[width=0.47\linewidth]{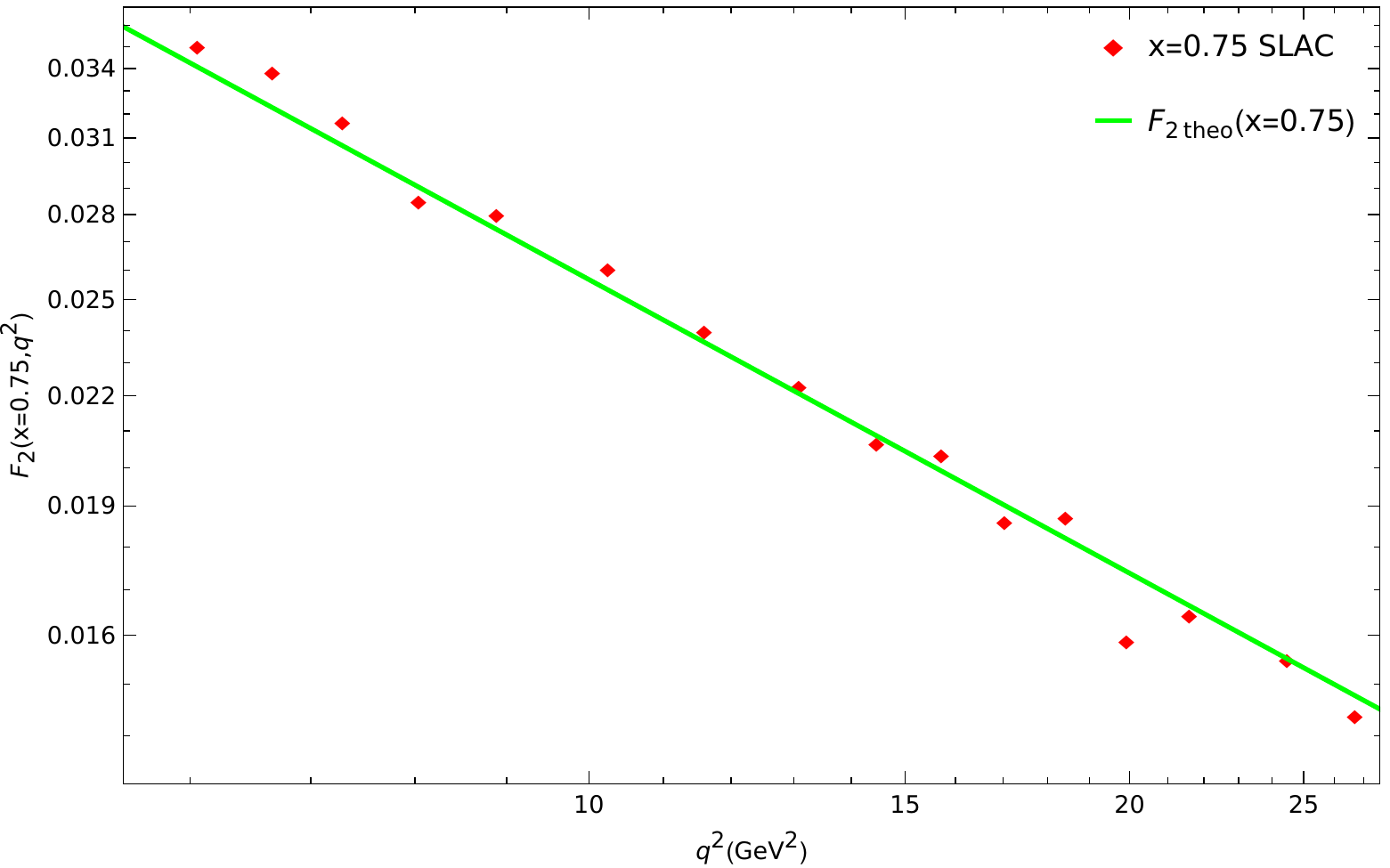}
    \includegraphics[width=0.47\linewidth]{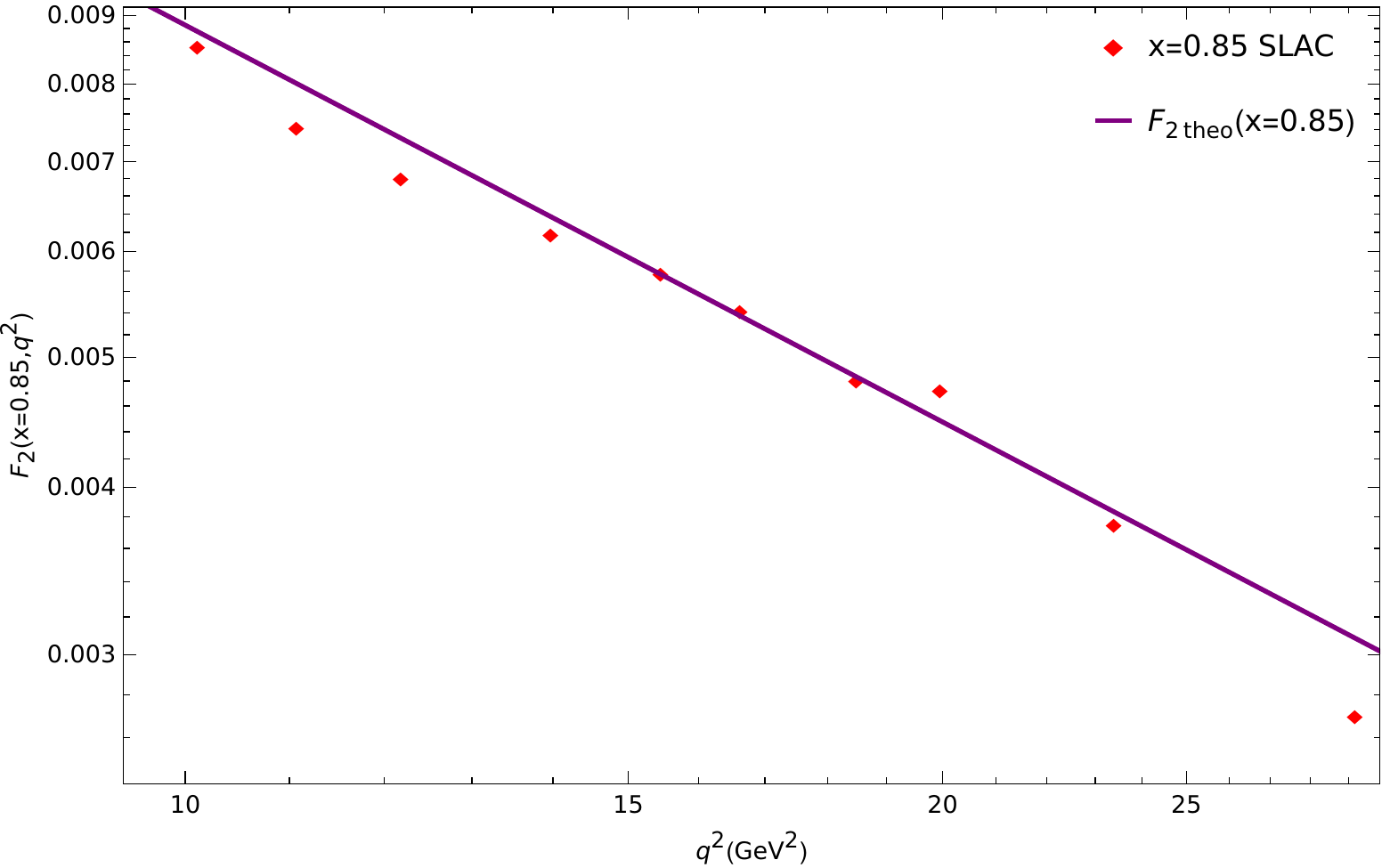}
    \caption{\sl Proton structure function $F_2$ as a function of $q^2$ for $x=0.55$ (upper left), $x=0.65$ (upper right), $x=0.75$ (lower left), and $x=0.85$ (lower right). The red dots and diamonds correspond to SLAC data \cite{Whitlow:1991uw}, while the solid lines are our results, from Eq.\ \eqref{F2}.}
    \label{F2075085}
\end{figure}

In our analysis of DIS in terms of the probe fields, there are two free parameters involved: $g_{\text{eff}}$ and $\tilde k$. The effective coupling constant $g_{\text{eff}}$ controls the overall magnitude of $F_2$ obtained from Eq.\ (\ref{F2}). For each value of $x$ we find distinct values of $g_{\rm eff}$ and $\tilde k$, defined in Eqs. \eqref{I_RL} and \eqref{geff}. These values are displayed in Table \ref{table:geff}. 
Our results for $F_2(x,q^2)$ with $x=0.55$, 0.65, 0.75 and 0.85 as functions of $q^2$ are shown in Figs. \ref{AllF2V3} and \ref{F2075085}.  With this simple setup, from our calculations for $F_2$ we obtain a very good agreement with the experimental data.


\section{Conclusions} \label{conc}

In this work, we presented a framework for computing proton structure functions in a class of Einstein-Dilaton holographic models. Our approach could also be applied to other choices of the dilaton potential $V(\phi)$. Unlike previous works, the background geometry is determined from a well-defined bulk action, which leads to equations of motion that can be numerically solved to determine the on-shell profiles of the metric and dilaton fields. Following Polchinski and Strassler, we introduce probe gauge and fermionic fields to discuss DIS from a holographic perspective. Using the metric and scalar fields self-consistently obtained from the Einstein-Dilaton model, one can then solve the corresponding equations of motion for the probe fields. With that information at hand, the structure functions can be directly computed. Therefore, from our framework, the dependence of the proton structure functions on $x$ and $q^2$ follow from the choices made for the dilaton potential $V(\phi)$, and the parameters $g_{\text{eff}}$ and $\tilde k$ associated with the interaction vertex of the probe fields.  

As an application, we use a particular realization of the Einstein-Dilaton model with a dilaton potential $V(\phi)$ constructed in such a way that the model displays linear confinement and quantitatively describes the glueball spectrum at zero temperature, in addition to describing the thermodynamic properties of the deconfined large $N$ pure Yang-Mills theory. In this case, all the parameters that define the background are fixed using a QCD input that is not connected to DIS physics. In the case of fermions, the situation is more subtle and we needed to introduce an extra potential to ensure their confinement via the definition of $\tilde m_5(\phi)$, in Eq. \eqref{m5}. Once this fixes the potential \eqref{PotSch1}, choosing the anomalous dimension $\gamma=0.817$, one finds the proton mass $\sim 0.937$ GeV associated with the ground state of the right mode, while the ground state of the left mode has a mass of 1.76 GeV. 

Using these fermionic fields with their excitations together with the photon field into the interaction vertex, Eq. \eqref{S_int}, with the appropriate choices of $g_{\rm eff}$ and $\tilde k$, shown in Table \ref{table:geff}, one finds very good agreement of our calculated $F_2$ with experimental data, as can be seen in Figs.\ \ref{AllF2V3} and \ref{F2075085}. In particular, we note that we find the same general qualitative behavior, i.e., a decreasing $F_2$ with increasing $q^2$, displayed by other holographic models that were tailored to describe $F_2$ in the same range of Bjorken $x$ considered here \cite{Braga:2011wa, FolcoCapossoli:2020pks}. 

It would be interesting to extend the present discussion to other DIS $x$ regimes, as well as DIS processes in a finite-temperature set-up, as the one encountered in quark-gluon plasmas.


\section*{Acknowledgments}

   We thank Adão Silva, Leonardo Pazito, and Eduardo Capossoli for their valuable comments on the making of this work. A.C.P.N. is supported by Coordenação de Aperfeiçoamento de Pessoal de Nível Superior (CAPES), and also acknowledges the hospitality extended to him at the Illinois Center for Advanced Studies of the Universe \& Department of Physics,
University of Illinois Urbana-Champaign, during his stay from April through September of 2024, supported by the CAPES PrInt program. H.B.-F. is partially supported by Conselho Nacional de Desenvolvimento Cient\'{\i}fico e Tecnol\'{o}gico (CNPq) under grant  310346/2023-1, and Fundação Carlos Chagas Filho de Amparo à Pesquisa do Estado do Rio de Janeiro (FAPERJ) under grant E-26/204.095/2024. 
 J.N. was partially
supported by the U.S. Department of Energy, Office of Science, Office for Nuclear Physics under Award No. DESC0023861. This research was partially supported by the National Science Foundation under grant No. NSF PHY-1748958.

\bibliography{references.bib}

\begin{thebibliography}{98}%
\makeatletter
\providecommand \@ifxundefined [1]{%
 \@ifx{#1\undefined}
}%
\providecommand \@ifnum [1]{%
 \ifnum #1\expandafter \@firstoftwo
 \else \expandafter \@secondoftwo
 \fi
}%
\providecommand \@ifx [1]{%
 \ifx #1\expandafter \@firstoftwo
 \else \expandafter \@secondoftwo
 \fi
}%
\providecommand \natexlab [1]{#1}%
\providecommand \enquote  [1]{``#1''}%
\providecommand \bibnamefont  [1]{#1}%
\providecommand \bibfnamefont [1]{#1}%
\providecommand \citenamefont [1]{#1}%
\providecommand \href@noop [0]{\@secondoftwo}%
\providecommand \href [0]{\begingroup \@sanitize@url \@href}%
\providecommand \@href[1]{\@@startlink{#1}\@@href}%
\providecommand \@@href[1]{\endgroup#1\@@endlink}%
\providecommand \@sanitize@url [0]{\catcode `\\12\catcode `\$12\catcode
  `\&12\catcode `\#12\catcode `\^12\catcode `\_12\catcode `\%12\relax}%
\providecommand \@@startlink[1]{}%
\providecommand \@@endlink[0]{}%
\providecommand \url  [0]{\begingroup\@sanitize@url \@url }%
\providecommand \@url [1]{\endgroup\@href {#1}{\urlprefix }}%
\providecommand \urlprefix  [0]{URL }%
\providecommand \Eprint [0]{\href }%
\providecommand \doibase [0]{http://dx.doi.org/}%
\providecommand \selectlanguage [0]{\@gobble}%
\providecommand \bibinfo  [0]{\@secondoftwo}%
\providecommand \bibfield  [0]{\@secondoftwo}%
\providecommand \translation [1]{[#1]}%
\providecommand \BibitemOpen [0]{}%
\providecommand \bibitemStop [0]{}%
\providecommand \bibitemNoStop [0]{.\EOS\space}%
\providecommand \EOS [0]{\spacefactor3000\relax}%
\providecommand \BibitemShut  [1]{\csname bibitem#1\endcsname}%
\let\auto@bib@innerbib\@empty
\bibitem [{\citenamefont {Breidenbach}\ \emph {et~al.}(1969)\citenamefont
  {Breidenbach}, \citenamefont {Friedman}, \citenamefont {Kendall},
  \citenamefont {Bloom}, \citenamefont {Coward}, \citenamefont {DeStaebler},
  \citenamefont {Drees}, \citenamefont {Mo},\ and\ \citenamefont
  {Taylor}}]{Breidenbach:1969kd}%
  \BibitemOpen
  \bibfield  {author} {\bibinfo {author} {\bibfnamefont {M.}~\bibnamefont
  {Breidenbach}}, \bibinfo {author} {\bibfnamefont {J.~I.}\ \bibnamefont
  {Friedman}}, \bibinfo {author} {\bibfnamefont {H.~W.}\ \bibnamefont
  {Kendall}}, \bibinfo {author} {\bibfnamefont {E.~D.}\ \bibnamefont {Bloom}},
  \bibinfo {author} {\bibfnamefont {D.~H.}\ \bibnamefont {Coward}}, \bibinfo
  {author} {\bibfnamefont {H.~C.}\ \bibnamefont {DeStaebler}}, \bibinfo
  {author} {\bibfnamefont {J.}~\bibnamefont {Drees}}, \bibinfo {author}
  {\bibfnamefont {L.~W.}\ \bibnamefont {Mo}}, \ and\ \bibinfo {author}
  {\bibfnamefont {R.~E.}\ \bibnamefont {Taylor}},\ }\href {\doibase
  10.1103/PhysRevLett.23.935} {\bibfield  {journal} {\bibinfo  {journal} {Phys.
  Rev. Lett.}\ }\textbf {\bibinfo {volume} {23}},\ \bibinfo {pages} {935}
  (\bibinfo {year} {1969})}\BibitemShut {NoStop}%
\bibitem [{\citenamefont {Bloom}\ \emph {et~al.}(1969)\citenamefont {Bloom}
  \emph {et~al.}}]{Bloom:1969kc}%
  \BibitemOpen
  \bibfield  {author} {\bibinfo {author} {\bibfnamefont {E.~D.}\ \bibnamefont
  {Bloom}} \emph {et~al.},\ }\href {\doibase 10.1103/PhysRevLett.23.930}
  {\bibfield  {journal} {\bibinfo  {journal} {Phys. Rev. Lett.}\ }\textbf
  {\bibinfo {volume} {23}},\ \bibinfo {pages} {930} (\bibinfo {year}
  {1969})}\BibitemShut {NoStop}%
\bibitem [{\citenamefont {Whitlow}\ \emph {et~al.}(1992)\citenamefont
  {Whitlow}, \citenamefont {Riordan}, \citenamefont {Dasu}, \citenamefont
  {Rock},\ and\ \citenamefont {Bodek}}]{Whitlow:1991uw}%
  \BibitemOpen
  \bibfield  {author} {\bibinfo {author} {\bibfnamefont {L.~W.}\ \bibnamefont
  {Whitlow}}, \bibinfo {author} {\bibfnamefont {E.~M.}\ \bibnamefont
  {Riordan}}, \bibinfo {author} {\bibfnamefont {S.}~\bibnamefont {Dasu}},
  \bibinfo {author} {\bibfnamefont {S.}~\bibnamefont {Rock}}, \ and\ \bibinfo
  {author} {\bibfnamefont {A.}~\bibnamefont {Bodek}},\ }\href {\doibase
  10.1016/0370-2693(92)90672-Q} {\bibfield  {journal} {\bibinfo  {journal}
  {Phys. Lett. B}\ }\textbf {\bibinfo {volume} {282}},\ \bibinfo {pages} {475}
  (\bibinfo {year} {1992})}\BibitemShut {NoStop}%
\bibitem [{\citenamefont {Gross}\ and\ \citenamefont
  {Wilczek}(1973)}]{Gross:1973id}%
  \BibitemOpen
  \bibfield  {author} {\bibinfo {author} {\bibfnamefont {D.~J.}\ \bibnamefont
  {Gross}}\ and\ \bibinfo {author} {\bibfnamefont {F.}~\bibnamefont
  {Wilczek}},\ }\href {\doibase 10.1103/PhysRevLett.30.1343} {\bibfield
  {journal} {\bibinfo  {journal} {Phys. Rev. Lett.}\ }\textbf {\bibinfo
  {volume} {30}},\ \bibinfo {pages} {1343} (\bibinfo {year}
  {1973})}\BibitemShut {NoStop}%
\bibitem [{\citenamefont {Politzer}(1973)}]{Politzer:1973fx}%
  \BibitemOpen
  \bibfield  {author} {\bibinfo {author} {\bibfnamefont {H.~D.}\ \bibnamefont
  {Politzer}},\ }\href {\doibase 10.1103/PhysRevLett.30.1346} {\bibfield
  {journal} {\bibinfo  {journal} {Phys. Rev. Lett.}\ }\textbf {\bibinfo
  {volume} {30}},\ \bibinfo {pages} {1346} (\bibinfo {year}
  {1973})}\BibitemShut {NoStop}%
\bibitem [{\citenamefont {Dokshitzer}(1977)}]{Dokshitzer:1977sg}%
  \BibitemOpen
  \bibfield  {author} {\bibinfo {author} {\bibfnamefont {Y.~L.}\ \bibnamefont
  {Dokshitzer}},\ }\href@noop {} {\bibfield  {journal} {\bibinfo  {journal}
  {Sov. Phys. JETP}\ }\textbf {\bibinfo {volume} {46}},\ \bibinfo {pages} {641}
  (\bibinfo {year} {1977})}\BibitemShut {NoStop}%
\bibitem [{\citenamefont {Gribov}\ and\ \citenamefont
  {Lipatov}(1972)}]{Gribov:1972ri}%
  \BibitemOpen
  \bibfield  {author} {\bibinfo {author} {\bibfnamefont {V.~N.}\ \bibnamefont
  {Gribov}}\ and\ \bibinfo {author} {\bibfnamefont {L.~N.}\ \bibnamefont
  {Lipatov}},\ }\href@noop {} {\bibfield  {journal} {\bibinfo  {journal} {Sov.
  J. Nucl. Phys.}\ }\textbf {\bibinfo {volume} {15}},\ \bibinfo {pages} {438}
  (\bibinfo {year} {1972})}\BibitemShut {NoStop}%
\bibitem [{\citenamefont {Altarelli}\ and\ \citenamefont
  {Parisi}(1977)}]{Altarelli:1977zs}%
  \BibitemOpen
  \bibfield  {author} {\bibinfo {author} {\bibfnamefont {G.}~\bibnamefont
  {Altarelli}}\ and\ \bibinfo {author} {\bibfnamefont {G.}~\bibnamefont
  {Parisi}},\ }\href {\doibase 10.1016/0550-3213(77)90384-4} {\bibfield
  {journal} {\bibinfo  {journal} {Nucl. Phys. B}\ }\textbf {\bibinfo {volume}
  {126}},\ \bibinfo {pages} {298} (\bibinfo {year} {1977})}\BibitemShut
  {NoStop}%
\bibitem [{\citenamefont {Gross}\ \emph {et~al.}(1977)\citenamefont {Gross},
  \citenamefont {Treiman},\ and\ \citenamefont {Wilczek}}]{Gross:1976xt}%
  \BibitemOpen
  \bibfield  {author} {\bibinfo {author} {\bibfnamefont {D.~J.}\ \bibnamefont
  {Gross}}, \bibinfo {author} {\bibfnamefont {S.~B.}\ \bibnamefont {Treiman}},
  \ and\ \bibinfo {author} {\bibfnamefont {F.~A.}\ \bibnamefont {Wilczek}},\
  }\href {\doibase 10.1103/PhysRevD.15.2486} {\bibfield  {journal} {\bibinfo
  {journal} {Phys. Rev. D}\ }\textbf {\bibinfo {volume} {15}},\ \bibinfo
  {pages} {2486} (\bibinfo {year} {1977})}\BibitemShut {NoStop}%
\bibitem [{\citenamefont {Jaroszewicz}(1982)}]{Jaroszewicz:1982gr}%
  \BibitemOpen
  \bibfield  {author} {\bibinfo {author} {\bibfnamefont {T.}~\bibnamefont
  {Jaroszewicz}},\ }\href {\doibase 10.1016/0370-2693(82)90345-8} {\bibfield
  {journal} {\bibinfo  {journal} {Phys. Lett. B}\ }\textbf {\bibinfo {volume}
  {116}},\ \bibinfo {pages} {291} (\bibinfo {year} {1982})}\BibitemShut
  {NoStop}%
\bibitem [{\citenamefont {Manohar}(1992)}]{Manohar:1992tz}%
  \BibitemOpen
  \bibfield  {author} {\bibinfo {author} {\bibfnamefont {A.~V.}\ \bibnamefont
  {Manohar}},\ }in\ \href@noop {} {\emph {\bibinfo {booktitle} {{Lake Louise
  Winter Institute: Symmetry and Spin in the Standard Model}}}}\ (\bibinfo
  {year} {1992})\ \Eprint {http://arxiv.org/abs/hep-ph/9204208}
  {arXiv:hep-ph/9204208} \BibitemShut {NoStop}%
\bibitem [{\citenamefont {Gelis}\ \emph {et~al.}(2010)\citenamefont {Gelis},
  \citenamefont {Iancu}, \citenamefont {Jalilian-Marian},\ and\ \citenamefont
  {Venugopalan}}]{Gelis:2010nm}%
  \BibitemOpen
  \bibfield  {author} {\bibinfo {author} {\bibfnamefont {F.}~\bibnamefont
  {Gelis}}, \bibinfo {author} {\bibfnamefont {E.}~\bibnamefont {Iancu}},
  \bibinfo {author} {\bibfnamefont {J.}~\bibnamefont {Jalilian-Marian}}, \ and\
  \bibinfo {author} {\bibfnamefont {R.}~\bibnamefont {Venugopalan}},\ }\href
  {\doibase 10.1146/annurev.nucl.010909.083629} {\bibfield  {journal} {\bibinfo
   {journal} {Ann. Rev. Nucl. Part. Sci.}\ }\textbf {\bibinfo {volume} {60}},\
  \bibinfo {pages} {463} (\bibinfo {year} {2010})},\ \Eprint
  {http://arxiv.org/abs/1002.0333} {arXiv:1002.0333 [hep-ph]} \BibitemShut
  {NoStop}%
\bibitem [{\citenamefont {Accardi}\ \emph {et~al.}(2016)\citenamefont {Accardi}
  \emph {et~al.}}]{Accardi:2012qut}%
  \BibitemOpen
  \bibfield  {author} {\bibinfo {author} {\bibfnamefont {A.}~\bibnamefont
  {Accardi}} \emph {et~al.},\ }\href {\doibase 10.1140/epja/i2016-16268-9}
  {\bibfield  {journal} {\bibinfo  {journal} {Eur. Phys. J. A}\ }\textbf
  {\bibinfo {volume} {52}},\ \bibinfo {pages} {268} (\bibinfo {year} {2016})},\
  \Eprint {http://arxiv.org/abs/1212.1701} {arXiv:1212.1701 [nucl-ex]}
  \BibitemShut {NoStop}%
\bibitem [{\citenamefont {Habib}(2010)}]{Habib:2010zz}%
  \BibitemOpen
  \bibfield  {author} {\bibinfo {author} {\bibfnamefont {S.}~\bibnamefont
  {Habib}} (\bibinfo {collaboration} {H1, ZEUS}),\ }\href {\doibase
  10.22323/1.106.0035} {\bibfield  {journal} {\bibinfo  {journal} {PoS}\
  }\textbf {\bibinfo {volume} {DIS2010}},\ \bibinfo {pages} {035} (\bibinfo
  {year} {2010})}\BibitemShut {NoStop}%
\bibitem [{\citenamefont {Maldacena}(1998)}]{Maldacena:1997re}%
  \BibitemOpen
  \bibfield  {author} {\bibinfo {author} {\bibfnamefont {J.~M.}\ \bibnamefont
  {Maldacena}},\ }\href {\doibase 10.4310/ATMP.1998.v2.n2.a1} {\bibfield
  {journal} {\bibinfo  {journal} {Adv. Theor. Math. Phys.}\ }\textbf {\bibinfo
  {volume} {2}},\ \bibinfo {pages} {231} (\bibinfo {year} {1998})},\ \Eprint
  {http://arxiv.org/abs/hep-th/9711200} {arXiv:hep-th/9711200} \BibitemShut
  {NoStop}%
\bibitem [{\citenamefont {Gubser}\ \emph {et~al.}(1998)\citenamefont {Gubser},
  \citenamefont {Klebanov},\ and\ \citenamefont {Polyakov}}]{Gubser:1998bc}%
  \BibitemOpen
  \bibfield  {author} {\bibinfo {author} {\bibfnamefont {S.~S.}\ \bibnamefont
  {Gubser}}, \bibinfo {author} {\bibfnamefont {I.~R.}\ \bibnamefont
  {Klebanov}}, \ and\ \bibinfo {author} {\bibfnamefont {A.~M.}\ \bibnamefont
  {Polyakov}},\ }\href {\doibase 10.1016/S0370-2693(98)00377-3} {\bibfield
  {journal} {\bibinfo  {journal} {Phys. Lett. B}\ }\textbf {\bibinfo {volume}
  {428}},\ \bibinfo {pages} {105} (\bibinfo {year} {1998})},\ \Eprint
  {http://arxiv.org/abs/hep-th/9802109} {arXiv:hep-th/9802109} \BibitemShut
  {NoStop}%
\bibitem [{\citenamefont {Witten}(1998)}]{Witten:1998qj}%
  \BibitemOpen
  \bibfield  {author} {\bibinfo {author} {\bibfnamefont {E.}~\bibnamefont
  {Witten}},\ }\href {\doibase 10.4310/ATMP.1998.v2.n2.a2} {\bibfield
  {journal} {\bibinfo  {journal} {Adv. Theor. Math. Phys.}\ }\textbf {\bibinfo
  {volume} {2}},\ \bibinfo {pages} {253} (\bibinfo {year} {1998})},\ \Eprint
  {http://arxiv.org/abs/hep-th/9802150} {arXiv:hep-th/9802150} \BibitemShut
  {NoStop}%
\bibitem [{\citenamefont {Aharony}\ \emph {et~al.}(2000)\citenamefont
  {Aharony}, \citenamefont {Gubser}, \citenamefont {Maldacena}, \citenamefont
  {Ooguri},\ and\ \citenamefont {Oz}}]{Aharony:1999ti}%
  \BibitemOpen
  \bibfield  {author} {\bibinfo {author} {\bibfnamefont {O.}~\bibnamefont
  {Aharony}}, \bibinfo {author} {\bibfnamefont {S.~S.}\ \bibnamefont {Gubser}},
  \bibinfo {author} {\bibfnamefont {J.~M.}\ \bibnamefont {Maldacena}}, \bibinfo
  {author} {\bibfnamefont {H.}~\bibnamefont {Ooguri}}, \ and\ \bibinfo {author}
  {\bibfnamefont {Y.}~\bibnamefont {Oz}},\ }\href {\doibase
  10.1016/S0370-1573(99)00083-6} {\bibfield  {journal} {\bibinfo  {journal}
  {Phys. Rept.}\ }\textbf {\bibinfo {volume} {323}},\ \bibinfo {pages} {183}
  (\bibinfo {year} {2000})},\ \Eprint {http://arxiv.org/abs/hep-th/9905111}
  {arXiv:hep-th/9905111} \BibitemShut {NoStop}%
\bibitem [{\citenamefont {Gubser}\ \emph {et~al.}(2002)\citenamefont {Gubser},
  \citenamefont {Klebanov},\ and\ \citenamefont {Polyakov}}]{Gubser:2002tv}%
  \BibitemOpen
  \bibfield  {author} {\bibinfo {author} {\bibfnamefont {S.~S.}\ \bibnamefont
  {Gubser}}, \bibinfo {author} {\bibfnamefont {I.~R.}\ \bibnamefont
  {Klebanov}}, \ and\ \bibinfo {author} {\bibfnamefont {A.~M.}\ \bibnamefont
  {Polyakov}},\ }\href {\doibase 10.1016/S0550-3213(02)00373-5} {\bibfield
  {journal} {\bibinfo  {journal} {Nucl. Phys. B}\ }\textbf {\bibinfo {volume}
  {636}},\ \bibinfo {pages} {99} (\bibinfo {year} {2002})},\ \Eprint
  {http://arxiv.org/abs/hep-th/0204051} {arXiv:hep-th/0204051} \BibitemShut
  {NoStop}%
\bibitem [{\citenamefont {Polchinski}\ and\ \citenamefont
  {Strassler}(2003)}]{Polchinski:2002jw}%
  \BibitemOpen
  \bibfield  {author} {\bibinfo {author} {\bibfnamefont {J.}~\bibnamefont
  {Polchinski}}\ and\ \bibinfo {author} {\bibfnamefont {M.~J.}\ \bibnamefont
  {Strassler}},\ }\href {\doibase 10.1088/1126-6708/2003/05/012} {\bibfield
  {journal} {\bibinfo  {journal} {JHEP}\ }\textbf {\bibinfo {volume} {05}},\
  \bibinfo {pages} {012} (\bibinfo {year} {2003})},\ \Eprint
  {http://arxiv.org/abs/hep-th/0209211} {arXiv:hep-th/0209211} \BibitemShut
  {NoStop}%
\bibitem [{\citenamefont {Borsa}\ \emph {et~al.}(2023)\citenamefont {Borsa},
  \citenamefont {Jorrin}, \citenamefont {Sassot},\ and\ \citenamefont
  {Schvellinger}}]{Borsa:2023tqr}%
  \BibitemOpen
  \bibfield  {author} {\bibinfo {author} {\bibfnamefont {I.}~\bibnamefont
  {Borsa}}, \bibinfo {author} {\bibfnamefont {D.}~\bibnamefont {Jorrin}},
  \bibinfo {author} {\bibfnamefont {R.}~\bibnamefont {Sassot}}, \ and\ \bibinfo
  {author} {\bibfnamefont {M.}~\bibnamefont {Schvellinger}},\ }\href {\doibase
  10.1103/PhysRevD.108.056024} {\bibfield  {journal} {\bibinfo  {journal}
  {Phys. Rev. D}\ }\textbf {\bibinfo {volume} {108}},\ \bibinfo {pages}
  {056024} (\bibinfo {year} {2023})},\ \Eprint
  {http://arxiv.org/abs/2308.01975} {arXiv:2308.01975 [hep-ph]} \BibitemShut
  {NoStop}%
\bibitem [{\citenamefont {Kovensky}\ \emph
  {et~al.}(2018{\natexlab{a}})\citenamefont {Kovensky}, \citenamefont
  {Michalski},\ and\ \citenamefont {Schvellinger}}]{Kovensky:2018xxa}%
  \BibitemOpen
  \bibfield  {author} {\bibinfo {author} {\bibfnamefont {N.}~\bibnamefont
  {Kovensky}}, \bibinfo {author} {\bibfnamefont {G.}~\bibnamefont {Michalski}},
  \ and\ \bibinfo {author} {\bibfnamefont {M.}~\bibnamefont {Schvellinger}},\
  }\href {\doibase 10.1007/JHEP10(2018)084} {\bibfield  {journal} {\bibinfo
  {journal} {JHEP}\ }\textbf {\bibinfo {volume} {10}},\ \bibinfo {pages} {084}
  (\bibinfo {year} {2018}{\natexlab{a}})},\ \Eprint
  {http://arxiv.org/abs/1807.11540} {arXiv:1807.11540 [hep-th]} \BibitemShut
  {NoStop}%
\bibitem [{\citenamefont {Hatta}\ \emph {et~al.}(2008)\citenamefont {Hatta},
  \citenamefont {Iancu},\ and\ \citenamefont {Mueller}}]{Hatta:2007he}%
  \BibitemOpen
  \bibfield  {author} {\bibinfo {author} {\bibfnamefont {Y.}~\bibnamefont
  {Hatta}}, \bibinfo {author} {\bibfnamefont {E.}~\bibnamefont {Iancu}}, \ and\
  \bibinfo {author} {\bibfnamefont {A.~H.}\ \bibnamefont {Mueller}},\ }\href
  {\doibase 10.1088/1126-6708/2008/01/026} {\bibfield  {journal} {\bibinfo
  {journal} {JHEP}\ }\textbf {\bibinfo {volume} {01}},\ \bibinfo {pages} {026}
  (\bibinfo {year} {2008})},\ \Eprint {http://arxiv.org/abs/0710.2148}
  {arXiv:0710.2148 [hep-th]} \BibitemShut {NoStop}%
\bibitem [{\citenamefont {Ballon~Bayona}\ \emph
  {et~al.}(2008{\natexlab{a}})\citenamefont {Ballon~Bayona}, \citenamefont
  {Boschi-Filho},\ and\ \citenamefont {Braga}}]{BallonBayona:2007rs}%
  \BibitemOpen
  \bibfield  {author} {\bibinfo {author} {\bibfnamefont {C.~A.}\ \bibnamefont
  {Ballon~Bayona}}, \bibinfo {author} {\bibfnamefont {H.}~\bibnamefont
  {Boschi-Filho}}, \ and\ \bibinfo {author} {\bibfnamefont {N.~R.~F.}\
  \bibnamefont {Braga}},\ }\href {\doibase 10.1088/1126-6708/2008/10/088}
  {\bibfield  {journal} {\bibinfo  {journal} {JHEP}\ }\textbf {\bibinfo
  {volume} {10}},\ \bibinfo {pages} {088} (\bibinfo {year}
  {2008}{\natexlab{a}})},\ \Eprint {http://arxiv.org/abs/0712.3530}
  {arXiv:0712.3530 [hep-th]} \BibitemShut {NoStop}%
\bibitem [{\citenamefont {Ballon~Bayona}\ \emph
  {et~al.}(2008{\natexlab{b}})\citenamefont {Ballon~Bayona}, \citenamefont
  {Boschi-Filho},\ and\ \citenamefont {Braga}}]{BallonBayona:2007qr}%
  \BibitemOpen
  \bibfield  {author} {\bibinfo {author} {\bibfnamefont {C.~A.}\ \bibnamefont
  {Ballon~Bayona}}, \bibinfo {author} {\bibfnamefont {H.}~\bibnamefont
  {Boschi-Filho}}, \ and\ \bibinfo {author} {\bibfnamefont {N.~R.~F.}\
  \bibnamefont {Braga}},\ }\href {\doibase 10.1088/1126-6708/2008/03/064}
  {\bibfield  {journal} {\bibinfo  {journal} {JHEP}\ }\textbf {\bibinfo
  {volume} {03}},\ \bibinfo {pages} {064} (\bibinfo {year}
  {2008}{\natexlab{b}})},\ \Eprint {http://arxiv.org/abs/0711.0221}
  {arXiv:0711.0221 [hep-th]} \BibitemShut {NoStop}%
\bibitem [{\citenamefont {Cornalba}\ and\ \citenamefont
  {Costa}(2008)}]{Cornalba:2008sp}%
  \BibitemOpen
  \bibfield  {author} {\bibinfo {author} {\bibfnamefont {L.}~\bibnamefont
  {Cornalba}}\ and\ \bibinfo {author} {\bibfnamefont {M.~S.}\ \bibnamefont
  {Costa}},\ }\href {\doibase 10.1103/PhysRevD.78.096010} {\bibfield  {journal}
  {\bibinfo  {journal} {Phys. Rev. D}\ }\textbf {\bibinfo {volume} {78}},\
  \bibinfo {pages} {096010} (\bibinfo {year} {2008})},\ \Eprint
  {http://arxiv.org/abs/0804.1562} {arXiv:0804.1562 [hep-ph]} \BibitemShut
  {NoStop}%
\bibitem [{\citenamefont {Pire}\ \emph {et~al.}(2008)\citenamefont {Pire},
  \citenamefont {Roiesnel}, \citenamefont {Szymanowski},\ and\ \citenamefont
  {Wallon}}]{Pire:2008zf}%
  \BibitemOpen
  \bibfield  {author} {\bibinfo {author} {\bibfnamefont {B.}~\bibnamefont
  {Pire}}, \bibinfo {author} {\bibfnamefont {C.}~\bibnamefont {Roiesnel}},
  \bibinfo {author} {\bibfnamefont {L.}~\bibnamefont {Szymanowski}}, \ and\
  \bibinfo {author} {\bibfnamefont {S.}~\bibnamefont {Wallon}},\ }\href
  {\doibase 10.1016/j.physletb.2008.10.026} {\bibfield  {journal} {\bibinfo
  {journal} {Phys. Lett. B}\ }\textbf {\bibinfo {volume} {670}},\ \bibinfo
  {pages} {84} (\bibinfo {year} {2008})},\ \Eprint
  {http://arxiv.org/abs/0805.4346} {arXiv:0805.4346 [hep-ph]} \BibitemShut
  {NoStop}%
\bibitem [{\citenamefont {Albacete}\ \emph {et~al.}(2008)\citenamefont
  {Albacete}, \citenamefont {Kovchegov},\ and\ \citenamefont
  {Taliotis}}]{Albacete:2008ze}%
  \BibitemOpen
  \bibfield  {author} {\bibinfo {author} {\bibfnamefont {J.~L.}\ \bibnamefont
  {Albacete}}, \bibinfo {author} {\bibfnamefont {Y.~V.}\ \bibnamefont
  {Kovchegov}}, \ and\ \bibinfo {author} {\bibfnamefont {A.}~\bibnamefont
  {Taliotis}},\ }\href {\doibase 10.1088/1126-6708/2008/07/074} {\bibfield
  {journal} {\bibinfo  {journal} {JHEP}\ }\textbf {\bibinfo {volume} {07}},\
  \bibinfo {pages} {074} (\bibinfo {year} {2008})},\ \Eprint
  {http://arxiv.org/abs/0806.1484} {arXiv:0806.1484 [hep-th]} \BibitemShut
  {NoStop}%
\bibitem [{\citenamefont {Ballon~Bayona}\ \emph
  {et~al.}(2008{\natexlab{c}})\citenamefont {Ballon~Bayona}, \citenamefont
  {Boschi-Filho},\ and\ \citenamefont {Braga}}]{BallonBayona:2008zi}%
  \BibitemOpen
  \bibfield  {author} {\bibinfo {author} {\bibfnamefont {C.~A.}\ \bibnamefont
  {Ballon~Bayona}}, \bibinfo {author} {\bibfnamefont {H.}~\bibnamefont
  {Boschi-Filho}}, \ and\ \bibinfo {author} {\bibfnamefont {N.~R.~F.}\
  \bibnamefont {Braga}},\ }\href {\doibase 10.1088/1126-6708/2008/09/114}
  {\bibfield  {journal} {\bibinfo  {journal} {JHEP}\ }\textbf {\bibinfo
  {volume} {09}},\ \bibinfo {pages} {114} (\bibinfo {year}
  {2008}{\natexlab{c}})},\ \Eprint {http://arxiv.org/abs/0807.1917}
  {arXiv:0807.1917 [hep-th]} \BibitemShut {NoStop}%
\bibitem [{\citenamefont {Gao}\ and\ \citenamefont {Xiao}(2009)}]{Gao:2009ze}%
  \BibitemOpen
  \bibfield  {author} {\bibinfo {author} {\bibfnamefont {J.-H.}\ \bibnamefont
  {Gao}}\ and\ \bibinfo {author} {\bibfnamefont {B.-W.}\ \bibnamefont {Xiao}},\
  }\href {\doibase 10.1103/PhysRevD.80.015025} {\bibfield  {journal} {\bibinfo
  {journal} {Phys. Rev. D}\ }\textbf {\bibinfo {volume} {80}},\ \bibinfo
  {pages} {015025} (\bibinfo {year} {2009})},\ \Eprint
  {http://arxiv.org/abs/0904.2870} {arXiv:0904.2870 [hep-ph]} \BibitemShut
  {NoStop}%
\bibitem [{\citenamefont {Taliotis}(2009)}]{Taliotis:2009ne}%
  \BibitemOpen
  \bibfield  {author} {\bibinfo {author} {\bibfnamefont {A.}~\bibnamefont
  {Taliotis}},\ }\href {\doibase 10.1016/j.nuclphysa.2009.10.026} {\bibfield
  {journal} {\bibinfo  {journal} {Nucl. Phys. A}\ }\textbf {\bibinfo {volume}
  {830}},\ \bibinfo {pages} {299C} (\bibinfo {year} {2009})},\ \Eprint
  {http://arxiv.org/abs/0907.4204} {arXiv:0907.4204 [hep-th]} \BibitemShut
  {NoStop}%
\bibitem [{\citenamefont {Yoshida}(2010)}]{Yoshida:2009dw}%
  \BibitemOpen
  \bibfield  {author} {\bibinfo {author} {\bibfnamefont {Y.}~\bibnamefont
  {Yoshida}},\ }\href {\doibase 10.1143/PTP.123.79} {\bibfield  {journal}
  {\bibinfo  {journal} {Prog. Theor. Phys.}\ }\textbf {\bibinfo {volume}
  {123}},\ \bibinfo {pages} {79} (\bibinfo {year} {2010})},\ \Eprint
  {http://arxiv.org/abs/0902.1015} {arXiv:0902.1015 [hep-th]} \BibitemShut
  {NoStop}%
\bibitem [{\citenamefont {Hatta}\ \emph {et~al.}(2009)\citenamefont {Hatta},
  \citenamefont {Ueda},\ and\ \citenamefont {Xiao}}]{Hatta:2009ra}%
  \BibitemOpen
  \bibfield  {author} {\bibinfo {author} {\bibfnamefont {Y.}~\bibnamefont
  {Hatta}}, \bibinfo {author} {\bibfnamefont {T.}~\bibnamefont {Ueda}}, \ and\
  \bibinfo {author} {\bibfnamefont {B.-W.}\ \bibnamefont {Xiao}},\ }\href
  {\doibase 10.1088/1126-6708/2009/08/007} {\bibfield  {journal} {\bibinfo
  {journal} {JHEP}\ }\textbf {\bibinfo {volume} {08}},\ \bibinfo {pages} {007}
  (\bibinfo {year} {2009})},\ \Eprint {http://arxiv.org/abs/0905.2493}
  {arXiv:0905.2493 [hep-ph]} \BibitemShut {NoStop}%
\bibitem [{\citenamefont {Avsar}\ \emph {et~al.}(2009)\citenamefont {Avsar},
  \citenamefont {Iancu}, \citenamefont {McLerran},\ and\ \citenamefont
  {Triantafyllopoulos}}]{Avsar:2009xf}%
  \BibitemOpen
  \bibfield  {author} {\bibinfo {author} {\bibfnamefont {E.}~\bibnamefont
  {Avsar}}, \bibinfo {author} {\bibfnamefont {E.}~\bibnamefont {Iancu}},
  \bibinfo {author} {\bibfnamefont {L.}~\bibnamefont {McLerran}}, \ and\
  \bibinfo {author} {\bibfnamefont {D.~N.}\ \bibnamefont
  {Triantafyllopoulos}},\ }\href {\doibase 10.1088/1126-6708/2009/11/105}
  {\bibfield  {journal} {\bibinfo  {journal} {JHEP}\ }\textbf {\bibinfo
  {volume} {11}},\ \bibinfo {pages} {105} (\bibinfo {year} {2009})},\ \Eprint
  {http://arxiv.org/abs/0907.4604} {arXiv:0907.4604 [hep-th]} \BibitemShut
  {NoStop}%
\bibitem [{\citenamefont {Cornalba}\ \emph
  {et~al.}(2010{\natexlab{a}})\citenamefont {Cornalba}, \citenamefont {Costa},\
  and\ \citenamefont {Penedones}}]{Cornalba:2009ax}%
  \BibitemOpen
  \bibfield  {author} {\bibinfo {author} {\bibfnamefont {L.}~\bibnamefont
  {Cornalba}}, \bibinfo {author} {\bibfnamefont {M.~S.}\ \bibnamefont {Costa}},
  \ and\ \bibinfo {author} {\bibfnamefont {J.}~\bibnamefont {Penedones}},\
  }\href {\doibase 10.1007/JHEP03(2010)133} {\bibfield  {journal} {\bibinfo
  {journal} {JHEP}\ }\textbf {\bibinfo {volume} {03}},\ \bibinfo {pages} {133}
  (\bibinfo {year} {2010}{\natexlab{a}})},\ \Eprint
  {http://arxiv.org/abs/0911.0043} {arXiv:0911.0043 [hep-th]} \BibitemShut
  {NoStop}%
\bibitem [{\citenamefont {Bayona}\ \emph {et~al.}(2010)\citenamefont {Bayona},
  \citenamefont {Boschi-Filho},\ and\ \citenamefont {Braga}}]{Bayona:2009qe}%
  \BibitemOpen
  \bibfield  {author} {\bibinfo {author} {\bibfnamefont {C.~A.~B.}\
  \bibnamefont {Bayona}}, \bibinfo {author} {\bibfnamefont {H.}~\bibnamefont
  {Boschi-Filho}}, \ and\ \bibinfo {author} {\bibfnamefont {N.~R.~F.}\
  \bibnamefont {Braga}},\ }\href {\doibase 10.1103/PhysRevD.81.086003}
  {\bibfield  {journal} {\bibinfo  {journal} {Phys. Rev. D}\ }\textbf {\bibinfo
  {volume} {81}},\ \bibinfo {pages} {086003} (\bibinfo {year} {2010})},\
  \Eprint {http://arxiv.org/abs/0912.0231} {arXiv:0912.0231 [hep-th]}
  \BibitemShut {NoStop}%
\bibitem [{\citenamefont {Cornalba}\ \emph
  {et~al.}(2010{\natexlab{b}})\citenamefont {Cornalba}, \citenamefont {Costa},\
  and\ \citenamefont {Penedones}}]{Cornalba:2010vk}%
  \BibitemOpen
  \bibfield  {author} {\bibinfo {author} {\bibfnamefont {L.}~\bibnamefont
  {Cornalba}}, \bibinfo {author} {\bibfnamefont {M.~S.}\ \bibnamefont {Costa}},
  \ and\ \bibinfo {author} {\bibfnamefont {J.}~\bibnamefont {Penedones}},\
  }\href {\doibase 10.1103/PhysRevLett.105.072003} {\bibfield  {journal}
  {\bibinfo  {journal} {Phys. Rev. Lett.}\ }\textbf {\bibinfo {volume} {105}},\
  \bibinfo {pages} {072003} (\bibinfo {year} {2010}{\natexlab{b}})},\ \Eprint
  {http://arxiv.org/abs/1001.1157} {arXiv:1001.1157 [hep-ph]} \BibitemShut
  {NoStop}%
\bibitem [{\citenamefont {Brower}\ \emph {et~al.}(2010)\citenamefont {Brower},
  \citenamefont {Djuric}, \citenamefont {Sarcevic},\ and\ \citenamefont
  {Tan}}]{Brower:2010wf}%
  \BibitemOpen
  \bibfield  {author} {\bibinfo {author} {\bibfnamefont {R.~C.}\ \bibnamefont
  {Brower}}, \bibinfo {author} {\bibfnamefont {M.}~\bibnamefont {Djuric}},
  \bibinfo {author} {\bibfnamefont {I.}~\bibnamefont {Sarcevic}}, \ and\
  \bibinfo {author} {\bibfnamefont {C.-I.}\ \bibnamefont {Tan}},\ }\href
  {\doibase 10.1007/JHEP11(2010)051} {\bibfield  {journal} {\bibinfo  {journal}
  {JHEP}\ }\textbf {\bibinfo {volume} {11}},\ \bibinfo {pages} {051} (\bibinfo
  {year} {2010})},\ \Eprint {http://arxiv.org/abs/1007.2259} {arXiv:1007.2259
  [hep-ph]} \BibitemShut {NoStop}%
\bibitem [{\citenamefont {Gao}\ and\ \citenamefont {Mou}(2010)}]{Gao:2010qk}%
  \BibitemOpen
  \bibfield  {author} {\bibinfo {author} {\bibfnamefont {J.-H.}\ \bibnamefont
  {Gao}}\ and\ \bibinfo {author} {\bibfnamefont {Z.-G.}\ \bibnamefont {Mou}},\
  }\href {\doibase 10.1103/PhysRevD.81.096006} {\bibfield  {journal} {\bibinfo
  {journal} {Phys. Rev. D}\ }\textbf {\bibinfo {volume} {81}},\ \bibinfo
  {pages} {096006} (\bibinfo {year} {2010})},\ \Eprint
  {http://arxiv.org/abs/1003.3066} {arXiv:1003.3066 [hep-ph]} \BibitemShut
  {NoStop}%
\bibitem [{\citenamefont {Ballon~Bayona}\ \emph {et~al.}(2010)\citenamefont
  {Ballon~Bayona}, \citenamefont {Boschi-Filho}, \citenamefont {Braga},\ and\
  \citenamefont {Torres}}]{BallonBayona:2010ae}%
  \BibitemOpen
  \bibfield  {author} {\bibinfo {author} {\bibfnamefont {C.~A.}\ \bibnamefont
  {Ballon~Bayona}}, \bibinfo {author} {\bibfnamefont {H.}~\bibnamefont
  {Boschi-Filho}}, \bibinfo {author} {\bibfnamefont {N.~R.~F.}\ \bibnamefont
  {Braga}}, \ and\ \bibinfo {author} {\bibfnamefont {M.~A.~C.}\ \bibnamefont
  {Torres}},\ }\href {\doibase 10.1007/JHEP10(2010)055} {\bibfield  {journal}
  {\bibinfo  {journal} {JHEP}\ }\textbf {\bibinfo {volume} {10}},\ \bibinfo
  {pages} {055} (\bibinfo {year} {2010})},\ \Eprint
  {http://arxiv.org/abs/1007.2448} {arXiv:1007.2448 [hep-th]} \BibitemShut
  {NoStop}%
\bibitem [{\citenamefont {Braga}\ and\ \citenamefont
  {Vega}(2012)}]{Braga:2011wa}%
  \BibitemOpen
  \bibfield  {author} {\bibinfo {author} {\bibfnamefont {N.~R.~F.}\
  \bibnamefont {Braga}}\ and\ \bibinfo {author} {\bibfnamefont
  {A.}~\bibnamefont {Vega}},\ }\href {\doibase 10.1140/epjc/s10052-012-2236-2}
  {\bibfield  {journal} {\bibinfo  {journal} {Eur. Phys. J. C}\ }\textbf
  {\bibinfo {volume} {72}},\ \bibinfo {pages} {2236} (\bibinfo {year}
  {2012})},\ \Eprint {http://arxiv.org/abs/1110.2548} {arXiv:1110.2548
  [hep-ph]} \BibitemShut {NoStop}%
\bibitem [{\citenamefont {Koile}\ \emph {et~al.}(2014)\citenamefont {Koile},
  \citenamefont {Macaluso},\ and\ \citenamefont
  {Schvellinger}}]{Koile:2013hba}%
  \BibitemOpen
  \bibfield  {author} {\bibinfo {author} {\bibfnamefont {E.}~\bibnamefont
  {Koile}}, \bibinfo {author} {\bibfnamefont {S.}~\bibnamefont {Macaluso}}, \
  and\ \bibinfo {author} {\bibfnamefont {M.}~\bibnamefont {Schvellinger}},\
  }\href {\doibase 10.1007/JHEP01(2014)166} {\bibfield  {journal} {\bibinfo
  {journal} {JHEP}\ }\textbf {\bibinfo {volume} {01}},\ \bibinfo {pages} {166}
  (\bibinfo {year} {2014})},\ \Eprint {http://arxiv.org/abs/1311.2601}
  {arXiv:1311.2601 [hep-th]} \BibitemShut {NoStop}%
\bibitem [{\citenamefont {Koile}\ \emph
  {et~al.}(2015{\natexlab{a}})\citenamefont {Koile}, \citenamefont {Kovensky},\
  and\ \citenamefont {Schvellinger}}]{Koile:2014vca}%
  \BibitemOpen
  \bibfield  {author} {\bibinfo {author} {\bibfnamefont {E.}~\bibnamefont
  {Koile}}, \bibinfo {author} {\bibfnamefont {N.}~\bibnamefont {Kovensky}}, \
  and\ \bibinfo {author} {\bibfnamefont {M.}~\bibnamefont {Schvellinger}},\
  }\href {\doibase 10.1007/JHEP05(2015)001} {\bibfield  {journal} {\bibinfo
  {journal} {JHEP}\ }\textbf {\bibinfo {volume} {05}},\ \bibinfo {pages} {001}
  (\bibinfo {year} {2015}{\natexlab{a}})},\ \Eprint
  {http://arxiv.org/abs/1412.6509} {arXiv:1412.6509 [hep-th]} \BibitemShut
  {NoStop}%
\bibitem [{\citenamefont {Gao}\ and\ \citenamefont {Mou}(2014)}]{Gao:2014nwa}%
  \BibitemOpen
  \bibfield  {author} {\bibinfo {author} {\bibfnamefont {J.-H.}\ \bibnamefont
  {Gao}}\ and\ \bibinfo {author} {\bibfnamefont {Z.-G.}\ \bibnamefont {Mou}},\
  }\href {\doibase 10.1103/PhysRevD.90.075018} {\bibfield  {journal} {\bibinfo
  {journal} {Phys. Rev. D}\ }\textbf {\bibinfo {volume} {90}},\ \bibinfo
  {pages} {075018} (\bibinfo {year} {2014})},\ \Eprint
  {http://arxiv.org/abs/1406.7576} {arXiv:1406.7576 [hep-ph]} \BibitemShut
  {NoStop}%
\bibitem [{\citenamefont {Folco~Capossoli}\ and\ \citenamefont
  {Boschi-Filho}(2015)}]{Capossoli:2015sfa}%
  \BibitemOpen
  \bibfield  {author} {\bibinfo {author} {\bibfnamefont {E.}~\bibnamefont
  {Folco~Capossoli}}\ and\ \bibinfo {author} {\bibfnamefont {H.}~\bibnamefont
  {Boschi-Filho}},\ }\href {\doibase 10.1103/PhysRevD.92.126012} {\bibfield
  {journal} {\bibinfo  {journal} {Phys. Rev. D}\ }\textbf {\bibinfo {volume}
  {92}},\ \bibinfo {pages} {126012} (\bibinfo {year} {2015})},\ \Eprint
  {http://arxiv.org/abs/1509.01761} {arXiv:1509.01761 [hep-th]} \BibitemShut
  {NoStop}%
\bibitem [{\citenamefont {Folco~Capossoli}\ \emph {et~al.}(2020)\citenamefont
  {Folco~Capossoli}, \citenamefont {Mart\'\i{}n~Contreras}, \citenamefont {Li},
  \citenamefont {Vega},\ and\ \citenamefont
  {Boschi-Filho}}]{FolcoCapossoli:2020pks}%
  \BibitemOpen
  \bibfield  {author} {\bibinfo {author} {\bibfnamefont {E.}~\bibnamefont
  {Folco~Capossoli}}, \bibinfo {author} {\bibfnamefont {M.~A.}\ \bibnamefont
  {Mart\'\i{}n~Contreras}}, \bibinfo {author} {\bibfnamefont {D.}~\bibnamefont
  {Li}}, \bibinfo {author} {\bibfnamefont {A.}~\bibnamefont {Vega}}, \ and\
  \bibinfo {author} {\bibfnamefont {H.}~\bibnamefont {Boschi-Filho}},\ }\href
  {\doibase 10.1103/PhysRevD.102.086004} {\bibfield  {journal} {\bibinfo
  {journal} {Phys. Rev. D}\ }\textbf {\bibinfo {volume} {102}},\ \bibinfo
  {pages} {086004} (\bibinfo {year} {2020})},\ \Eprint
  {http://arxiv.org/abs/2007.09283} {arXiv:2007.09283 [hep-ph]} \BibitemShut
  {NoStop}%
\bibitem [{\citenamefont {Koile}\ \emph
  {et~al.}(2015{\natexlab{b}})\citenamefont {Koile}, \citenamefont {Kovensky},\
  and\ \citenamefont {Schvellinger}}]{Koile:2015qsa}%
  \BibitemOpen
  \bibfield  {author} {\bibinfo {author} {\bibfnamefont {E.}~\bibnamefont
  {Koile}}, \bibinfo {author} {\bibfnamefont {N.}~\bibnamefont {Kovensky}}, \
  and\ \bibinfo {author} {\bibfnamefont {M.}~\bibnamefont {Schvellinger}},\
  }\href {\doibase 10.1007/JHEP12(2015)009} {\bibfield  {journal} {\bibinfo
  {journal} {JHEP}\ }\textbf {\bibinfo {volume} {12}},\ \bibinfo {pages} {009}
  (\bibinfo {year} {2015}{\natexlab{b}})},\ \Eprint
  {http://arxiv.org/abs/1507.07942} {arXiv:1507.07942 [hep-th]} \BibitemShut
  {NoStop}%
\bibitem [{\citenamefont {Jorrin}\ \emph
  {et~al.}(2016{\natexlab{a}})\citenamefont {Jorrin}, \citenamefont
  {Kovensky},\ and\ \citenamefont {Schvellinger}}]{Jorrin:2016rbx}%
  \BibitemOpen
  \bibfield  {author} {\bibinfo {author} {\bibfnamefont {D.}~\bibnamefont
  {Jorrin}}, \bibinfo {author} {\bibfnamefont {N.}~\bibnamefont {Kovensky}}, \
  and\ \bibinfo {author} {\bibfnamefont {M.}~\bibnamefont {Schvellinger}},\
  }\href {\doibase 10.1007/JHEP04(2016)113} {\bibfield  {journal} {\bibinfo
  {journal} {JHEP}\ }\textbf {\bibinfo {volume} {04}},\ \bibinfo {pages} {113}
  (\bibinfo {year} {2016}{\natexlab{a}})},\ \Eprint
  {http://arxiv.org/abs/1601.01627} {arXiv:1601.01627 [hep-th]} \BibitemShut
  {NoStop}%
\bibitem [{\citenamefont {Jorrin}\ \emph
  {et~al.}(2016{\natexlab{b}})\citenamefont {Jorrin}, \citenamefont
  {Schvellinger},\ and\ \citenamefont {Kovensky}}]{Kovensky:2016ryy}%
  \BibitemOpen
  \bibfield  {author} {\bibinfo {author} {\bibfnamefont {D.}~\bibnamefont
  {Jorrin}}, \bibinfo {author} {\bibfnamefont {M.}~\bibnamefont
  {Schvellinger}}, \ and\ \bibinfo {author} {\bibfnamefont {N.}~\bibnamefont
  {Kovensky}},\ }\href {\doibase 10.1007/JHEP12(2016)003} {\bibfield  {journal}
  {\bibinfo  {journal} {JHEP}\ }\textbf {\bibinfo {volume} {12}},\ \bibinfo
  {pages} {003} (\bibinfo {year} {2016}{\natexlab{b}})},\ \Eprint
  {http://arxiv.org/abs/1609.01202} {arXiv:1609.01202 [hep-th]} \BibitemShut
  {NoStop}%
\bibitem [{\citenamefont {Kovensky}\ \emph
  {et~al.}(2018{\natexlab{b}})\citenamefont {Kovensky}, \citenamefont
  {Michalski},\ and\ \citenamefont {Schvellinger}}]{Kovensky:2017oqs}%
  \BibitemOpen
  \bibfield  {author} {\bibinfo {author} {\bibfnamefont {N.}~\bibnamefont
  {Kovensky}}, \bibinfo {author} {\bibfnamefont {G.}~\bibnamefont {Michalski}},
  \ and\ \bibinfo {author} {\bibfnamefont {M.}~\bibnamefont {Schvellinger}},\
  }\href {\doibase 10.1007/JHEP04(2018)118} {\bibfield  {journal} {\bibinfo
  {journal} {JHEP}\ }\textbf {\bibinfo {volume} {04}},\ \bibinfo {pages} {118}
  (\bibinfo {year} {2018}{\natexlab{b}})},\ \Eprint
  {http://arxiv.org/abs/1711.06171} {arXiv:1711.06171 [hep-th]} \BibitemShut
  {NoStop}%
\bibitem [{\citenamefont {Amorim}\ \emph {et~al.}(2018)\citenamefont {Amorim},
  \citenamefont {Carcass\'es~Quevedo},\ and\ \citenamefont
  {Costa}}]{Amorim:2018yod}%
  \BibitemOpen
  \bibfield  {author} {\bibinfo {author} {\bibfnamefont {A.}~\bibnamefont
  {Amorim}}, \bibinfo {author} {\bibfnamefont {R.}~\bibnamefont
  {Carcass\'es~Quevedo}}, \ and\ \bibinfo {author} {\bibfnamefont {M.~S.}\
  \bibnamefont {Costa}},\ }\href {\doibase 10.1103/PhysRevD.98.026016}
  {\bibfield  {journal} {\bibinfo  {journal} {Phys. Rev. D}\ }\textbf {\bibinfo
  {volume} {98}},\ \bibinfo {pages} {026016} (\bibinfo {year} {2018})},\
  \Eprint {http://arxiv.org/abs/1804.07778} {arXiv:1804.07778 [hep-ph]}
  \BibitemShut {NoStop}%
\bibitem [{\citenamefont {Watanabe}\ \emph {et~al.}(2020)\citenamefont
  {Watanabe}, \citenamefont {Sawada},\ and\ \citenamefont
  {Huang}}]{Watanabe:2019zny}%
  \BibitemOpen
  \bibfield  {author} {\bibinfo {author} {\bibfnamefont {A.}~\bibnamefont
  {Watanabe}}, \bibinfo {author} {\bibfnamefont {T.}~\bibnamefont {Sawada}}, \
  and\ \bibinfo {author} {\bibfnamefont {M.}~\bibnamefont {Huang}},\ }\href
  {\doibase 10.1016/j.physletb.2020.135470} {\bibfield  {journal} {\bibinfo
  {journal} {Phys. Lett. B}\ }\textbf {\bibinfo {volume} {805}},\ \bibinfo
  {pages} {135470} (\bibinfo {year} {2020})},\ \Eprint
  {http://arxiv.org/abs/1910.10008} {arXiv:1910.10008 [hep-ph]} \BibitemShut
  {NoStop}%
\bibitem [{\citenamefont {Jorrin}\ \emph {et~al.}(2020)\citenamefont {Jorrin},
  \citenamefont {Michalski},\ and\ \citenamefont
  {Schvellinger}}]{Jorrin:2020cil}%
  \BibitemOpen
  \bibfield  {author} {\bibinfo {author} {\bibfnamefont {D.}~\bibnamefont
  {Jorrin}}, \bibinfo {author} {\bibfnamefont {G.}~\bibnamefont {Michalski}}, \
  and\ \bibinfo {author} {\bibfnamefont {M.}~\bibnamefont {Schvellinger}},\
  }\href {\doibase 10.1007/JHEP06(2020)063} {\bibfield  {journal} {\bibinfo
  {journal} {JHEP}\ }\textbf {\bibinfo {volume} {06}},\ \bibinfo {pages} {063}
  (\bibinfo {year} {2020})},\ \Eprint {http://arxiv.org/abs/2004.02909}
  {arXiv:2004.02909 [hep-th]} \BibitemShut {NoStop}%
\bibitem [{\citenamefont {Amorim}\ and\ \citenamefont
  {Costa}(2021)}]{Amorim:2021ffr}%
  \BibitemOpen
  \bibfield  {author} {\bibinfo {author} {\bibfnamefont {A.}~\bibnamefont
  {Amorim}}\ and\ \bibinfo {author} {\bibfnamefont {M.~S.}\ \bibnamefont
  {Costa}},\ }\href {\doibase 10.1103/PhysRevD.103.026007} {\bibfield
  {journal} {\bibinfo  {journal} {Phys. Rev. D}\ }\textbf {\bibinfo {volume}
  {103}},\ \bibinfo {pages} {026007} (\bibinfo {year} {2021})}\BibitemShut
  {NoStop}%
\bibitem [{\citenamefont {Mamo}\ and\ \citenamefont
  {Zahed}(2021)}]{Mamo:2021cle}%
  \BibitemOpen
  \bibfield  {author} {\bibinfo {author} {\bibfnamefont {K.~A.}\ \bibnamefont
  {Mamo}}\ and\ \bibinfo {author} {\bibfnamefont {I.}~\bibnamefont {Zahed}},\
  }\href {\doibase 10.1103/PhysRevD.104.066010} {\bibfield  {journal} {\bibinfo
   {journal} {Phys. Rev. D}\ }\textbf {\bibinfo {volume} {104}},\ \bibinfo
  {pages} {066010} (\bibinfo {year} {2021})},\ \Eprint
  {http://arxiv.org/abs/2102.00608} {arXiv:2102.00608 [hep-ph]} \BibitemShut
  {NoStop}%
\bibitem [{\citenamefont {Tahery}\ \emph {et~al.}(2023)\citenamefont {Tahery},
  \citenamefont {Wang},\ and\ \citenamefont {Chen}}]{Tahery:2021xsj}%
  \BibitemOpen
  \bibfield  {author} {\bibinfo {author} {\bibfnamefont {S.}~\bibnamefont
  {Tahery}}, \bibinfo {author} {\bibfnamefont {X.}~\bibnamefont {Wang}}, \ and\
  \bibinfo {author} {\bibfnamefont {X.}~\bibnamefont {Chen}},\ }\href {\doibase
  10.1088/1674-1137/ac936b} {\bibfield  {journal} {\bibinfo  {journal} {Chin.
  Phys. C}\ }\textbf {\bibinfo {volume} {47}},\ \bibinfo {pages} {013101}
  (\bibinfo {year} {2023})},\ \Eprint {http://arxiv.org/abs/2111.06616}
  {arXiv:2111.06616 [hep-ph]} \BibitemShut {NoStop}%
\bibitem [{\citenamefont {Jorrin}\ and\ \citenamefont
  {Schvellinger}(2022)}]{Jorrin:2022lua}%
  \BibitemOpen
  \bibfield  {author} {\bibinfo {author} {\bibfnamefont {D.}~\bibnamefont
  {Jorrin}}\ and\ \bibinfo {author} {\bibfnamefont {M.}~\bibnamefont
  {Schvellinger}},\ }\href {\doibase 10.1103/PhysRevD.106.066024} {\bibfield
  {journal} {\bibinfo  {journal} {Phys. Rev. D}\ }\textbf {\bibinfo {volume}
  {106}},\ \bibinfo {pages} {066024} (\bibinfo {year} {2022})},\ \Eprint
  {http://arxiv.org/abs/2207.02984} {arXiv:2207.02984 [hep-ph]} \BibitemShut
  {NoStop}%
\bibitem [{\citenamefont {Bigazzi}\ and\ \citenamefont
  {Castellani}(2024)}]{Bigazzi:2023odl}%
  \BibitemOpen
  \bibfield  {author} {\bibinfo {author} {\bibfnamefont {F.}~\bibnamefont
  {Bigazzi}}\ and\ \bibinfo {author} {\bibfnamefont {F.}~\bibnamefont
  {Castellani}},\ }\href {\doibase 10.1007/JHEP04(2024)037} {\bibfield
  {journal} {\bibinfo  {journal} {JHEP}\ }\textbf {\bibinfo {volume} {04}},\
  \bibinfo {pages} {037} (\bibinfo {year} {2024})},\ \Eprint
  {http://arxiv.org/abs/2308.16833} {arXiv:2308.16833 [hep-ph]} \BibitemShut
  {NoStop}%
\bibitem [{\citenamefont {Mayrhofer}(2024)}]{Mayrhofer:2024vnb}%
  \BibitemOpen
  \bibfield  {author} {\bibinfo {author} {\bibfnamefont {C.}~\bibnamefont
  {Mayrhofer}},\ }\emph {\bibinfo {title} {{Deep Inelastic Scattering and the
  WSS Model: A Study of Holographic Pomeron Exchange}}},\ \href {\doibase
  10.34726/hss.2024.126061} {Ph.D. thesis},\ \bibinfo  {school} {Vienna, Tech.
  U.} (\bibinfo {year} {2024})\BibitemShut {NoStop}%
\bibitem [{\citenamefont {Gursoy}\ and\ \citenamefont
  {Kiritsis}(2008)}]{Gursoy:2007cb}%
  \BibitemOpen
  \bibfield  {author} {\bibinfo {author} {\bibfnamefont {U.}~\bibnamefont
  {Gursoy}}\ and\ \bibinfo {author} {\bibfnamefont {E.}~\bibnamefont
  {Kiritsis}},\ }\href {\doibase 10.1088/1126-6708/2008/02/032} {\bibfield
  {journal} {\bibinfo  {journal} {JHEP}\ }\textbf {\bibinfo {volume} {02}},\
  \bibinfo {pages} {032} (\bibinfo {year} {2008})},\ \Eprint
  {http://arxiv.org/abs/0707.1324} {arXiv:0707.1324 [hep-th]} \BibitemShut
  {NoStop}%
\bibitem [{\citenamefont {Gursoy}\ \emph
  {et~al.}(2008{\natexlab{a}})\citenamefont {Gursoy}, \citenamefont
  {Kiritsis},\ and\ \citenamefont {Nitti}}]{Gursoy:2007er}%
  \BibitemOpen
  \bibfield  {author} {\bibinfo {author} {\bibfnamefont {U.}~\bibnamefont
  {Gursoy}}, \bibinfo {author} {\bibfnamefont {E.}~\bibnamefont {Kiritsis}}, \
  and\ \bibinfo {author} {\bibfnamefont {F.}~\bibnamefont {Nitti}},\ }\href
  {\doibase 10.1088/1126-6708/2008/02/019} {\bibfield  {journal} {\bibinfo
  {journal} {JHEP}\ }\textbf {\bibinfo {volume} {02}},\ \bibinfo {pages} {019}
  (\bibinfo {year} {2008}{\natexlab{a}})},\ \Eprint
  {http://arxiv.org/abs/0707.1349} {arXiv:0707.1349 [hep-th]} \BibitemShut
  {NoStop}%
\bibitem [{\citenamefont {Erlich}\ \emph {et~al.}(2005)\citenamefont {Erlich},
  \citenamefont {Katz}, \citenamefont {Son},\ and\ \citenamefont
  {Stephanov}}]{Erlich:2005qh}%
  \BibitemOpen
  \bibfield  {author} {\bibinfo {author} {\bibfnamefont {J.}~\bibnamefont
  {Erlich}}, \bibinfo {author} {\bibfnamefont {E.}~\bibnamefont {Katz}},
  \bibinfo {author} {\bibfnamefont {D.~T.}\ \bibnamefont {Son}}, \ and\
  \bibinfo {author} {\bibfnamefont {M.~A.}\ \bibnamefont {Stephanov}},\ }\href
  {\doibase 10.1103/PhysRevLett.95.261602} {\bibfield  {journal} {\bibinfo
  {journal} {Phys. Rev. Lett.}\ }\textbf {\bibinfo {volume} {95}},\ \bibinfo
  {pages} {261602} (\bibinfo {year} {2005})},\ \Eprint
  {http://arxiv.org/abs/hep-ph/0501128} {arXiv:hep-ph/0501128} \BibitemShut
  {NoStop}%
\bibitem [{\citenamefont {Karch}\ \emph {et~al.}(2006)\citenamefont {Karch},
  \citenamefont {Katz}, \citenamefont {Son},\ and\ \citenamefont
  {Stephanov}}]{Karch:2006pv}%
  \BibitemOpen
  \bibfield  {author} {\bibinfo {author} {\bibfnamefont {A.}~\bibnamefont
  {Karch}}, \bibinfo {author} {\bibfnamefont {E.}~\bibnamefont {Katz}},
  \bibinfo {author} {\bibfnamefont {D.~T.}\ \bibnamefont {Son}}, \ and\
  \bibinfo {author} {\bibfnamefont {M.~A.}\ \bibnamefont {Stephanov}},\ }\href
  {\doibase 10.1103/PhysRevD.74.015005} {\bibfield  {journal} {\bibinfo
  {journal} {Phys. Rev. D}\ }\textbf {\bibinfo {volume} {74}},\ \bibinfo
  {pages} {015005} (\bibinfo {year} {2006})},\ \Eprint
  {http://arxiv.org/abs/hep-ph/0602229} {arXiv:hep-ph/0602229} \BibitemShut
  {NoStop}%
\bibitem [{\citenamefont {Ballon-Bayona}\ \emph {et~al.}(2018)\citenamefont
  {Ballon-Bayona}, \citenamefont {Boschi-Filho}, \citenamefont {Mamani},
  \citenamefont {Miranda},\ and\ \citenamefont
  {Zanchin}}]{Ballon-Bayona:2017sxa}%
  \BibitemOpen
  \bibfield  {author} {\bibinfo {author} {\bibfnamefont {A.}~\bibnamefont
  {Ballon-Bayona}}, \bibinfo {author} {\bibfnamefont {H.}~\bibnamefont
  {Boschi-Filho}}, \bibinfo {author} {\bibfnamefont {L.~A.~H.}\ \bibnamefont
  {Mamani}}, \bibinfo {author} {\bibfnamefont {A.~S.}\ \bibnamefont {Miranda}},
  \ and\ \bibinfo {author} {\bibfnamefont {V.~T.}\ \bibnamefont {Zanchin}},\
  }\href {\doibase 10.1103/PhysRevD.97.046001} {\bibfield  {journal} {\bibinfo
  {journal} {Phys. Rev. D}\ }\textbf {\bibinfo {volume} {97}},\ \bibinfo
  {pages} {046001} (\bibinfo {year} {2018})},\ \Eprint
  {http://arxiv.org/abs/1708.08968} {arXiv:1708.08968 [hep-th]} \BibitemShut
  {NoStop}%
\bibitem [{\citenamefont {Ballon-Bayona}\ \emph {et~al.}(2023)\citenamefont
  {Ballon-Bayona}, \citenamefont {Frederico}, \citenamefont {Mamani},\ and\
  \citenamefont {de~Paula}}]{Ballon-Bayona:2023zal}%
  \BibitemOpen
  \bibfield  {author} {\bibinfo {author} {\bibfnamefont {A.}~\bibnamefont
  {Ballon-Bayona}}, \bibinfo {author} {\bibfnamefont {T.}~\bibnamefont
  {Frederico}}, \bibinfo {author} {\bibfnamefont {L.~A.~H.}\ \bibnamefont
  {Mamani}}, \ and\ \bibinfo {author} {\bibfnamefont {W.}~\bibnamefont
  {de~Paula}},\ }\href {\doibase 10.1103/PhysRevD.108.106016} {\bibfield
  {journal} {\bibinfo  {journal} {Phys. Rev. D}\ }\textbf {\bibinfo {volume}
  {108}},\ \bibinfo {pages} {106016} (\bibinfo {year} {2023})},\ \Eprint
  {http://arxiv.org/abs/2308.07503} {arXiv:2308.07503 [hep-ph]} \BibitemShut
  {NoStop}%
\bibitem [{\citenamefont {Ballon-Bayona}\ and\ \citenamefont
  {Junior}(2024)}]{Ballon-Bayona:2024yuz}%
  \BibitemOpen
  \bibfield  {author} {\bibinfo {author} {\bibfnamefont {A.}~\bibnamefont
  {Ballon-Bayona}}\ and\ \bibinfo {author} {\bibfnamefont {A.~a. S. d.~S.}\
  \bibnamefont {Junior}},\ }\href {\doibase 10.1103/PhysRevD.109.094050}
  {\bibfield  {journal} {\bibinfo  {journal} {Phys. Rev. D}\ }\textbf {\bibinfo
  {volume} {109}},\ \bibinfo {pages} {094050} (\bibinfo {year} {2024})},\
  \Eprint {http://arxiv.org/abs/2402.17950} {arXiv:2402.17950 [hep-ph]}
  \BibitemShut {NoStop}%
\bibitem [{\citenamefont {Gursoy}\ \emph {et~al.}(2009)\citenamefont {Gursoy},
  \citenamefont {Kiritsis}, \citenamefont {Mazzanti},\ and\ \citenamefont
  {Nitti}}]{Gursoy:2008za}%
  \BibitemOpen
  \bibfield  {author} {\bibinfo {author} {\bibfnamefont {U.}~\bibnamefont
  {Gursoy}}, \bibinfo {author} {\bibfnamefont {E.}~\bibnamefont {Kiritsis}},
  \bibinfo {author} {\bibfnamefont {L.}~\bibnamefont {Mazzanti}}, \ and\
  \bibinfo {author} {\bibfnamefont {F.}~\bibnamefont {Nitti}},\ }\href
  {\doibase 10.1088/1126-6708/2009/05/033} {\bibfield  {journal} {\bibinfo
  {journal} {JHEP}\ }\textbf {\bibinfo {volume} {05}},\ \bibinfo {pages} {033}
  (\bibinfo {year} {2009})},\ \Eprint {http://arxiv.org/abs/0812.0792}
  {arXiv:0812.0792 [hep-th]} \BibitemShut {NoStop}%
\bibitem [{\citenamefont {Gursoy}\ \emph
  {et~al.}(2008{\natexlab{b}})\citenamefont {Gursoy}, \citenamefont {Kiritsis},
  \citenamefont {Mazzanti},\ and\ \citenamefont {Nitti}}]{Gursoy:2008bu}%
  \BibitemOpen
  \bibfield  {author} {\bibinfo {author} {\bibfnamefont {U.}~\bibnamefont
  {Gursoy}}, \bibinfo {author} {\bibfnamefont {E.}~\bibnamefont {Kiritsis}},
  \bibinfo {author} {\bibfnamefont {L.}~\bibnamefont {Mazzanti}}, \ and\
  \bibinfo {author} {\bibfnamefont {F.}~\bibnamefont {Nitti}},\ }\href
  {\doibase 10.1103/PhysRevLett.101.181601} {\bibfield  {journal} {\bibinfo
  {journal} {Phys. Rev. Lett.}\ }\textbf {\bibinfo {volume} {101}},\ \bibinfo
  {pages} {181601} (\bibinfo {year} {2008}{\natexlab{b}})},\ \Eprint
  {http://arxiv.org/abs/0804.0899} {arXiv:0804.0899 [hep-th]} \BibitemShut
  {NoStop}%
\bibitem [{\citenamefont {Gubser}\ and\ \citenamefont
  {Nellore}(2008)}]{Gubser:2008ny}%
  \BibitemOpen
  \bibfield  {author} {\bibinfo {author} {\bibfnamefont {S.~S.}\ \bibnamefont
  {Gubser}}\ and\ \bibinfo {author} {\bibfnamefont {A.}~\bibnamefont
  {Nellore}},\ }\href {\doibase 10.1103/PhysRevD.78.086007} {\bibfield
  {journal} {\bibinfo  {journal} {Phys. Rev. D}\ }\textbf {\bibinfo {volume}
  {78}},\ \bibinfo {pages} {086007} (\bibinfo {year} {2008})},\ \Eprint
  {http://arxiv.org/abs/0804.0434} {arXiv:0804.0434 [hep-th]} \BibitemShut
  {NoStop}%
\bibitem [{\citenamefont {Gubser}\ \emph
  {et~al.}(2008{\natexlab{a}})\citenamefont {Gubser}, \citenamefont {Nellore},
  \citenamefont {Pufu},\ and\ \citenamefont {Rocha}}]{Gubser:2008yx}%
  \BibitemOpen
  \bibfield  {author} {\bibinfo {author} {\bibfnamefont {S.~S.}\ \bibnamefont
  {Gubser}}, \bibinfo {author} {\bibfnamefont {A.}~\bibnamefont {Nellore}},
  \bibinfo {author} {\bibfnamefont {S.~S.}\ \bibnamefont {Pufu}}, \ and\
  \bibinfo {author} {\bibfnamefont {F.~D.}\ \bibnamefont {Rocha}},\ }\href
  {\doibase 10.1103/PhysRevLett.101.131601} {\bibfield  {journal} {\bibinfo
  {journal} {Phys. Rev. Lett.}\ }\textbf {\bibinfo {volume} {101}},\ \bibinfo
  {pages} {131601} (\bibinfo {year} {2008}{\natexlab{a}})},\ \Eprint
  {http://arxiv.org/abs/0804.1950} {arXiv:0804.1950 [hep-th]} \BibitemShut
  {NoStop}%
\bibitem [{\citenamefont {Gubser}\ \emph
  {et~al.}(2008{\natexlab{b}})\citenamefont {Gubser}, \citenamefont {Pufu},\
  and\ \citenamefont {Rocha}}]{Gubser:2008sz}%
  \BibitemOpen
  \bibfield  {author} {\bibinfo {author} {\bibfnamefont {S.~S.}\ \bibnamefont
  {Gubser}}, \bibinfo {author} {\bibfnamefont {S.~S.}\ \bibnamefont {Pufu}}, \
  and\ \bibinfo {author} {\bibfnamefont {F.~D.}\ \bibnamefont {Rocha}},\ }\href
  {\doibase 10.1088/1126-6708/2008/08/085} {\bibfield  {journal} {\bibinfo
  {journal} {JHEP}\ }\textbf {\bibinfo {volume} {08}},\ \bibinfo {pages} {085}
  (\bibinfo {year} {2008}{\natexlab{b}})},\ \Eprint
  {http://arxiv.org/abs/0806.0407} {arXiv:0806.0407 [hep-th]} \BibitemShut
  {NoStop}%
\bibitem [{\citenamefont {Noronha}(2010)}]{Noronha:2009ud}%
  \BibitemOpen
  \bibfield  {author} {\bibinfo {author} {\bibfnamefont {J.}~\bibnamefont
  {Noronha}},\ }\href {\doibase 10.1103/PhysRevD.81.045011} {\bibfield
  {journal} {\bibinfo  {journal} {Phys. Rev. D}\ }\textbf {\bibinfo {volume}
  {81}},\ \bibinfo {pages} {045011} (\bibinfo {year} {2010})},\ \Eprint
  {http://arxiv.org/abs/0910.1261} {arXiv:0910.1261 [hep-th]} \BibitemShut
  {NoStop}%
\bibitem [{\citenamefont {Gursoy}\ \emph {et~al.}(2011)\citenamefont {Gursoy},
  \citenamefont {Kiritsis}, \citenamefont {Mazzanti}, \citenamefont
  {Michalogiorgakis},\ and\ \citenamefont {Nitti}}]{Gursoy:2010fj}%
  \BibitemOpen
  \bibfield  {author} {\bibinfo {author} {\bibfnamefont {U.}~\bibnamefont
  {Gursoy}}, \bibinfo {author} {\bibfnamefont {E.}~\bibnamefont {Kiritsis}},
  \bibinfo {author} {\bibfnamefont {L.}~\bibnamefont {Mazzanti}}, \bibinfo
  {author} {\bibfnamefont {G.}~\bibnamefont {Michalogiorgakis}}, \ and\
  \bibinfo {author} {\bibfnamefont {F.}~\bibnamefont {Nitti}},\ }\href
  {\doibase 10.1007/978-3-642-04864-7_4} {\bibfield  {journal} {\bibinfo
  {journal} {Lect. Notes Phys.}\ }\textbf {\bibinfo {volume} {828}},\ \bibinfo
  {pages} {79} (\bibinfo {year} {2011})},\ \Eprint
  {http://arxiv.org/abs/1006.5461} {arXiv:1006.5461 [hep-th]} \BibitemShut
  {NoStop}%
\bibitem [{\citenamefont {Finazzo}\ and\ \citenamefont
  {Noronha}(2014{\natexlab{a}})}]{Finazzo:2013efa}%
  \BibitemOpen
  \bibfield  {author} {\bibinfo {author} {\bibfnamefont {S.~I.}\ \bibnamefont
  {Finazzo}}\ and\ \bibinfo {author} {\bibfnamefont {J.}~\bibnamefont
  {Noronha}},\ }\href {\doibase 10.1103/PhysRevD.89.106008} {\bibfield
  {journal} {\bibinfo  {journal} {Phys. Rev. D}\ }\textbf {\bibinfo {volume}
  {89}},\ \bibinfo {pages} {106008} (\bibinfo {year} {2014}{\natexlab{a}})},\
  \Eprint {http://arxiv.org/abs/1311.6675} {arXiv:1311.6675 [hep-th]}
  \BibitemShut {NoStop}%
\bibitem [{\citenamefont {Finazzo}\ and\ \citenamefont
  {Noronha}(2014{\natexlab{b}})}]{Finazzo:2014zga}%
  \BibitemOpen
  \bibfield  {author} {\bibinfo {author} {\bibfnamefont {S.~I.}\ \bibnamefont
  {Finazzo}}\ and\ \bibinfo {author} {\bibfnamefont {J.}~\bibnamefont
  {Noronha}},\ }\href {\doibase 10.1103/PhysRevD.90.115028} {\bibfield
  {journal} {\bibinfo  {journal} {Phys. Rev. D}\ }\textbf {\bibinfo {volume}
  {90}},\ \bibinfo {pages} {115028} (\bibinfo {year} {2014}{\natexlab{b}})},\
  \Eprint {http://arxiv.org/abs/1411.4330} {arXiv:1411.4330 [hep-th]}
  \BibitemShut {NoStop}%
\bibitem [{\citenamefont {DeWolfe}\ \emph
  {et~al.}(2011{\natexlab{a}})\citenamefont {DeWolfe}, \citenamefont {Gubser},\
  and\ \citenamefont {Rosen}}]{DeWolfe:2010he}%
  \BibitemOpen
  \bibfield  {author} {\bibinfo {author} {\bibfnamefont {O.}~\bibnamefont
  {DeWolfe}}, \bibinfo {author} {\bibfnamefont {S.~S.}\ \bibnamefont {Gubser}},
  \ and\ \bibinfo {author} {\bibfnamefont {C.}~\bibnamefont {Rosen}},\ }\href
  {\doibase 10.1103/PhysRevD.83.086005} {\bibfield  {journal} {\bibinfo
  {journal} {Phys. Rev. D}\ }\textbf {\bibinfo {volume} {83}},\ \bibinfo
  {pages} {086005} (\bibinfo {year} {2011}{\natexlab{a}})},\ \Eprint
  {http://arxiv.org/abs/1012.1864} {arXiv:1012.1864 [hep-th]} \BibitemShut
  {NoStop}%
\bibitem [{\citenamefont {DeWolfe}\ \emph
  {et~al.}(2011{\natexlab{b}})\citenamefont {DeWolfe}, \citenamefont {Gubser},\
  and\ \citenamefont {Rosen}}]{DeWolfe:2011ts}%
  \BibitemOpen
  \bibfield  {author} {\bibinfo {author} {\bibfnamefont {O.}~\bibnamefont
  {DeWolfe}}, \bibinfo {author} {\bibfnamefont {S.~S.}\ \bibnamefont {Gubser}},
  \ and\ \bibinfo {author} {\bibfnamefont {C.}~\bibnamefont {Rosen}},\ }\href
  {\doibase 10.1103/PhysRevD.84.126014} {\bibfield  {journal} {\bibinfo
  {journal} {Phys. Rev. D}\ }\textbf {\bibinfo {volume} {84}},\ \bibinfo
  {pages} {126014} (\bibinfo {year} {2011}{\natexlab{b}})},\ \Eprint
  {http://arxiv.org/abs/1108.2029} {arXiv:1108.2029 [hep-th]} \BibitemShut
  {NoStop}%
\bibitem [{\citenamefont {Rougemont}\ \emph {et~al.}(2016)\citenamefont
  {Rougemont}, \citenamefont {Ficnar}, \citenamefont {Finazzo},\ and\
  \citenamefont {Noronha}}]{Rougemont:2015wca}%
  \BibitemOpen
  \bibfield  {author} {\bibinfo {author} {\bibfnamefont {R.}~\bibnamefont
  {Rougemont}}, \bibinfo {author} {\bibfnamefont {A.}~\bibnamefont {Ficnar}},
  \bibinfo {author} {\bibfnamefont {S.}~\bibnamefont {Finazzo}}, \ and\
  \bibinfo {author} {\bibfnamefont {J.}~\bibnamefont {Noronha}},\ }\href
  {\doibase 10.1007/JHEP04(2016)102} {\bibfield  {journal} {\bibinfo  {journal}
  {JHEP}\ }\textbf {\bibinfo {volume} {04}},\ \bibinfo {pages} {102} (\bibinfo
  {year} {2016})},\ \Eprint {http://arxiv.org/abs/1507.06556} {arXiv:1507.06556
  [hep-th]} \BibitemShut {NoStop}%
\bibitem [{\citenamefont {Rougemont}\ \emph {et~al.}(2015)\citenamefont
  {Rougemont}, \citenamefont {Noronha},\ and\ \citenamefont
  {Noronha-Hostler}}]{Rougemont:2015ona}%
  \BibitemOpen
  \bibfield  {author} {\bibinfo {author} {\bibfnamefont {R.}~\bibnamefont
  {Rougemont}}, \bibinfo {author} {\bibfnamefont {J.}~\bibnamefont {Noronha}},
  \ and\ \bibinfo {author} {\bibfnamefont {J.}~\bibnamefont
  {Noronha-Hostler}},\ }\href {\doibase 10.1103/PhysRevLett.115.202301}
  {\bibfield  {journal} {\bibinfo  {journal} {Phys. Rev. Lett.}\ }\textbf
  {\bibinfo {volume} {115}},\ \bibinfo {pages} {202301} (\bibinfo {year}
  {2015})},\ \Eprint {http://arxiv.org/abs/1507.06972} {arXiv:1507.06972
  [hep-ph]} \BibitemShut {NoStop}%
\bibitem [{\citenamefont {Critelli}\ \emph {et~al.}(2017)\citenamefont
  {Critelli}, \citenamefont {Noronha}, \citenamefont {Noronha-Hostler},
  \citenamefont {Portillo}, \citenamefont {Ratti},\ and\ \citenamefont
  {Rougemont}}]{Critelli:2017oub}%
  \BibitemOpen
  \bibfield  {author} {\bibinfo {author} {\bibfnamefont {R.}~\bibnamefont
  {Critelli}}, \bibinfo {author} {\bibfnamefont {J.}~\bibnamefont {Noronha}},
  \bibinfo {author} {\bibfnamefont {J.}~\bibnamefont {Noronha-Hostler}},
  \bibinfo {author} {\bibfnamefont {I.}~\bibnamefont {Portillo}}, \bibinfo
  {author} {\bibfnamefont {C.}~\bibnamefont {Ratti}}, \ and\ \bibinfo {author}
  {\bibfnamefont {R.}~\bibnamefont {Rougemont}},\ }\href {\doibase
  10.1103/PhysRevD.96.096026} {\bibfield  {journal} {\bibinfo  {journal} {Phys.
  Rev. D}\ }\textbf {\bibinfo {volume} {96}},\ \bibinfo {pages} {096026}
  (\bibinfo {year} {2017})},\ \Eprint {http://arxiv.org/abs/1706.00455}
  {arXiv:1706.00455 [nucl-th]} \BibitemShut {NoStop}%
\bibitem [{\citenamefont {Ballon-Bayona}\ \emph {et~al.}(2020)\citenamefont
  {Ballon-Bayona}, \citenamefont {Boschi-Filho}, \citenamefont {Capossoli},\
  and\ \citenamefont {Rodrigues}}]{Ballon-Bayona:2020xls}%
  \BibitemOpen
  \bibfield  {author} {\bibinfo {author} {\bibfnamefont {A.}~\bibnamefont
  {Ballon-Bayona}}, \bibinfo {author} {\bibfnamefont {H.}~\bibnamefont
  {Boschi-Filho}}, \bibinfo {author} {\bibfnamefont {E.~F.}\ \bibnamefont
  {Capossoli}}, \ and\ \bibinfo {author} {\bibfnamefont {D.~M.}\ \bibnamefont
  {Rodrigues}},\ }\href {\doibase 10.1103/PhysRevD.102.126003} {\bibfield
  {journal} {\bibinfo  {journal} {Phys. Rev. D}\ }\textbf {\bibinfo {volume}
  {102}},\ \bibinfo {pages} {126003} (\bibinfo {year} {2020})},\ \Eprint
  {http://arxiv.org/abs/2006.08810} {arXiv:2006.08810 [hep-th]} \BibitemShut
  {NoStop}%
\bibitem [{\citenamefont {Grefa}\ \emph {et~al.}(2021)\citenamefont {Grefa},
  \citenamefont {Noronha}, \citenamefont {Noronha-Hostler}, \citenamefont
  {Portillo}, \citenamefont {Ratti},\ and\ \citenamefont
  {Rougemont}}]{Grefa:2021qvt}%
  \BibitemOpen
  \bibfield  {author} {\bibinfo {author} {\bibfnamefont {J.}~\bibnamefont
  {Grefa}}, \bibinfo {author} {\bibfnamefont {J.}~\bibnamefont {Noronha}},
  \bibinfo {author} {\bibfnamefont {J.}~\bibnamefont {Noronha-Hostler}},
  \bibinfo {author} {\bibfnamefont {I.}~\bibnamefont {Portillo}}, \bibinfo
  {author} {\bibfnamefont {C.}~\bibnamefont {Ratti}}, \ and\ \bibinfo {author}
  {\bibfnamefont {R.}~\bibnamefont {Rougemont}},\ }\href {\doibase
  10.1103/PhysRevD.104.034002} {\bibfield  {journal} {\bibinfo  {journal}
  {Phys. Rev. D}\ }\textbf {\bibinfo {volume} {104}},\ \bibinfo {pages}
  {034002} (\bibinfo {year} {2021})},\ \Eprint
  {http://arxiv.org/abs/2102.12042} {arXiv:2102.12042 [nucl-th]} \BibitemShut
  {NoStop}%
\bibitem [{\citenamefont {Grefa}\ \emph {et~al.}(2022)\citenamefont {Grefa},
  \citenamefont {Hippert}, \citenamefont {Noronha}, \citenamefont
  {Noronha-Hostler}, \citenamefont {Portillo}, \citenamefont {Ratti},\ and\
  \citenamefont {Rougemont}}]{Grefa:2022sav}%
  \BibitemOpen
  \bibfield  {author} {\bibinfo {author} {\bibfnamefont {J.}~\bibnamefont
  {Grefa}}, \bibinfo {author} {\bibfnamefont {M.}~\bibnamefont {Hippert}},
  \bibinfo {author} {\bibfnamefont {J.}~\bibnamefont {Noronha}}, \bibinfo
  {author} {\bibfnamefont {J.}~\bibnamefont {Noronha-Hostler}}, \bibinfo
  {author} {\bibfnamefont {I.}~\bibnamefont {Portillo}}, \bibinfo {author}
  {\bibfnamefont {C.}~\bibnamefont {Ratti}}, \ and\ \bibinfo {author}
  {\bibfnamefont {R.}~\bibnamefont {Rougemont}},\ }\href {\doibase
  10.1103/PhysRevD.106.034024} {\bibfield  {journal} {\bibinfo  {journal}
  {Phys. Rev. D}\ }\textbf {\bibinfo {volume} {106}},\ \bibinfo {pages}
  {034024} (\bibinfo {year} {2022})},\ \Eprint
  {http://arxiv.org/abs/2203.00139} {arXiv:2203.00139 [nucl-th]} \BibitemShut
  {NoStop}%
\bibitem [{\citenamefont {Hippert}\ \emph {et~al.}(2024)\citenamefont
  {Hippert}, \citenamefont {Grefa}, \citenamefont {Manning}, \citenamefont
  {Noronha}, \citenamefont {Noronha-Hostler}, \citenamefont {Portillo~Vazquez},
  \citenamefont {Ratti}, \citenamefont {Rougemont},\ and\ \citenamefont
  {Trujillo}}]{Hippert:2023bel}%
  \BibitemOpen
  \bibfield  {author} {\bibinfo {author} {\bibfnamefont {M.}~\bibnamefont
  {Hippert}}, \bibinfo {author} {\bibfnamefont {J.}~\bibnamefont {Grefa}},
  \bibinfo {author} {\bibfnamefont {T.~A.}\ \bibnamefont {Manning}}, \bibinfo
  {author} {\bibfnamefont {J.}~\bibnamefont {Noronha}}, \bibinfo {author}
  {\bibfnamefont {J.}~\bibnamefont {Noronha-Hostler}}, \bibinfo {author}
  {\bibfnamefont {I.}~\bibnamefont {Portillo~Vazquez}}, \bibinfo {author}
  {\bibfnamefont {C.}~\bibnamefont {Ratti}}, \bibinfo {author} {\bibfnamefont
  {R.}~\bibnamefont {Rougemont}}, \ and\ \bibinfo {author} {\bibfnamefont
  {M.}~\bibnamefont {Trujillo}},\ }\href {\doibase 10.1103/PhysRevD.110.094006}
  {\bibfield  {journal} {\bibinfo  {journal} {Phys. Rev. D}\ }\textbf {\bibinfo
  {volume} {110}},\ \bibinfo {pages} {094006} (\bibinfo {year} {2024})},\
  \Eprint {http://arxiv.org/abs/2309.00579} {arXiv:2309.00579 [nucl-th]}
  \BibitemShut {NoStop}%
\bibitem [{\citenamefont {Rougemont}\ \emph {et~al.}(2024)\citenamefont
  {Rougemont}, \citenamefont {Grefa}, \citenamefont {Hippert}, \citenamefont
  {Noronha}, \citenamefont {Noronha-Hostler}, \citenamefont {Portillo},\ and\
  \citenamefont {Ratti}}]{Rougemont:2023gfz}%
  \BibitemOpen
  \bibfield  {author} {\bibinfo {author} {\bibfnamefont {R.}~\bibnamefont
  {Rougemont}}, \bibinfo {author} {\bibfnamefont {J.}~\bibnamefont {Grefa}},
  \bibinfo {author} {\bibfnamefont {M.}~\bibnamefont {Hippert}}, \bibinfo
  {author} {\bibfnamefont {J.}~\bibnamefont {Noronha}}, \bibinfo {author}
  {\bibfnamefont {J.}~\bibnamefont {Noronha-Hostler}}, \bibinfo {author}
  {\bibfnamefont {I.}~\bibnamefont {Portillo}}, \ and\ \bibinfo {author}
  {\bibfnamefont {C.}~\bibnamefont {Ratti}},\ }\href {\doibase
  10.1016/j.ppnp.2023.104093} {\bibfield  {journal} {\bibinfo  {journal} {Prog.
  Part. Nucl. Phys.}\ }\textbf {\bibinfo {volume} {135}},\ \bibinfo {pages}
  {104093} (\bibinfo {year} {2024})},\ \Eprint
  {http://arxiv.org/abs/2307.03885} {arXiv:2307.03885 [nucl-th]} \BibitemShut
  {NoStop}%
\bibitem [{\citenamefont {Jarvinen}\ and\ \citenamefont
  {Kiritsis}(2012)}]{Jarvinen:2011qe}%
  \BibitemOpen
  \bibfield  {author} {\bibinfo {author} {\bibfnamefont {M.}~\bibnamefont
  {Jarvinen}}\ and\ \bibinfo {author} {\bibfnamefont {E.}~\bibnamefont
  {Kiritsis}},\ }\href {\doibase 10.1007/JHEP03(2012)002} {\bibfield  {journal}
  {\bibinfo  {journal} {JHEP}\ }\textbf {\bibinfo {volume} {03}},\ \bibinfo
  {pages} {002} (\bibinfo {year} {2012})},\ \Eprint
  {http://arxiv.org/abs/1112.1261} {arXiv:1112.1261 [hep-ph]} \BibitemShut
  {NoStop}%
\bibitem [{\citenamefont {Deng}\ and\ \citenamefont
  {Hou}(2025)}]{Deng:2025fpq}%
  \BibitemOpen
  \bibfield  {author} {\bibinfo {author} {\bibfnamefont {J.}~\bibnamefont
  {Deng}}\ and\ \bibinfo {author} {\bibfnamefont {D.}~\bibnamefont {Hou}},\
  }\href@noop {} {\bibfield  {journal} {\bibinfo  {journal} {To be published}\
  } (\bibinfo {year} {2025})},\ \Eprint {http://arxiv.org/abs/2502.00771}
  {arXiv:2502.00771 [nucl-th]} \BibitemShut {NoStop}%
\bibitem [{\citenamefont {Lucini}\ and\ \citenamefont
  {Teper}(2001)}]{Lucini:2001ej}%
  \BibitemOpen
  \bibfield  {author} {\bibinfo {author} {\bibfnamefont {B.}~\bibnamefont
  {Lucini}}\ and\ \bibinfo {author} {\bibfnamefont {M.}~\bibnamefont {Teper}},\
  }\href {\doibase 10.1088/1126-6708/2001/06/050} {\bibfield  {journal}
  {\bibinfo  {journal} {JHEP}\ }\textbf {\bibinfo {volume} {06}},\ \bibinfo
  {pages} {050} (\bibinfo {year} {2001})},\ \Eprint
  {http://arxiv.org/abs/hep-lat/0103027} {arXiv:hep-lat/0103027} \BibitemShut
  {NoStop}%
\bibitem [{\citenamefont {Lucini}\ \emph {et~al.}(2010)\citenamefont {Lucini},
  \citenamefont {Rago},\ and\ \citenamefont {Rinaldi}}]{Lucini:2010nv}%
  \BibitemOpen
  \bibfield  {author} {\bibinfo {author} {\bibfnamefont {B.}~\bibnamefont
  {Lucini}}, \bibinfo {author} {\bibfnamefont {A.}~\bibnamefont {Rago}}, \ and\
  \bibinfo {author} {\bibfnamefont {E.}~\bibnamefont {Rinaldi}},\ }\href
  {\doibase 10.1007/JHEP08(2010)119} {\bibfield  {journal} {\bibinfo  {journal}
  {JHEP}\ }\textbf {\bibinfo {volume} {08}},\ \bibinfo {pages} {119} (\bibinfo
  {year} {2010})},\ \Eprint {http://arxiv.org/abs/1007.3879} {arXiv:1007.3879
  [hep-lat]} \BibitemShut {NoStop}%
\bibitem [{\citenamefont {Panero}(2009)}]{Panero:2009tv}%
  \BibitemOpen
  \bibfield  {author} {\bibinfo {author} {\bibfnamefont {M.}~\bibnamefont
  {Panero}},\ }\href {\doibase 10.1103/PhysRevLett.103.232001} {\bibfield
  {journal} {\bibinfo  {journal} {Phys. Rev. Lett.}\ }\textbf {\bibinfo
  {volume} {103}},\ \bibinfo {pages} {232001} (\bibinfo {year} {2009})},\
  \Eprint {http://arxiv.org/abs/0907.3719} {arXiv:0907.3719 [hep-lat]}
  \BibitemShut {NoStop}%
\bibitem [{\citenamefont {Abidin}\ and\ \citenamefont
  {Carlson}(2009)}]{Abidin:2009hr}%
  \BibitemOpen
  \bibfield  {author} {\bibinfo {author} {\bibfnamefont {Z.}~\bibnamefont
  {Abidin}}\ and\ \bibinfo {author} {\bibfnamefont {C.~E.}\ \bibnamefont
  {Carlson}},\ }\href {\doibase 10.1103/PhysRevD.79.115003} {\bibfield
  {journal} {\bibinfo  {journal} {Phys. Rev. D}\ }\textbf {\bibinfo {volume}
  {79}},\ \bibinfo {pages} {115003} (\bibinfo {year} {2009})},\ \Eprint
  {http://arxiv.org/abs/0903.4818} {arXiv:0903.4818 [hep-ph]} \BibitemShut
  {NoStop}%
\bibitem [{\citenamefont {Gubser}(2000)}]{Gubser:2000nd}%
  \BibitemOpen
  \bibfield  {author} {\bibinfo {author} {\bibfnamefont {S.~S.}\ \bibnamefont
  {Gubser}},\ }\href {\doibase 10.4310/ATMP.2000.v4.n3.a6} {\bibfield
  {journal} {\bibinfo  {journal} {Adv. Theor. Math. Phys.}\ }\textbf {\bibinfo
  {volume} {4}},\ \bibinfo {pages} {679} (\bibinfo {year} {2000})},\ \Eprint
  {http://arxiv.org/abs/hep-th/0002160} {arXiv:hep-th/0002160} \BibitemShut
  {NoStop}%
\bibitem [{\citenamefont {Breitenlohner}\ and\ \citenamefont
  {Freedman}(1982{\natexlab{a}})}]{Breitenlohner:1982bm}%
  \BibitemOpen
  \bibfield  {author} {\bibinfo {author} {\bibfnamefont {P.}~\bibnamefont
  {Breitenlohner}}\ and\ \bibinfo {author} {\bibfnamefont {D.~Z.}\ \bibnamefont
  {Freedman}},\ }\href {\doibase 10.1016/0370-2693(82)90643-8} {\bibfield
  {journal} {\bibinfo  {journal} {Phys. Lett. B}\ }\textbf {\bibinfo {volume}
  {115}},\ \bibinfo {pages} {197} (\bibinfo {year}
  {1982}{\natexlab{a}})}\BibitemShut {NoStop}%
\bibitem [{\citenamefont {Breitenlohner}\ and\ \citenamefont
  {Freedman}(1982{\natexlab{b}})}]{Breitenlohner:1982jf}%
  \BibitemOpen
  \bibfield  {author} {\bibinfo {author} {\bibfnamefont {P.}~\bibnamefont
  {Breitenlohner}}\ and\ \bibinfo {author} {\bibfnamefont {D.~Z.}\ \bibnamefont
  {Freedman}},\ }\href {\doibase 10.1016/0003-4916(82)90116-6} {\bibfield
  {journal} {\bibinfo  {journal} {Annals Phys.}\ }\textbf {\bibinfo {volume}
  {144}},\ \bibinfo {pages} {249} (\bibinfo {year}
  {1982}{\natexlab{b}})}\BibitemShut {NoStop}%
\bibitem [{\citenamefont {Mezincescu}\ and\ \citenamefont
  {Townsend}(1985)}]{Mezincescu:1984ev}%
  \BibitemOpen
  \bibfield  {author} {\bibinfo {author} {\bibfnamefont {L.}~\bibnamefont
  {Mezincescu}}\ and\ \bibinfo {author} {\bibfnamefont {P.~K.}\ \bibnamefont
  {Townsend}},\ }\href {\doibase 10.1016/0003-4916(85)90150-2} {\bibfield
  {journal} {\bibinfo  {journal} {Annals Phys.}\ }\textbf {\bibinfo {volume}
  {160}},\ \bibinfo {pages} {406} (\bibinfo {year} {1985})}\BibitemShut
  {NoStop}%
\bibitem [{\citenamefont {Henningson}\ and\ \citenamefont
  {Sfetsos}(1998)}]{Henningson:1998cd}%
  \BibitemOpen
  \bibfield  {author} {\bibinfo {author} {\bibfnamefont {M.}~\bibnamefont
  {Henningson}}\ and\ \bibinfo {author} {\bibfnamefont {K.}~\bibnamefont
  {Sfetsos}},\ }\href {\doibase 10.1016/S0370-2693(98)00559-0} {\bibfield
  {journal} {\bibinfo  {journal} {Phys. Lett. B}\ }\textbf {\bibinfo {volume}
  {431}},\ \bibinfo {pages} {63} (\bibinfo {year} {1998})},\ \Eprint
  {http://arxiv.org/abs/hep-th/9803251} {arXiv:hep-th/9803251} \BibitemShut
  {NoStop}%
\bibitem [{\citenamefont {Mueck}\ and\ \citenamefont
  {Viswanathan}(1998)}]{Mueck:1998iz}%
  \BibitemOpen
  \bibfield  {author} {\bibinfo {author} {\bibfnamefont {W.}~\bibnamefont
  {Mueck}}\ and\ \bibinfo {author} {\bibfnamefont {K.~S.}\ \bibnamefont
  {Viswanathan}},\ }\href {\doibase 10.1103/PhysRevD.58.106006} {\bibfield
  {journal} {\bibinfo  {journal} {Phys. Rev. D}\ }\textbf {\bibinfo {volume}
  {58}},\ \bibinfo {pages} {106006} (\bibinfo {year} {1998})},\ \Eprint
  {http://arxiv.org/abs/hep-th/9805145} {arXiv:hep-th/9805145} \BibitemShut
  {NoStop}%
\bibitem [{\citenamefont {Kirsch}(2006)}]{Kirsch:2006he}%
  \BibitemOpen
  \bibfield  {author} {\bibinfo {author} {\bibfnamefont {I.}~\bibnamefont
  {Kirsch}},\ }\href {\doibase 10.1088/1126-6708/2006/09/052} {\bibfield
  {journal} {\bibinfo  {journal} {JHEP}\ }\textbf {\bibinfo {volume} {09}},\
  \bibinfo {pages} {052} (\bibinfo {year} {2006})},\ \Eprint
  {http://arxiv.org/abs/hep-th/0607205} {arXiv:hep-th/0607205} \BibitemShut
  {NoStop}%
\end{thebibliography}%

\end{document}